\documentclass[prb,twocolumn,notitlepage,superscriptaddress]{revtex4-2}
\usepackage{amsmath,amssymb}
\usepackage[hidelinks,colorlinks,linkcolor=blue,
citecolor=blue,urlcolor=blue]{hyperref}
\usepackage{graphicx,siunitx}
\usepackage[dvipsnames]{xcolor}

\usepackage{pifont}
\newcommand{\xmark}{\ding{55}}%

\usepackage[normalem]{ulem}

\usepackage{xcolor}
\DeclareUnicodeCharacter{3000}{\textcolor{red}{BAD!!}}

\newcommand{\be}{\begin{equation}}
\newcommand{\ee}{\end{equation}}
\newcommand{\bea}{\begin{equation} \begin{aligned}}
\newcommand{\eea}{\end{aligned} \end{equation} }
\newcommand{\bi}{\begin{itemize}}
\newcommand{\ei}{\end{itemize}}

\renewcommand{\be}{\beta}

\newcommand{\bpm}{\begin{pmatrix}}
\newcommand{\epm}{\end{pmatrix}}
\newcommand{\eps}{\epsilon}

\renewcommand{\th}{\theta}

\newcommand{\del}{\partial}

\newcommand{\Tr}{\text{Tr} \ }

\DeclareRobustCommand{\Sec}[1]{Sec.~\ref{#1}}

\DeclareRobustCommand{\App}[1]{App.~\ref{#1}}

\DeclareRobustCommand{\Tab}[1]{Table~\ref{#1}}

\DeclareRobustCommand{\Fig}[1]{Fig.~\ref{#1}}
\DeclareRobustCommand{\Figs}[2]{Figs.~\ref{#1} and \ref{#2}}
\DeclareRobustCommand{\Eq}[1]{Eq.~(\ref{#1})}

\usepackage{amsmath,amsfonts,amssymb,amsthm,epsfig,array}
\usepackage{dsfont}
\usepackage{slashed}
\usepackage{graphics}

\usepackage{float}

\usepackage{verbatim}%\begin{comment} ... \end{comment}

\usepackage{color}

\usepackage{tabularx}
\usepackage[mathscr]{euscript}
\usepackage{mathtools}

\newcommand{\bsl}[1]{\boldsymbol{#1}}
\newcommand{\mbf}[1]{\boldsymbol{#1}}

\renewcommand{\mod}{\,\mathrm{mod}\,}

\newcommand{\bra}[1]{\langle #1|}
\newcommand{\ket}[1]{|#1 \rangle}

\newcommand{\ii}{\mathrm{i}}

\newcommand{\dsR}{\mathbb{R}}

\newcommand{\U}{\mathrm{U}}

\newcommand{\eqnref}[1]{Eq.\,\eqref{#1}}
\newcommand{\figref}[1]{Fig.\,\ref{#1}}

\newcommand{\tabref}[1]{Tab.\,\ref{#1}}
\newcommand{\secref}[1]{Sec.\,\ref{#1}}
\newcommand{\appref}[1]{Appendix\,\ref{#1}}
\newcommand{\refcite}[1]{Ref.\,\cite{#1}}

\newcommand{\eq}[1]{\begin{equation} #1 \end{equation}}

\newcommand{\eqa}[1]{\begin{align}\begin{split} #1 \end{split}\end{align}}

\usepackage{environ}
\NewEnviron{eqs}{%
\begin{equation}\begin{split}
\BODY
\end{split}\end{equation}
}

\let\oldAA\AA
\renewcommand{\AA}{\text{\normalfont\oldAA}}

\newcommand{\ie}{{\emph{i.e.}}}
\newcommand{\eg}{{\emph{e.g.}}}
\newcommand{\TR}{\mathcal{T}}
\newcommand{\cc}{\mathcal{K}}

\newcommand{\C}{\mathcal{C}}

\newcommand{\tmt}{\text{$t$MoTe$_2$}}

\newcommand{\Ch}{\text{Ch}}
\newcommand{\K}{\text{K}}

\newtheorem{proposition}{Proposition}
\newcommand{\propref}[1]{Prop.\,\ref{#1}}

\newtheorem{definition}{Definition}

\begin{document}

\title{Fractional Chern Insulators vs. Non-Magnetic States in Twisted Bilayer MoTe$_2$}

\author{Jiabin Yu}
\affiliation{Department of Physics, Princeton University, Princeton, New Jersey 08544, USA}

\author{Jonah Herzog-Arbeitman}
\affiliation{Department of Physics, Princeton University, Princeton, New Jersey 08544, USA}

\author{Minxuan Wang}
\affiliation{National High Magnetic Field Laboratory, Tallahassee, Florida, 32310, USA}

\author{Oskar Vafek}
\affiliation{National High Magnetic Field Laboratory, Tallahassee, Florida, 32310, USA}
\affiliation{Department of Physics, Florida State University, Tallahassee, Florida 32306, USA}

\author{B. Andrei Bernevig}
\affiliation{Department of Physics, Princeton University, Princeton, New Jersey 08544, USA}
\affiliation{Donostia International Physics Center, P. Manuel de Lardizabal 4, 20018 Donostia-San Sebastian, Spain}
\affiliation{IKERBASQUE, Basque Foundation for Science, Bilbao, Spain}

\author{Nicolas Regnault}
\affiliation{Laboratoire de Physique de l'Ecole normale sup\'{e}rieure, ENS, Universit\'{e} PSL, CNRS, Sorbonne Universit\'{e}, Universit\'{e} Paris-Diderot, Sorbonne Paris Cit\'{e}, 75005 Paris, France}
\affiliation{Department of Physics, Princeton University, Princeton, New Jersey 08544, USA}

\date{\today}

\begin{abstract}

Fractionally filled Chern bands with strong interactions may give rise to fractional Chern insulator (FCI) states, the zero-field analogue of the fractional quantum Hall effect. Recent experiments have demonstrated the existence of FCIs in twisted bilayer MoTe$_2$ without external magnetic fields---most robust at $\nu=-2/3$---as well as Chern insulators (CIs) at $\nu=-1$. Although the appearance of both of these states is theoretically natural in an interacting topological system, experiments repeatedly observe nonmagnetic states (lacking FCIs) at $\nu=-1/3$ and $-4/3$, a puzzling result which has not been fully theoretically explained.
In this work, we perform Hartree-Fock and exact diagonalization calculations to test whether the standard MoTe$_2$ moir\'e model with the (greatly varying) parameter values available in the literature can reproduce the non-magnetic states at $\nu=-1/3$ and $-4/3$ in unison with the FCI at $\nu=-2/3$ and CI state at $\nu = -1$. We focus on the experimentally relevant twist angles and, crucially, include remote bands. We find that the parameters proposed in [Wang et al. (2023)] can nearly capture the experimental phenomena at $\nu=-1/3,-2/3,-1,-4/3$ simultaneously, though the predicted ground states at $\nu=-1/3$ are still mostly fully-spin-polarized and a larger dielectric constant $\epsilon>10$ than is typical of hexagonal boron nitride (h-BN) substrate $\epsilon\sim 6$ is required. Our results show the importance of remote bands in identifying the competing magnetic orders and lay the groundwork for further study of the realistic phase diagram.
\end{abstract}

\maketitle

\section{Introduction}

As proposed more than a decade ago~\cite{neupert, sheng, regnault}, interactions can induce fractional Chern insulator (FCI) states when nearly flat Chern bands~\cite{Tang11,Sun2011} (in zero magnetic field) are fractionally filled.
Owing to the development of moir\'e materials~\cite{Cao2018TBGMott,Cao2018TBGSC}, there has been extensive theoretical~\cite{Bergholtz13, Parameswaran13,Abouelkomsan2020FCIMoire,Vishwanath2020FCITBG,RepellinFCITBG,Parker2021fieldtunedFCITBG,Wilhelm2021FCITBG,Stern2021FCITBG,Li2021FCItTMD,Crepel2022FCITMD,Fu2022FCIDirac,Abouelkomsan2023FCITBG,Cano2023FCItTMD,2023arXiv230907222W}
interest in the realization of FCI states in this platform. First steps towards FCIs were taken through the experimental observation of fractional quantum Hall (FQH)-like states in twisted bilayer graphene~\cite{Xie2021TBGFCIfiniteB} and in bilayer graphene–hexagonal boron nitride (h-BN) heterotructures~\cite{Young2018FCIBLGMoire}, both -- especially the latter -- requiring a large external magnetic field.
Remarkably, recent experiments~\cite{cai2023signatures,zeng2023integer,park2023observation,Xu2023FCItMoTe2} demonstrated true FCI states without external magnetic fields in twisted bilayer MoTe$_2$ (\tmt) at fractional fillings $\nu=-2/3,-3/5$, as well as a Chern insulator (CI) exhibiting the integer quantum anomalous Hall effect at $\nu=-1$, spurring immediate theoretical interest ~\cite{reddy2023fractional,wang2023fractional,Wu2023IntegerFillingstMoTe2,Dong2023CFLtMoTe2,Goldman2023CFLtMoTe2,MacDonald2023MagicAngletTMD,Reddy2023GlobalPDFCI,Song2023tTMDFCI,Xu2023MLWOFCItTMD,Zaletel2023tMoTe2FCI,Liu2023HFtMoTe2num2}.
($\nu$ is the electron filling measured from the charge neutrality point.)
The FCI and CI states observed in the experiments have been reproduced theoretically~\cite{reddy2023fractional,wang2023fractional,Dong2023CFLtMoTe2,Goldman2023CFLtMoTe2,Reddy2023GlobalPDFCI,Xu2023MLWOFCItTMD,Zaletel2023tMoTe2FCI} based on the continuum model proposed in \refcite{Wu2019TIintTMD}.

\refcite{Wu2019TIintTMD} already showed that the valley-filtered bands exhibit a Chern number at the single-particle level. Fractionally filling them ought to produce FCIs, as Refs.\,\cite{neupert, sheng, regnault} showed at multiple rational fillings of Chern bands. However, there is a puzzling and essential departure from the typical FCI phase diagram~\cite{neupert, sheng, regnault}. Unlike at $\nu=-2/3$, experiments~\cite{cai2023signatures,zeng2023integer,park2023observation,Xu2023FCItMoTe2} repeatedly found non-magnetic (and non-FCI) states at $\nu=-1/3$, although several existing theoretical works~\cite{wang2023fractional,Dong2023CFLtMoTe2} have predicted the experimentally non-existent $\nu=-1/3$ FCI at experimentally relevant angles.
The magnetic ordering at $\nu=-1/3$ in comparison to $\nu=-2/3$ was studied in Refs.\,\cite{reddy2023fractional,wang2023fractional}, but remote bands were neglected.
Robust non-magnetic states were also found at $\nu=-4/3$ in experiments, despite $\nu=-4/3$ being the particle-hole (PH) partner of $\nu=-2/3$ within the lowest-energy \emph{spinful} bands. The ground state at $\nu=-4/3$ has not been theoretically studied.
The full phase diagram poses an unavoidable question for theory: can the known continuum model of \refcite{Wu2019TIintTMD} with the widely used parameter values~\cite{reddy2023fractional,wang2023fractional,Dong2023CFLtMoTe2,Goldman2023CFLtMoTe2,Reddy2023GlobalPDFCI,Xu2023MLWOFCItTMD,Zaletel2023tMoTe2FCI} capture the nonmagnetic states at $\nu=-1/3,-4/3$, FCI states at $\nu=-2/3$, and the CI state at $\nu=-1$?
A more comprehensive test of the model should include the FCI state at $\nu=-3/5$. (The appearance of the FCI state at $\nu=-3/5$ is consistent with the initial theory~\cite{neupert, sheng, regnault}.) Nevertheless, in this paper we limit ourselves to the more robust and pronounced $\nu=-2/3$ FCI and focus on explaining this key puzzle and other essential features of the theoretical model.

\begin{table}[h]
\centering
\begin{tabular}{c|cccc|c}
$\nu =$&  $-1/3$   & $-2/3$ & $ -1$ &$  -4/3$ & $\eps$  \\
\hline
Parameters $\backslash$ experiment & N-M  & FCI & CI & N-M & \\
\hline
\hline
Ref.\,\cite{reddy2023fractional} param. (1BPV) & $\sim \checkmark^{\text{\cite{reddy2023fractional}}}$  & $\checkmark^{\text{\cite{reddy2023fractional}}}$ & $\checkmark$ & \xmark & [5,6.25] \\
\hline
Ref.\,\cite{reddy2023fractional} param. (2BPV) & $\checkmark$  & ? & $\checkmark^{\text{\cite{Zaletel2023tMoTe2FCI}}}$ & \xmark?  & [5,6.25]\\
\hline
Ref.\,\cite{wang2023fractional} param. (1BPV) & $\sim$ $\checkmark^{\text{\cite{wang2023fractional}}}$  & $\checkmark^{\text{\cite{wang2023fractional}}}$ & $\checkmark$  & \xmark & [10,25] \\
\hline
Ref.\,\cite{wang2023fractional} param. (2BPV) & $\sim\checkmark$  &   $\checkmark$ & $\checkmark$  & $\approx\checkmark$  & [10,25]  \\
\hline
\end{tabular}
\caption{Minimal phase diagram at $\th \sim 3.7^\circ$ (determined with HF at $\nu = -1$ and ED otherwise). N-M stands for non-magnetic. Checkmarks indicate a match to experiment for a considerable part of values of $\eps$ in the indicated range, and crosses indicate a mismatch. The question mark indicates an uncertainty due to finite size effects, ``$\sim \checkmark$" for $\nu=-1/3$ means a fully spin polarized state with much smaller spin gap compared to $\nu = -2/3$,  and ``$\approx \checkmark$" for $\nu=-4/3$ means a ground state with small (but nonzero) total spin.  Citations to earlier work are shown if they are consistent with our work in the same range of parameters. We only consider calculations including \emph{both} valleys, a prerequisite for a prediction of magnetism. Note that \refcite{Xu2023MLWOFCItTMD} performed two-band ED calculations at $\nu = -2/3$ on a related model with different irreps in the remote bands due to a truncation of the Wannier basis, but only in the fully-spin-polarized sector.}
\label{tab:resultstable}
\end{table}

To answer the question, we perform self-consistent Hartree-Fock (HF) and exact-diagonalization (ED) calculations for the experimental angles~\cite{cai2023signatures,zeng2023integer,park2023observation,Xu2023FCItMoTe2} $\theta\in [3.5^\circ,4.0^\circ]$. We only focus on the two mainly used sets of parameter values in Refs.\,\cite{wang2023fractional,reddy2023fractional}. As will be discussed, the two other sets of parameters proposed in the literature~\cite{Wu2019TIintTMD,Xu2023MLWOFCItTMD} are similar to \refcite{reddy2023fractional}. Crucially, we include remote bands in our calculation, but restrict ourselves to at most two bands per valley where the model is thought to be most faithful~\cite{wang2023fractional,reddy2023fractional}.

Our two-band-per-valley (2BPV) HF calculation shows that a CI at $\nu=-1$ occurs for dielectric constants $\epsilon>10$ using the parameter values in \refcite{wang2023fractional}. (This is in contrast to \refcite{Dong2023CFLtMoTe2} which finds a CI at $\epsilon=8$.) The required value of the dielectric constant is considerably larger than $\epsilon\sim 6$ estimated from the h-BN substrate~\cite{laturia_dielectric_2018}. On the other hand, similar to what was found in \refcite{Zaletel2023tMoTe2FCI}, dielectric constants as small as $\epsilon=5$ yield CIs for the parameters of \refcite{reddy2023fractional}.
On the other hand, our 1BPV HF gives CI at $\nu=-1$ for dielectric constants as small as $\epsilon=5$ for both sets of parameters~\cite{reddy2023fractional,wang2023fractional}; the difference in the 1BPV and 2BPV HF results (most notably for the parameters in \cite{wang2023fractional}) indicates the importance of the remote bands.
Despite this enlarged CI region at $\nu=-1$, we do not see clear signatures that the parameters in \refcite{reddy2023fractional} can fully capture the experimental phenomena at the fractional fillings of interest, as discussed in the following.

We further perform 1BPV and 2BPV ED calculations for the parameters of Refs.\,\cite{reddy2023fractional,wang2023fractional} for the $\epsilon$ range that covers the CI region at $\nu=-1$ in HF calculations (\ie, $\epsilon\in[5,25]$ for \refcite{reddy2023fractional} and $\epsilon\in[10,25]$ for \refcite{wang2023fractional}).

1BPV ED calculations were performed in Refs.\cite{reddy2023fractional,wang2023fractional}, and weaker ferromagnetism was found at $\nu=-1/3$ than at $\nu=-2/3$, for values of $\epsilon$ yielding FCIs at $\nu=-2/3$.
In our 1BPV ED calculations (consistent with \refcite{reddy2023fractional} and \refcite{wang2023fractional} on the same system sizes), the parameter values in \refcite{reddy2023fractional} require a much stronger interaction ($ \epsilon \in [5,6.25]$) to give FCIs at $\nu=-2/3$ than the parameters in \refcite{wang2023fractional} where $ \epsilon $ can be as large as $25$.
Refs.\,\cite{reddy2023fractional,wang2023fractional} found the weaker ferromagnetism at $\nu=-1/3$ concomitant with the  $\nu=-2/3$ FCI for 1BVP calculations on 12 unit cells. We further consider 15, 18, and 24 unit cell systems. Though larger system sizes do reduce the difference in ferromagnetic stability between $\nu=-1/3$ and $\nu=-2/3$, we find that the finite size effects are not significant enough to eliminate the trend that ferromagnetism at $\nu=-1/3$ is considerably weaker than that at $\nu=-2/3$.
On the other hand, our 1BPV ED results show that when the interaction is strong enough to give FCI at $\nu=-2/3$, $\nu=-4/3$ (not considered in earlier work) is ferromagnetic and always has a similar or stronger magnetic stability as $\nu=-2/3$, which is inconsistent with the experiments. To resolve this inconsistency, we will include remote bands to the ED calculation.

Our 2BPV ED calculations include one more remote band per valley. Generally, we find that their inclusion typically reduces the spin gaps at $\nu=-1/3$ and $-2/3$ (or even causing a sign change) and changes the spin of the ground state at $\nu=-4/3$.
We note that although the remote bands were included in the study of $\nu=-2/3$ in the fully-spin-polarized sector in \refcite{Xu2023MLWOFCItTMD}, the effects of remote bands on the spin gap have not yet been studied.
In particular for the parameter values in \refcite{wang2023fractional}, when FCIs appear at $\nu=-2/3$, the spin gap at $\nu=-1/3$, albeit still fully-spin-polarized across most of the region, is dramatically reduced, and nearly spin-unpolarized states (\ie, small total spin) are now favored at $\nu=-4/3$. Taken together, the inclusion of remote bands greatly ameliorates the agreement with the experimental phase diagram for the parameters in \refcite{wang2023fractional}, although we caution that this is contingent on a dielectric constant $\epsilon > 10$.

On the other hand, the same 2BPV ED calculations for the parameter values in \refcite{reddy2023fractional} show large-spin states are energetically favored at $\nu=-4/3$ and (for the system sizes currently accessible to us) have even larger spin-1 gap than that at $\nu=-2/3$ for $\epsilon \in [5,6.25]$. Thus, we have not seen clear signatures that the parameters in \refcite{reddy2023fractional} can capture the significant experimental difference in magnetism between $\nu=-2/3$ and $\nu=-4/3$ for the experimental angles $\theta\in[3.5^\circ,4.0^\circ]$ for the system sizes currently accessible to us, although $\nu=-1/3$ becomes nonmagnetic in agreement with experiments. A summary of our results in comparison to experiment and the literature may be found in \Tab{tab:resultstable}.

These results show the need for the analog of the Landau level mixing to explain the experimental facts. Since the single-particle bands have different Chern number sequences depending on the twist angle, the physics of $t$MoTe$_2$ is likely to be very rich.

In the rest of this paper, we review the single-particle model in \secref{sec:SP} and discuss the interaction and the PH symmetry in \secref{sec:interaction}. We further discuss the HF, 1BPV ED, and 2BPV ED results in \secref{sec:HFsection}, \secref{sec:1BPV_ED} and \secref{sec:2BPV_ED}, respectively. We eventually conclude the paper in \secref{sec:conclusion}, and provide more details in a series of appendices.

\section{Single-Particle model}
\label{sec:SP}

\begin{figure}
\centering
\includegraphics[width=\columnwidth]{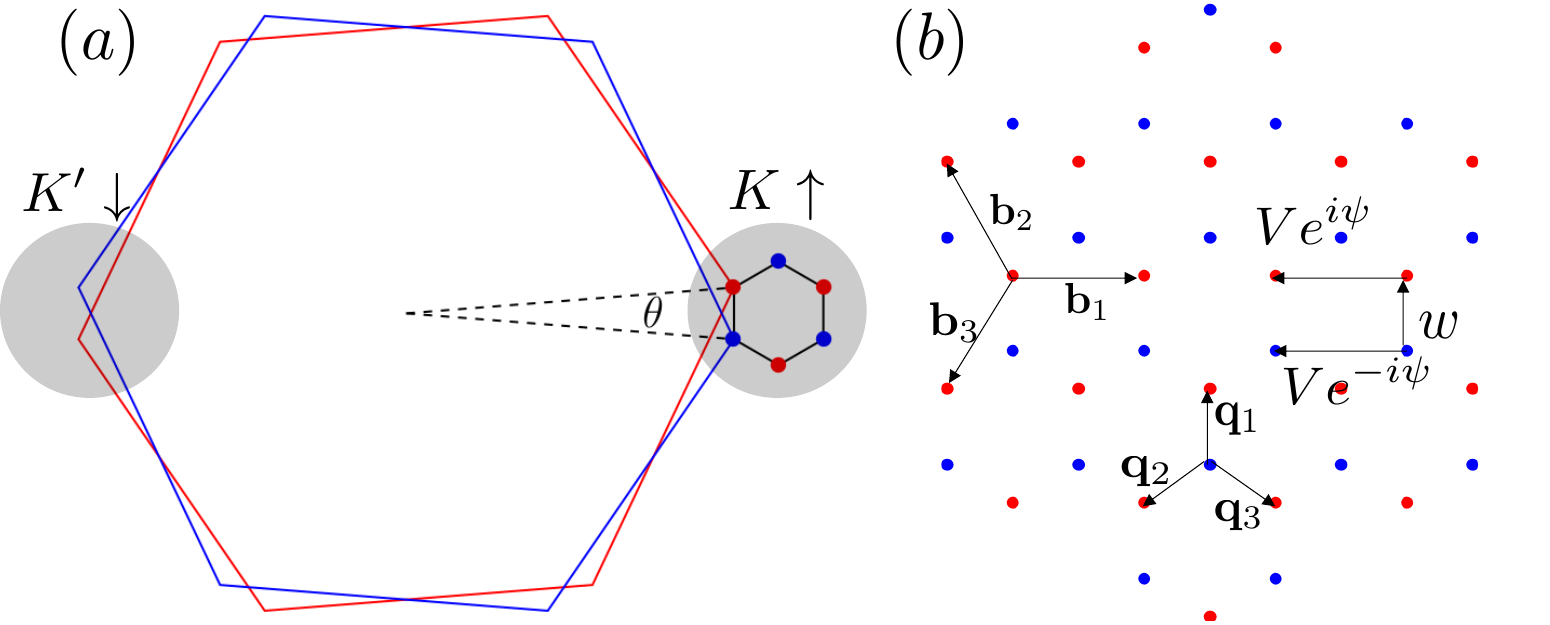}
\caption{(a) Twisted BZs of the TMD bilayer, with oppositely spin-polarized low-energy states at $K$ and $K'$. Twisting creates a moir\'e BZ shown (for a single valley) in (b) where blue/red represent moir\'e reciprocal lattice points in the top/bottom layers. The momentum space hoppings $V e^{\pm  i \psi},w$ of the moir\'e potential are marked.
}
\label{fig:modelmain}
\end{figure}

The moir\'e physics of the TMDs was originally proposed by \refcite{Wu2019TIintTMD} to be captured by a two-valley continuum model formed from low-energy states at the $K$ and $K'$ valleys, where the strong spin-orbit coupling splits the spin up and down states. By time-reversal, the states at valley $K$ have the opposite spin as the states at valley $K’$. Thus spin is effectively locked to valley, and the only global symmetries of the model are charge $U(1)$ and  valley/spin ($S_z$) $U(1)$. Because MoTe$_2$ is a semiconductor, the effective single-particle moir\'e Hamiltonian around the quadratic band edge takes the form (in the $K$ valley)
\bea
\label{eq:Hkmacdonald}
H_K(\bsl{r}) &= \bpm \frac{\hbar^2\nabla^2}{2m_*} + V_{+}(\bsl{r}) & t(\bsl{r}) \\ t^*(\bsl{r}) &  \frac{\hbar^2\nabla^2}{2m_*} + V_{-}(\bsl{r}) \epm
\eea
where the matrix acts on the \emph{electron} wavefunction $(\psi_+(\bsl{r}),\psi_-(\bsl{r}))^T$ with support on momenta near $R(\pm \frac{\th}{2}) \bsl{K}$ in the top/bottom layer (see \Fig{fig:modelmain}a), $V_\pm(\bsl{r})$ are moir\'e potentials in the top/bottom layer, and $t(\bsl{r})$ is the inter-layer moir\'e tunneling. Here $R(\theta)$ is a rotation matrix, $m_*$ is the effective mass, and $\bsl{K} = \frac{4\pi}{3a_0} \hat{x}$ is the untwisted TMD $K$ point with $a_0 = 3.52\AA$~\cite{Wu2019TIintTMD}. Note that $\nabla^2$ is unbounded below and \Eq{eq:Hkmacdonald} describes the electron valence bands below the charge neutrality point. Spinful time-reversal yields the $K'$ valley model $H_{K'}(\bsl{r}) = H_K(\bsl{r})^*$.

\subsection{Moir\'e Potentials}

Ref.~\cite{Wu2019TIintTMD} obtained expressions for $V_\pm(\bsl{r})$ and $t(\bsl{r})$ by using the lowest order symmetry-allowed Fourier modes with parameters fit to bilayer band structures.
The point group symmetry of AA-stacked MoTe$_2$~\cite{PhysRevLett.108.196802,PhysRevB.94.155425} is generated by mirror $z\rightarrow -z$, three-fold rotation about the $z$-axis  $C_{3z}$, two-fold rotation about the $y$-axis $C_{2y}$, and spinful time-reversal $\mathcal{T}$. Note that $C_{3z}$ acts locally on each layer, whereas $C_{2y}\mathcal{T}$ flips the layers. Twisting the layers in opposite directions preserves only the $C_{3z}$ and $C_{2y}\mathcal{T}$ symmetries at the $K$ point, yielding the magnetic point group $3m'$ as the intra-valley symmetry group of the moir\'e model $H_K(\bsl{r})$. Their explicit representations are given in \App{app:SP}.

The inter-layer coupling $t(\bsl{r})$ can be expanded in terms of the moir\'e scattering momentum
\bea
\bsl{q}_1 &= R(\th/2)\bsl{K} - R(-\th/2)\bsl{K} = 2 \sin \frac{\th}{2} \, |\bsl{K}| \hat{y}  \\
\eea
and $\bsl{q}_{i+1} = R(2\pi/3) \bsl{q}_i$ \cite{2011PNAS..10812233B}. Keeping only the lowest order Fourier modes gives
\bea
t(\bsl{r}) = w \sum_{n=1}^3 e^{i \bsl{q}_n \cdot \bsl{r}}
\eea
where $C_{3z}$ ensures all three modes have equal amplitude $w$. The overall phase of $t(\bsl{r})$ is not observable since it depends on the arbitrary relative phase choice between the top and bottom layers, and is not constrained by $C_{2y}\mathcal{T}$. By convention we take $w < 0$.

Next we consider the potentials $V_\pm(\bsl{r})$. Because they are intra-layer, they are supported on the moir\'e reciprocal lattice spanned by the vectors
\bea
\label{eq:b_1_b_2}
\mbf{b}_1 = \mbf{q}_3 - \mbf{q}_2 = \sqrt{3}|\mbf{q}_1|\hat{x}, \quad \mbf{b}_2 = R(2\pi/3) \mbf{b}_1
\eea
and we define $\mbf{b}_3 = -(\mbf{b}_1+\mbf{b}_2) = R(2\pi/3) \mbf{b}_2$ for convenience. The lowest order reciprocal lattice points are $\mbf{0}$ and the first shell $\mbf{G} = R(2\pi n/6)\mbf{b}_1$ for $n= 0,\dots, 5$. Keeping only these harmonics and imposing $C_{3z}, C_{2y}\mathcal{T}$, the most general form of the potential can be written in terms of an amplitude $V$ and phase $\psi$ as
\bea
V_\pm(\mbf{r}) =2V \sum_{n=1}^3 \cos(\mbf{b}_n \cdot \mbf{r}\pm \psi)
\eea
up to an overall chemical potential. \Fig{fig:modelmain} depicts these next-nearest neighbor hoppings on the momentum space lattice~\cite{2018arXiv180710676S}.

Within the approximation of keeping the lowest harmonics only, an emergent intra-valley pseudo-inversion symmetry of $H_K(\mbf{r})$ appears (see \App{app:SP}). Although the  pseudo-inversion can be broken by $C_{3z}, C_{2y}\mathcal{T}$-preserving higher order terms (see \App{app:SP}), we focus on the original, lowest order model. As the model stands, the pseudo-inversion symmetry relates $\mbf{k}$ and $-\mbf{k}$ so that the band structures in the $K$ and $K'$ valleys are identical. The \emph{ab-initio} calculations in \refcite{wang2023fractional} are in agreement with this to good accuracy.

\begin{table}[h]
\centering
\begin{tabular}{c|ccccc}
MoTe$_2$& $m_* \text{ ($m_e$)}$ & $w \text{ (meV)}$  &  $V \text{ (meV)} $ &$ \psi$  \\
\hline
\text{Ref.\,}\cite{reddy2023fractional} & $0.62$ & $-13.3$  &  $11.2$& $-91^\circ$ \\
\hline
\text{Ref.\,}\cite{Wu2019TIintTMD} & $0.62$ & $-8.5$  &  $8$& $-89.6^\circ$ \\
\hline
\text{Ref.\,} \cite{2023arXiv230809697X} &$  0.62$ &$  -11.2 $& $9.2 $ & $-99^\circ$ \\
\hline
\text{Ref.\,}\cite{wang2023fractional} &$  0.6$ &$  -23.8 $& $20.8 $ & $-107.7^\circ$ \\
\end{tabular}
\caption{Proposed parameter values for the single-particle moir\'e model \Eq{eq:Hkmacdonald}.
}
\label{tab:paramtable}
\end{table}

\begin{figure}[h]
\centering
\includegraphics[width=\columnwidth]{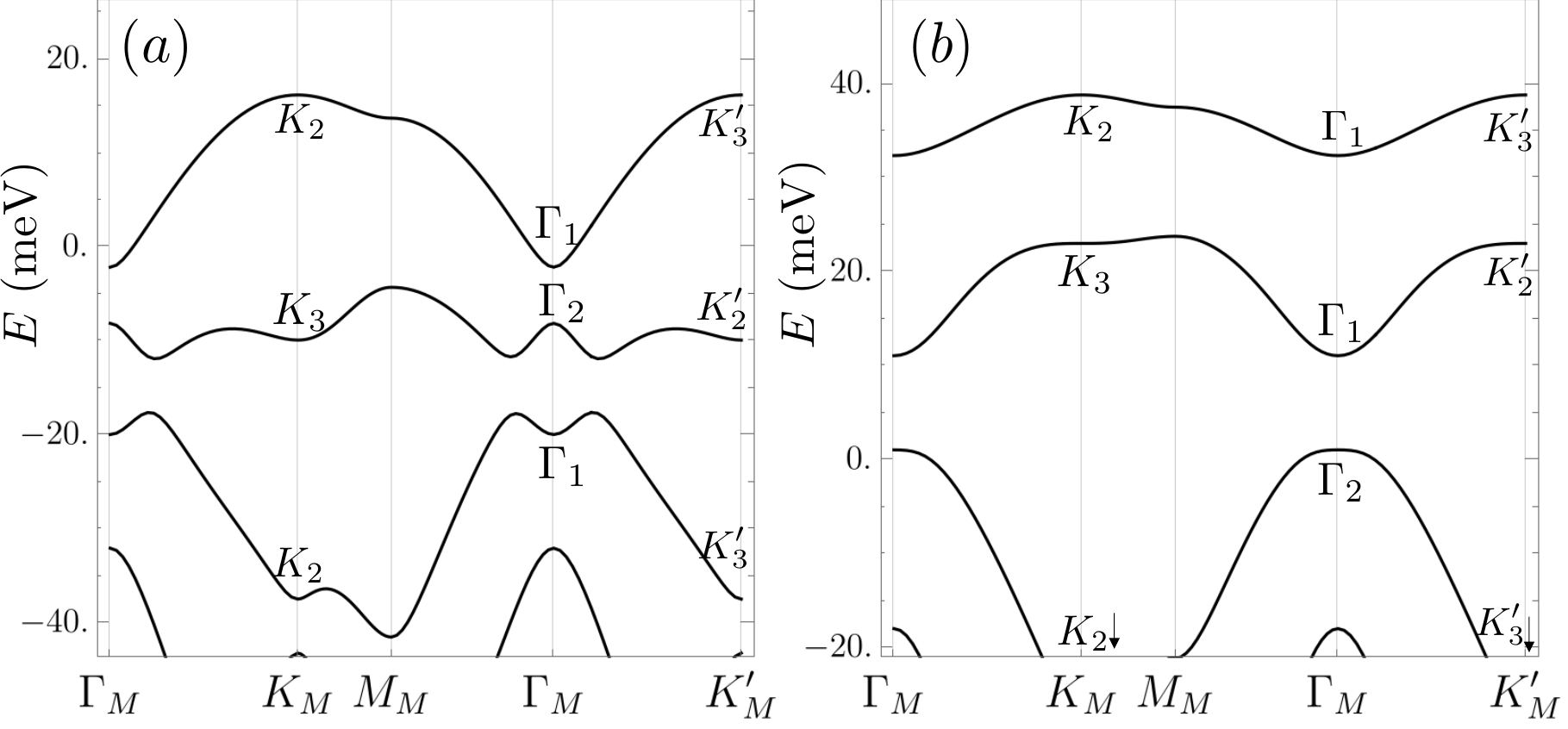}
\caption{Single-particle band structures at $\th = 3.7^\circ$ for the parameters in Ref.\,\cite{reddy2023fractional} (a) and Ref.\,\cite{wang2023fractional} (b), along with their $C_{3z}$ irreps. In \Fig{fig:phasediagrammain}, $(a)$ is marked with a circle and $(b)$ with a star.}
\label{fig:bandsmain}
\end{figure}

\begin{figure}[H]
\centering
\includegraphics[width=0.9\columnwidth]{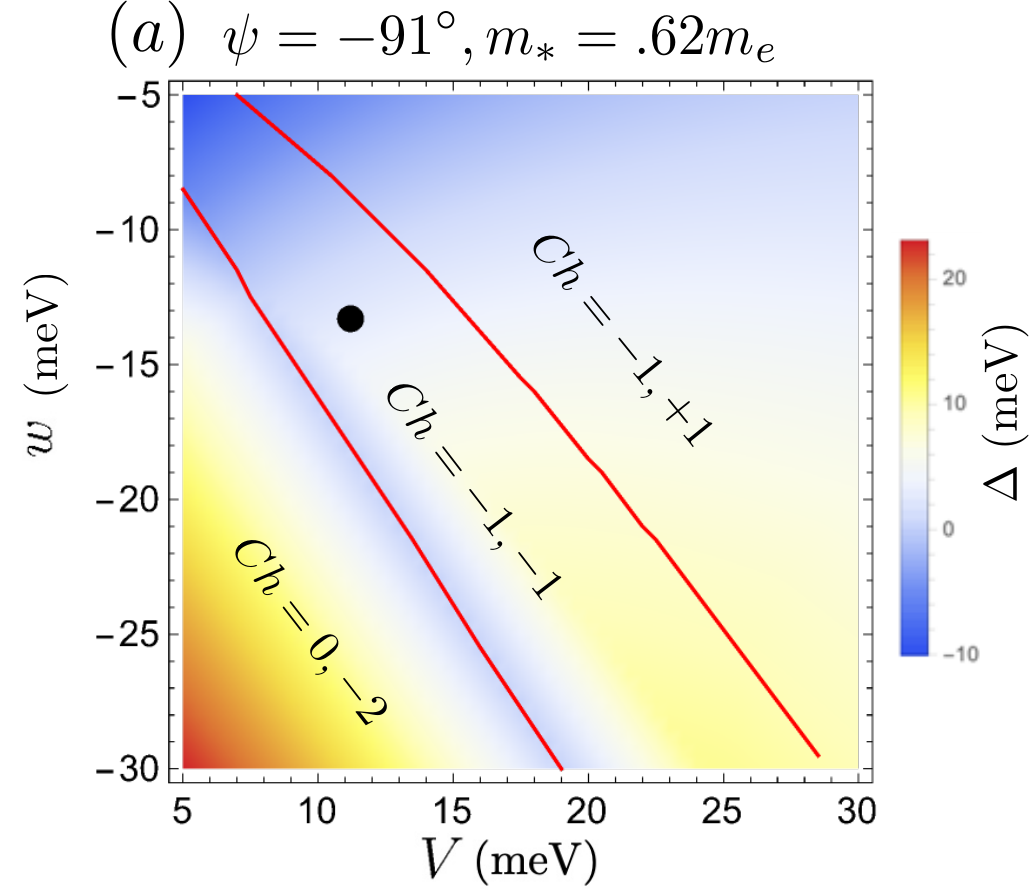}
\includegraphics[width=0.9\columnwidth]{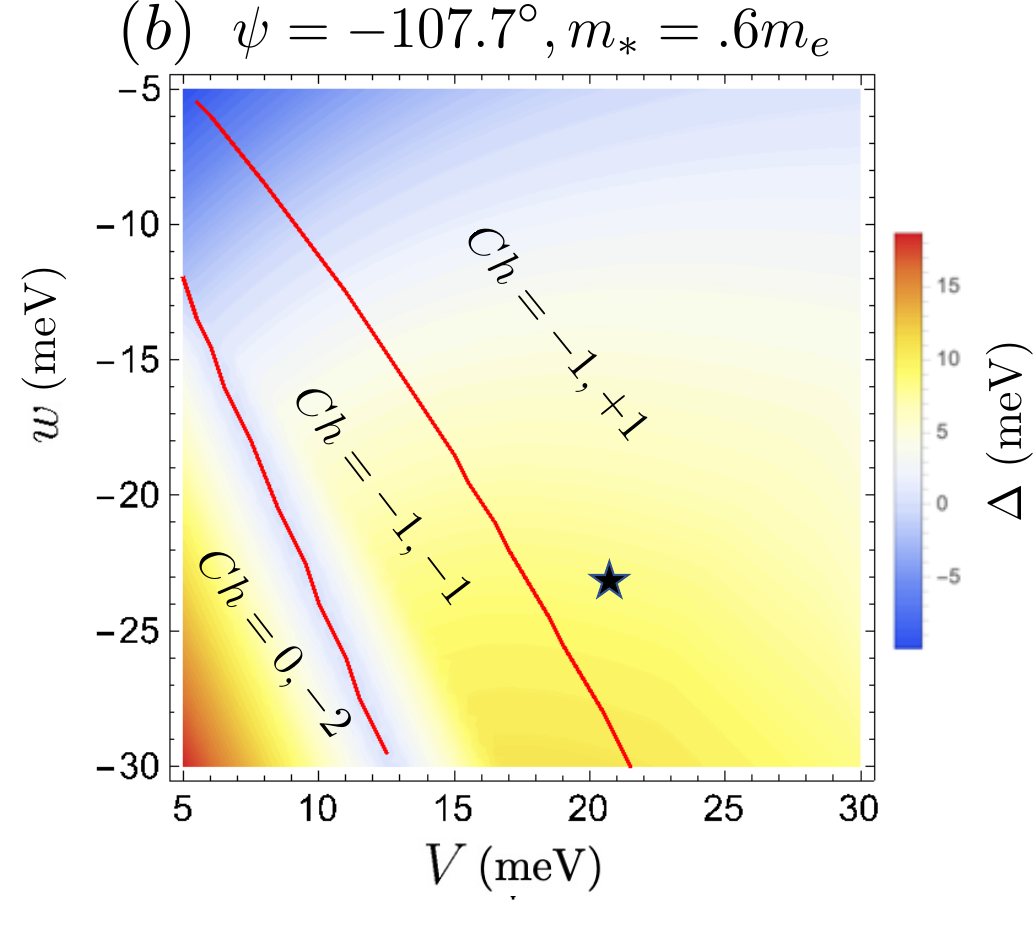}
\caption{Single-Particle phase diagram at $\th = 3.7^\circ$. The parameter space of the model contains three topologically distinct phases delineated by direct gap closings (red lines). In each phase, the Chern numbers $\Ch$ of the two bands nearest the Fermi energy are labeled and the indirect gap $\Delta$ right below the highest valence band is colored. (a) Parameters of Ref. \cite{reddy2023fractional} are marked with a black dot. (b) Parameters of Ref. \cite{wang2023fractional} are marked with a black star.}
\label{fig:phasediagrammain}
\end{figure}

The values of model parameters have been determined by matching the DFT band structure (either at AA stacking~\cite{Wu2019TIintTMD} or at commensurate twist angles~\cite{wang2023fractional,reddy2023fractional,2023arXiv230809697X}), which are summarized in \tabref{tab:paramtable}. Among the four sets of parameter values in \tabref{tab:paramtable}, the first three sets of parameter values are fairly similar: they exhibit dispersive bands structures (see \figref{fig:bandsmain}a), peaked quantum geometry, and $\Ch = -1$ topology in both bands closest to the charge neutrality per valley (see \App{app:SP}). In contrast, the 4th set of parameter values in \tabref{tab:paramtable} is qualitatively different, showing a flatter active band (see \figref{fig:bandsmain}b), nearly ideal quantum geometry \cite{Roy14,2023arXiv230401251E,2022arXiv220915023L,PhysRevLett.127.246403,PhysRevResearch.5.023167,MacDonald2023MagicAngletTMD}, and opposite Chern numbers $\Ch = \pm 1$ for the bands closest to the charge neutrality per valley.
Therefore, in the following, we will focus on the 1st (Ref.\,\cite{reddy2023fractional}) and 4th (Ref.\,\cite{wang2023fractional}) sets of parameter values in \tabref{tab:paramtable}. Since experiments repeatedly show FCIs around the twist angle $3.7^\circ$, we also restrict our attention to the range of angles $\th \in [3.5^\circ,4^\circ]$.
\Fig{fig:phasediagrammain} maps out the single-particle phase diagram of the active bands across $V,w$ for two values of $\psi$ proposed in Refs. \cite{wang2023fractional,reddy2023fractional}, finding three topologically distinct regimes when the lowest two bands are considered.  \Fig{fig:phasediagrammain} shows that the parameters of Ref.\,\cite{reddy2023fractional} and Ref.\,\cite{wang2023fractional} fall in different phases, and can be expected to yield different many-body phase diagrams.

We note that \refcite{reddy2023fractional} and \refcite{wang2023fractional} extracted their parameters from different commensurate twist angles, $4.4^\circ$ and $3.89^\circ$ respectively. Thus the parameters used in \refcite{wang2023fractional} are perhaps more reliable for describing the experiments at the angle of $3.7^\circ$. Though, it is likely that the model parameters depend on twist angle due to relaxation effects \cite{PhysRevB.107.075123,PhysRevB.107.075408}.

\subsection{Topology}

The full magnetic wallpaper group (in one valley) is $p31m'$ (or 157.55 in the BNS setting) generated by $C_{3z}, C_{2y}\mathcal{T}$ and moir\'e translations. Band structures can be labeled by their $C_{3z}$ eigenvalues at the high-symmetry points $\Gamma_M, K_M, K'_M$ in the moir\'e Brilloin zone (BZ) (for a fixed valley). \Fig{fig:bandsmain} depicts two example band structures. The active (highest valence) band in both cases has the same eigenvalues, but the lower two bands differ by a band inversion at the $\Gamma_M$ point, as can be seen from the interchange of the $\Gamma_1$ and $\Gamma_2$ irrep. A minimal character table for the spin-less irreps (we choose $C_{3z}^3 = +1$ since it is intra-valley) at the $C_{3z}$-symmetric points in one valley (giving the notation in real space and momentum space) is
\bea
\begin{array}{c|c|cc}
\mbf{r} & \mbf{k} &1& C_{3z}  \\
\hline
A &\Gamma_1, K_1, K'_1 &1&1 \\
{}^2E &\Gamma_2, K_2, K'_3 &1 & e^{\frac{2\pi i}{3}}  \\
{}^1E &\Gamma_3, K_3, K'_2 & 1 & e^{- \frac{2\pi i}{3}} \\
\end{array} \ .
\eea
We see that all irreps are one-dimensional, showing that there are no symmetry-protected degeneracies. Secondly, the Chern number $\Ch$ of an individual band is related to the symmetry eigenvalues by \cite{PhysRevB.86.115112}
\bea
\label{eq:symC3}
\exp \left[ \frac{2\pi i}{3} \Ch \right]= \prod_{\mbf{k}=\Gamma_M,K_M,K'_M} \mathcal{D}_{\mbf{k}}[C_{3z}]
\eea
where $\mathcal{D}_{\mbf{k}}[g]$ is the representation of $g$ at $\mbf{k}$, which can be used to determine $\Ch \mod 3$. Lastly, we remark that this space group hosts decomposable elementary band representations (see \App{app:SP}) which generically give rise to $\Ch = \pm 1 \mod 3$ Chern bands \cite{PhysRevE.96.023310,2018PhRvB..97c5139C}. Thus there is a symmetry principle \cite{2017Natur.547..298B} that predicts the appearance of topology.

\subsection{Remarks on the Continuous Model}
\label{sec:applicability_continuous_model}

According to previous DFT calculations~\cite{wang2023fractional,reddy2023fractional}, the current single-particle model~\refcite{Wu2019TIintTMD} can faithfully capture the top two bands in the $K,K'$ valleys, but fails farther from the charge neutrality.
Specifically, the \emph{ab-initio} calculations \cite{wang2023fractional} show low-energy moir\'e bands from the TMD $\Gamma$ valley appearing right below the second highest $K,K'$-valley bands, which are absent from the Hamiltonian. Therefore, in the following sections, we include at most two bands per valley in our calculations. The $\Gamma$ bands will be studied in our forthcoming work.

The single-particle Hamiltonian can be written in second quantized notation as
\bea
H_0 &= \int d^2 r \sum_{\eta,ll'} c^\dagger_{\eta,l,\bsl{r}}   \left[ H_{\eta} (\bsl{r}) \right]_{ll'}  c_{\eta,l',\bsl{r}}
\eea
where $c_{\eta, l, \bsl{r}}^\dagger$ creates an electron at position $\bsl{r}$ in the $l$th layer and the $\eta$ valley. The global symmetries of $H_0$, as well as the interacting Hamiltonians to be introduced in the following section, are
\bea
N &= \int d^2 r \sum_{\eta,l} c_{\eta, l, \bsl{r}}^\dagger c_{\eta, l, \bsl{r}} \\
S_z &= \frac{1}{2} \int d^2 r \sum_{\eta,l} \eta c_{\eta, l, \bsl{r}}^\dagger c_{\eta, l, \bsl{r}} \\
\eea
where $\eta = \pm$ denotes the valley $\eta K$ quantum number, which is equivalently the spin quantum number $S_z = \eta/2$.
In this work, we only consider states with $S_z\geq 0$ since those with negative $S_z$ are related by time-reversal.

\section{Interaction and particle-hole symmetry}
\label{sec:interaction}

We now consider the many-body Hamiltonian obtained from adding the Coulomb interaction to the single-particle model discussed above. We first discuss the different interactions suggested in the literature and then clarify their behavior under particle-hole transformations.

\subsection{Interaction among electrons and holes.}
\label{sec:inteh}
In the previous theoretical works
~\cite{Li2021FCItTMD,reddy2023fractional,wang2023fractional,Dong2023CFLtMoTe2,Goldman2023CFLtMoTe2,Reddy2023GlobalPDFCI,Xu2023MLWOFCItTMD,Zaletel2023tMoTe2FCI} on FCIs in {\tmt}, two different interactions are used. One choice is the Coulomb interaction among holes, employed by \refcite{reddy2023fractional} in the full Hamiltonian
\eqa{
\label{eq:H_h_unprojected}
H_{h} & = \int d^2 r \sum_{\eta,ll'} \widetilde{c}^\dagger_{\eta,l,\bsl{r}} \left[ - H_{\eta} (\bsl{r}) \right]_{ll'} \widetilde{c}_{\eta,l',\bsl{r}} \\
& + \frac{1}{2}\sum_{l \eta,l' \eta'} \int d^2 r d^2 r' V(\bsl{r}-\bsl{r}')  \widetilde{c}^\dagger_{\eta, l, \bsl{r}} \widetilde{c}^\dagger_{\eta', l', \bsl{r}'} \widetilde{c}_{\eta', l', \bsl{r}'} \widetilde{c}_{\eta, l, \bsl{r}}  \ ,
}
where $\widetilde{c}_{\eta,l,\bsl{r}}^\dagger$ creates a hole at position $\bsl{r}$ in the $l$th layer and $\eta$ valley and $V(\bsl{r})$ is the double-gated screened Coulomb potential with gate distance $\xi$ and dielectric constant $\eps$ (see \appref{app:int} for explicit formulas). Note that $\widetilde{c}_{\eta,l,\bsl{r}}^\dagger$ and $c_{\eta,l,\bsl{r}}$ are related by a complex conjugation operator, \ie, $\widetilde{c}_{\eta,l,\bsl{r}}^\dagger = \cc c_{\eta,l,\bsl{r}} \cc^{-1}$ with $\cc$ the complex conjugate (see \App{app:int}). $H_h$ annihilates the charge neutrality point $\nu = 0$, which is the hole vacuum. This is because the charge neutrality ($\ket{\nu = 0}$) is the product state of all valence electron bands occupied (all holes unoccupied) which is the maximally filled state in the Hilbert space:
\bea
\label{eq:CNPvacuum}
\ket{\nu = 0} = \prod_{\bsl{r},\eta,l} c^\dag_{\eta,l,\bsl{r}} \ket{0} \ ,
\eea
with $c_{\eta,l,\bsl{r}} \ket{0} = 0$,
leading to  $\widetilde{c}_{\eta, l, \bsl{r}} \ket{\nu = 0} = 0$.

The other choice is the Coulomb interaction among electrons, which is used in Refs.\,\cite{Li2021FCItTMD,wang2023fractional}:
\eqa{
\label{eq:H_e}
H_{e} & = \int d^2 r \sum_{\eta,ll'} c^\dagger_{\eta,l,\bsl{r}}   \left[ H_{\eta} (\bsl{r}) \right]_{ll'}  c_{\eta,l',\bsl{r}} \\
& + \frac{1}{2}\sum_{l\eta,l' \eta'} \int d^2 r d^2 r' V(\bsl{r}-\bsl{r}')   c^\dagger_{\eta, l, \bsl{r}} c^\dagger_{\eta', l', \bsl{r}'} c_{\eta', l', \bsl{r}'} c_{\eta, l, \bsl{r}}  \ ,
}
recalling that $c_{\eta, l, \bsl{r}}^\dagger$ creates an electron at position $\bsl{r}$ in the $l$th layer and the $\eta$ valley. As is apparent from \Eq{eq:CNPvacuum}, the electron interaction $H_e$ does not annihilate $\ket{\nu = 0}$, the state around which the single-particle Hamiltonian was derived,

In our HF and ED calculations, we use a projected Hamiltonian in the band basis. In this basis, $H_h$ can be written as (see \appref{app:int})
\eqa{
\label{eq:H_h}
H_{h} & = - \sum_{\bsl{k},\eta,n}\widetilde{\gamma}^\dagger_{\eta,n,\bsl{k}} \widetilde{\gamma}_{\eta,n,\bsl{k}}  E_{\eta,n}(\bsl{k}) + H_{h,int} \ ,
}
where
\eqa{
H_{h,int}  & = \frac{1}{2} \sum_{\bsl{k},\bsl{k}',\bsl{q}} \sum_{\eta,\eta'} \sum_{m, m', n', n}V_{\eta \eta', m m' n' n}(\bsl{k},\bsl{k}',\bsl{q})\\
&  \qquad \times \widetilde{\gamma}^\dagger_{\eta,m,\bsl{k}+\bsl{q}} \widetilde{\gamma}^\dagger_{\eta',m',\bsl{k}'-\bsl{q}} \widetilde{\gamma}_{\eta', n',\bsl{k}'}
\widetilde{\gamma}_{\eta,n,\bsl{k}}\ ,
}
and $V_{\eta \eta', m m' n' n}(\bsl{k},\bsl{k}',\bsl{q})$ labels the projected interaction and $m,m',n',n$ are the band indices. Note that $-E_{\eta,n}(\bsl{k})$ is the $n$th hole band energy (which is bounded below) in valley $\eta$, and
$\widetilde{\gamma}^\dagger_{\eta,n, \bsl{k}}$ creates a hole in the $n$th hole band in $\eta$ valley at $\bsl{k}$.
The projection of $H_{e, int}$ is analogous (see \appref{app:int}).

We note that if we do not perform the band projection and keep the whole continuous Hamiltonian, $H_h$ (\eqnref{eq:H_h_unprojected}) and $H_e$ (\eqnref{eq:H_e}) only differ by a chemical potential-like term that is proportional to the particle number operator, which means that the unprojected \eqnref{eq:H_h_unprojected} and \eqnref{eq:H_e} should give the same results (up to a overall constant) in a fixed particle-number sector.
However, in practice, we always keep a finite set of bands, and the projected $H_e$ and $H_h$ are not guaranteed to give the same results in a fixed particle-number sector anymore (see \appref{app:int}).
In the rest of this paper, we will always use the projected Hamiltonains with a finite set of bands.
\secref{sec:HFsection} will show that the $H_{e}$ and $H_{h}$ give different results at $\nu=-1$ in projected HF calculation.

\subsection{Particle-Hole Symmetry}

\label{sec:PH_sym}

Any single Landau level (LL) under the projected Coulomb interaction possesses exact particle-hole (PH) symmetry, which was widely used in study of the fractional quantum Hall effect~\cite{Girvin1984PHFQHE}. However, this is not the case in general for FCIs~\cite{Moessner2013FCIPHBreaking}, and we will give a brief summary of the role of particle-hole symmetry in $H_h$ and $H_e$.

In close analogy to the PH symmetry of a single LL, we consider the following intra-band PH transformation

\eq{
\label{eq:PHtransformation}
\C \gamma^\dagger_{\eta,n,\bsl{k}} \C^{-1} = \gamma_{\eta,n,\bsl{k}}\ ,
}
where $\C$ is an anti-unitary operator.
Under this PH transformation, $H_{e}$ acquires two extra terms:
\eqa{
\label{eq:H_e_PH}
\C H_{e} \C^{-1} &  = H_e  + \sum_{\eta,\bsl{k},n}\gamma_{\eta,n,\bsl{k}}^\dagger \gamma_{\eta,n,\bsl{k}} (-2 E_{\eta,n}(\bsl{k})) \\
& + \sum_{\eta,\bsl{k},n m}\gamma_{\eta,n,\bsl{k}}^\dagger \gamma_{\eta,m,\bsl{k}} \epsilon_{\eta,n m}(\bsl{k}) + const.\ ,
}
where $(-2 E_{\eta,n}(\bsl{k}))$ accounts for the sign flipping of the single-particle dispersion, and $\epsilon_{\eta,n m}(\bsl{k}) $ is an effective one-body term arising from the interaction matrix elements. Its explicit form is given in \appref{app:int}. The hole Hamiltonian $H_h$ transforms similarly, with $\gamma_{\eta,n,\bsl{k}} \to \widetilde{\gamma}_{\eta,n,\bsl{k}}$ and $E_{\eta,n}(\bsl{k})\rightarrow - E_{\eta,n}(\bsl{k})$ in \eqnref{eq:H_e_PH}.
The case of single LL can be recovered by dropping the valley and band index, and then PH invariance follows from the flat kinetic energy ($E(\bsl{k}) =\text{const.}$) and the flat one-body term $(\epsilon(\bsl{k})= \text{const.})$ due to the unique LL wavefunction \cite{2023arXiv230400866M} which has uniform quantum geometry~\cite{2022PhRvL.129g6401H}.

Note that, in the transformation of $H_e$ and $H_h$, we have kept the number of bands in the projected Hamiltonian fully general.
As mentioned at the end of \Sec{sec:inteh}, only a finite number of bands (and spin-valley flavors) can be kept in numerical calculations, and importantly the PH transformation acts differently depending on this truncation.
For instance, one could keep only one active band in a single valley (corresponding a fully-spin-polarized Hilbert space); in this case, the PH transformation transforms the electron filling $\nu$ to $-1-\nu$ (\eg, $-1/3$ to $-2/3$, recalling that $\nu$ is the electron filling measured from the charge neutrality point).
This case has been studied in \refcite{Reddy2023GlobalPDFCI}, which shows that the PH symmetry is not exact (e.g. the spectra at $-1/3$ and $-2/3$ will not be identical) but might be approximately correct outsides the experimental range of angles considered in this work.

In this work, we will consider PH symmetry breaking beyond the fully-spin-polarized sector, which is a basic requirement for studying the non-magnetized states observed in experiment.
In particular, if we keep one band in each valley, PH maps $\nu$ to $-2-\nu$ (\eg, $-2/3$ to $-4/3$). For this case it is then possible to have approximate PH symmetry between $-2/3$ and $-4/3$.
However, if the remote bands are heavily involved in the low-energy physics, they might strongly break any approximate PH symmetry (if it exists) between $\nu$ and $-2-\nu$, akin to the situation of the fractional quantum Hall effect at $\nu=5/2$~\cite{Rezayi2023FQHE5over2,Loic2023FQHE5over2}.
Strictly speaking, if we keep two valleys and two bands per valley, we can only have exact PH symmetry between $\nu$ and $-4-\nu$.
We will see in \secref{sec:1BPV_ED}-\ref{sec:2BPV_ED} that these effects are indeed relevant for the parameter values in \refcite{wang2023fractional}.

\section{Self-Consistent Hartree-Fock Calculations at $\nu=-1$}
\label{sec:HFsection}

\begin{figure}[t]
\centering
\includegraphics[width=\columnwidth]{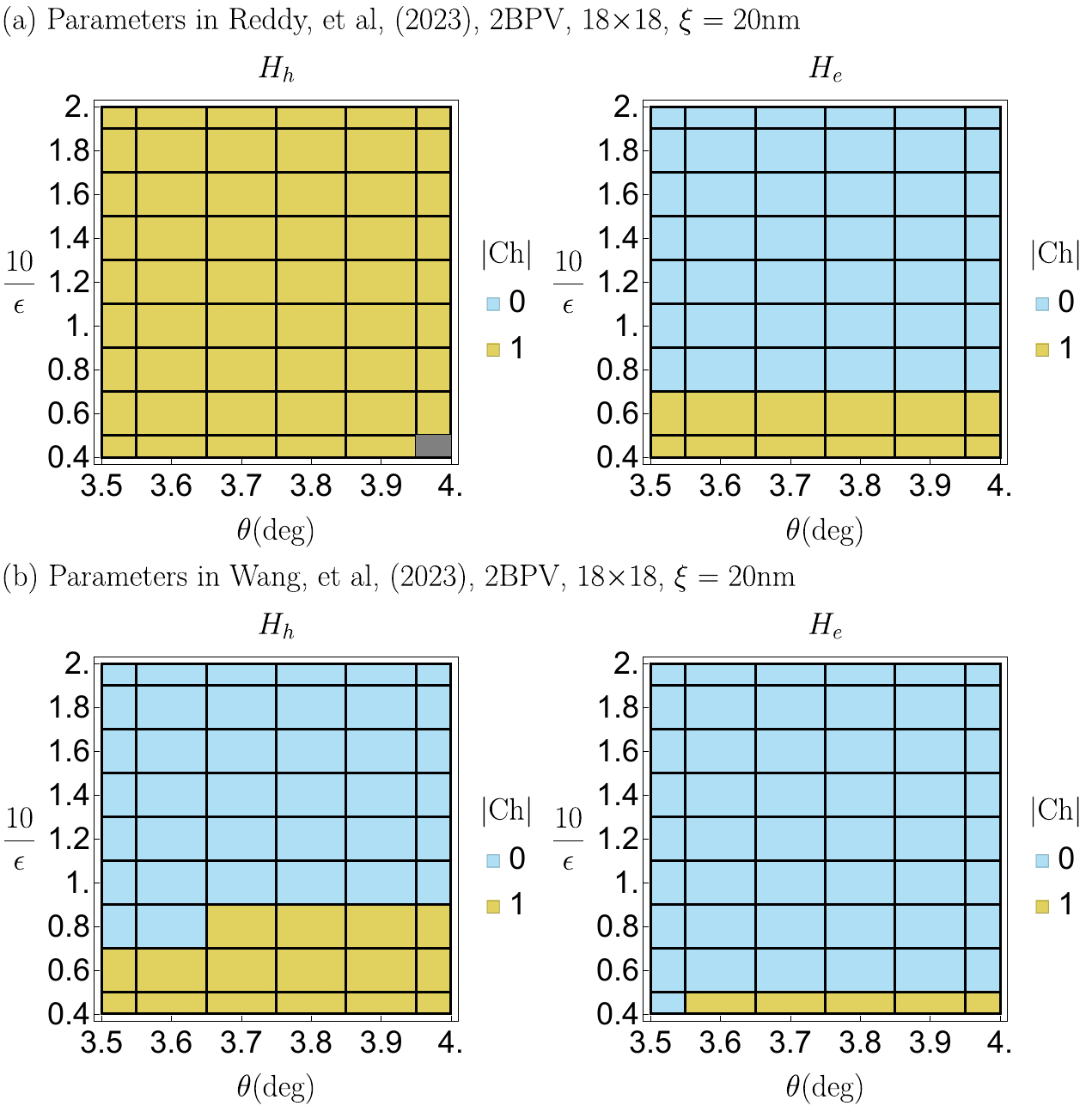}
\caption{
The 2BPV Hartree-Fock results for $\xi=20nm$ at $\nu=-1$.
$H_h$ and $H_e$ refer to the Hamiltonian used, and $\Ch$ refers to the Chern number.
The gray region in (a) indicates a \emph{non}-ferromagnetic ground state, which is inter-valley coherent translationally-breaking state with wavevector $\K_M$ and which has zero Chern number.
$18\times 18$ labels the system size.
In general, the system size $L_1\times L_2$ means that the momenta included in the calculation are $(n/L_1) \bsl{b}_1 + (m/L_2) \bsl{b}_2$ with $n=0,...,L_1-1$ and $m=0,...,L_2-1$, where $\bsl{b}_1$ and $\bsl{b}_2$ are defined in \eqnref{eq:b_1_b_2}.
}
\label{fig:HF_x18y18_main}
\end{figure}

\begin{figure*}[t]
\centering
\includegraphics[width=1.9\columnwidth]{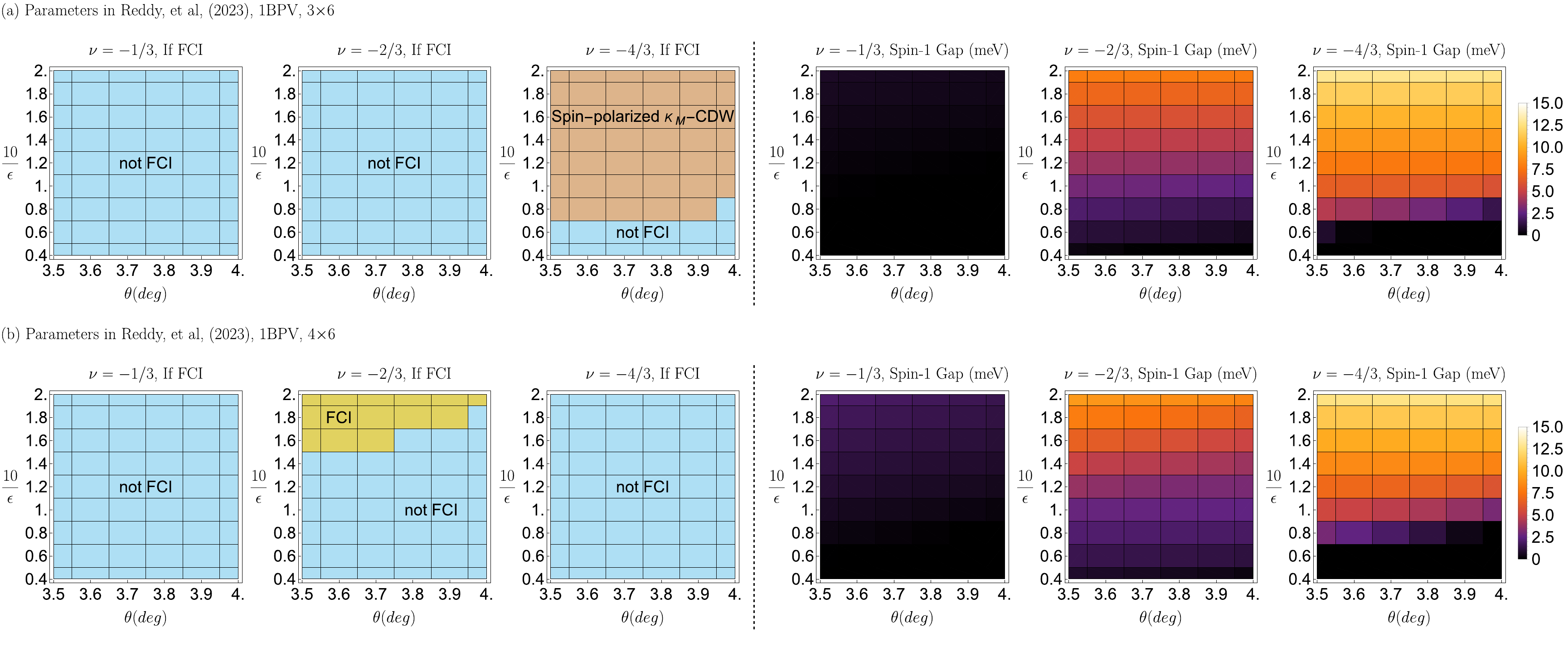}
\caption{1BPV ED calculations on $H_h$ for parameter values in \refcite{reddy2023fractional} on $3\times 6$ $(a)$ and $4 \times 6$ $(b)$  systems.
$3\times 6$ in (a) and $4 \times 6$ in (b) stand for different system sizes.
In the left most three figures of (a,b), green (``FCI") labels the region that satisfies the criterion in \propref{prop:FCI}, and blue (``not FCI") means that we do not see clear signatures of FCI or maximally-spin-polarized CDW.
For $\nu=-4/3$ in (a) for the system size of $3\times 6$, 1BPV results show that the system is in a maximally-spin-polarized $K_M$-CDW phase for relatively large interaction (brown) and for experimental angles $\theta\in[3.5^\circ,4.0^\circ]$; the $K_M$-CDW phase is suppressed for the system size of $4\times 6$ since its momentum mesh do not include the $K_M$ or $K_M'$ point.
The rightmost three figures of (a,b) give the spin-1 gaps, which are shown with the same color scale for all plots.
If the spin-1 gap is negative, it is set to zero in the plot.
}
\label{fig:ED_1BPV_Main_Reddy_etal}
\end{figure*}

\begin{figure*}[t]
\centering
\includegraphics[width=1.9\columnwidth]{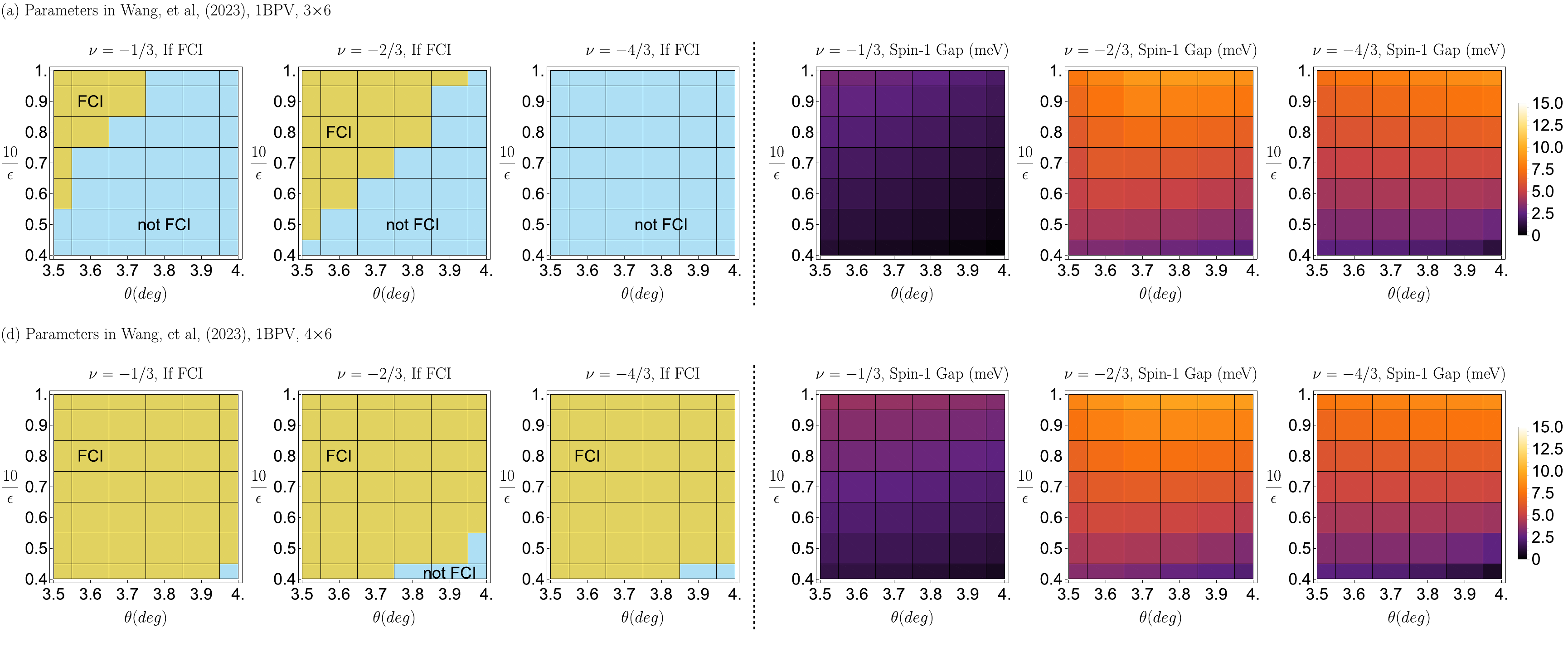}
\caption{1BPV ED calculations on $H_h$ for parameter values in \refcite{wang2023fractional} on $3\times 6$ $(a)$ and $4 \times 6$ $(b)$  systems.
In the left most three figures of (a,b), green (``FCI") labels the region that satisfies the criterion in \propref{prop:FCI}, and blue (``not FCI") means that we do not see clear signatures of FCI or maximally-spin-polarized CDW.
The rightmost three figures of (a,b)given the spin-1 gaps, which are shown with the same color scale for all plots.
If the spin-1 gap is negative, it is set to zero in the plot.
In the ``not FCI" region, the properties of the ground state(s) (other than the momentum and spin polarization) are not conclusively determined.
}
\label{fig:ED_1BPV_Main_Wang_etal}
\end{figure*}

Self-consistent HF calculations at $\nu=-1$ for {\tmt} have been performed in previous works~\cite{Li2021FCItTMD,Dong2023CFLtMoTe2,Zaletel2023tMoTe2FCI,Xu2023FCItMoTe2}, among which Refs.\,\cite{Wu2023IntegerFillingstMoTe2,Zaletel2023tMoTe2FCI,Xu2023FCItMoTe2} demonstrated different phases induced by band mixing.
In this section, we provide HF results at $\nu=-1$ for sets of parameter values in \refcite{reddy2023fractional} and \refcite{wang2023fractional} for both hole and electron interactions $H_h$ and $H_e$.
These HF calculations will serve to demonstrate the difference between the interactions $H_h$ and $H_e$ and will allow us to screen out the choices of parameters for which no CI appears at $\nu=-1$ in preparation for more time-consuming ED calculations.

As discussed in \secref{sec:applicability_continuous_model}, we include two bands per valley---the two highest electron bands or two lowest hole bands per valley---as done in \refcite{Zaletel2023tMoTe2FCI}.
In the end of this section, we will mention the 1BPV HF results.
Our 2BPV HF calculation is different from the 3BPV HF calculations done in \refcite{Wu2023IntegerFillingstMoTe2}, and different from the 2BPV calculation in \refcite{Xu2023FCItMoTe2} which replaces part of the states in the second band in each valley by those in the third band due to Wannierization.

Our 2BPV HF results are summarized in \figref{fig:HF_x18y18_main} where we show the phase diagram for the experimentally-relevant angles $\theta\in[3.5^\circ, 4.0^\circ]$ and a range of dielectric constants $\epsilon\in[5,25]$.
For both interactions and both parameter values, we find that the ground state is ferromagnetic (fully-spin-polarized) over nearly the entire phase diagram. The only exception is the appearance of an inter-valley coherent (IVC) translation-breaking state with wavevector $\K_M$, called IVC-$\K_M$~\cite{Zaletel2023tMoTe2FCI}, which appears at $\epsilon$ close to 25 for $H_h$ (grey region in \figref{fig:HF_x18y18_main}(a)).
We note that although the IVC-$\K_M$ state becomes the ground state only at $\theta=4.0^\circ$ in our $(\theta,\epsilon)$ mesh for $H_h$ in \figref{fig:HF_x18y18_main}(a), it has energy very close to the $|\Ch|=1$ state for smaller angles at $\epsilon=25$, \eg, for $(\theta,\epsilon)=(3.8^\circ,25)$, the $|\Ch|=1$ state only wins over the IVC-$\K_M$ state by about $0.1$meV per unit cell.
Such IVC-$\K_M$ states were also found in \refcite{Zaletel2023tMoTe2FCI} for $\epsilon$ close to 25.
Intuitively, the translationally-invariant IVC state that mixes the lowest-energy bands in the two valleys is not energetically favored because it is topologically obstructed---the opposite Chern numbers of the lowest-energy bands in the two valleys require the order parameter to have zeros \cite{Murakami2003BerryPhaseMSC,Li2018WSMObstructedPairing,Zaletel2020AHTBG,PhysRevB.107.L201106}.

Comparing $H_h$ to $H_e$ in Figs.\,\ref{fig:HF_x18y18_main}$(a,b)$, the hole interaction $H_h$ in \eqnref{eq:H_h} gives a much larger region where $|\Ch| = 1$, \ie, the CI phase, than $H_e$. The $\Ch = 0 $ region that occurs at stronger interaction for $H_e$ in \figref{fig:HF_x18y18_main}(a) and for both $H_h$ and $H_e$ in \figref{fig:HF_x18y18_main}(b) comes from the band inversion between the active band and the remote band in one valley (typically, a single band inversion at $\K_M$ or $\K_M'$ due to spontaneous breaking of $C_{2y}\TR$ in each valley). This effect is only possible due to the inclusion of remote bands, and underscores their importance.
Since only $|\Ch| = 1$ is consistent with the CI state observed at $\nu=-1$~\cite{cai2023signatures,zeng2023integer,park2023observation,Xu2023FCItMoTe2}, our HF results suggest that the hole interaction $H_h$ is more suitable to realize CI effect at $\nu=-1$ than $H_e$.

Notably for $H_h$, \figref{fig:HF_x18y18_main}(a) shows that the $|\Ch| = 1$ region persists all the way to $\epsilon=5$ using the parameters of \refcite{reddy2023fractional} (consistent with the results of \refcite{Zaletel2023tMoTe2FCI}) but only to $\epsilon=10$ for the parameter of \refcite{wang2023fractional} (different from the results in \refcite{Dong2023CFLtMoTe2} which find a CI at $\epsilon=8$). Lastly, to check the dependence on screening length, we calculated the phase diagrams for two other screening lengths, $\xi=60nm$ and $\xi=150nm$, and find no changes in the CI region for the $(\theta,\epsilon)$ mesh that we choose though the gap of the CI does increase as the screening length increases (see \appref{app:HF_results}).

Physically, $H_e$ is not normal ordered with respect to the charge neutrality point. For this reason, we will not use $H_e$ but will focus on $H_h$ in the remainder of the paper.
The CI regions then roughly correspond to $\epsilon\in [5,25]$ for the parameter values in \refcite{reddy2023fractional} and $\epsilon\in [10,25]$ for the parameter values in \refcite{wang2023fractional}.

We also performed the 2BPV HF calculations with the different values of $V$ and $w$ for the twist angle $3.7^\circ$ and for typical values of $\epsilon$, and found that the CI state is robust as long as $V,w$ do not differ too much from those in \refcite{wang2023fractional} and \refcite{reddy2023fractional} (see \appref{app:HF_results}).

As a comparison to the 2BPV case, we perform 1BPV HF calculations for $H_h$ at $\xi=20$nm which give CI at $\nu=-1$ for the entire $\theta\in[3.5^\circ,4.0^\circ]$ and $\epsilon\in[5,25]$ for both sets of parameters~\cite{reddy2023fractional,wang2023fractional} (see \appref{app:HF_results}); CI at $\nu=-1$ was also found in the 1BPV HF calculation in \refcite{Li2021FCItTMD} outsides the experimental angle range.
So the 1BPV calculations missed the IVC-$\K_M$ states for parameters in \refcite{reddy2023fractional} and the $\Ch=0$ states for the parameters in \refcite{wang2023fractional} in the 2BPV results (\figref{fig:HF_x18y18_main}) indicating the importance of the remote bands.

To summarize, at the level of 2BPV HF at $\nu = -1$, the CI regions of the phase diagram differ considerably for the parameters of \refcite{reddy2023fractional} and \refcite{wang2023fractional}. To realize the CI phase for the parameters in \refcite{wang2023fractional}, roughly $\eps \in [10,25]$ is required, compared to the larger range $\eps \in [5,25]$ for \refcite{reddy2023fractional}. We will show in the following section that ED calculations at $\nu=-1/3,-2/3, -4/3$ show a better match to experiment using the parameters of \refcite{wang2023fractional} compared to \refcite{reddy2023fractional}.

\section{One-band-per-valley Exact Diagonalization Calculations}
\label{sec:1BPV_ED}

1BPV ED calculations have been performed at fractional fillings in previous works~\cite{reddy2023fractional,wang2023fractional,Dong2023CFLtMoTe2,Goldman2023CFLtMoTe2,Reddy2023GlobalPDFCI,Xu2023MLWOFCItTMD,Zaletel2023tMoTe2FCI} restricting to the top electron band in each valley.
In particular, Refs.\,\cite{reddy2023fractional,wang2023fractional} studied the magnetic properties at $\nu=-1/3,-2/3$ within 1BPV ED calculations and found weaker ferromagnetism at $\nu=-1/3$ than that at $\nu=-2/3$. The difference in the magnetic properties between $\nu=-2/3$ and $\nu=-4/3$ was not studied.

In general, a measure of magnetic stability is the spinful gap~\cite{reddy2023fractional,wang2023fractional}, \ie, the energy difference between the lowest-energy state in the  $S_z\neq S_{\rm max}$ sectors and the lowest-energy state with the same particle number in the $S_z = S_{\rm max}$ (ferromagnetic) sector (former minus latter). Recall that $S_z$ is the total spin of the ground state for which we only consider $S_z\geq 0$ owing to the TR symmetry; note that $S_{\rm max}$ is the spin of the maximally-spin-polarized state in the Hilbert space.
If the spinful gap is positive (negative), the ground state is (is not) maximally-spin-polarized. When the ground states are maximally-spin-polarized, a larger spinful gap means increased stability.
We note that in the 1BPV case, a maximally-spin-polarized state of a given filling $\nu$ is a fully-spin-polarized state only if $\nu>-1$.
For $\nu<-1$, we cannot have all holes in one valley since one valley has only a single band. Taking $\nu=-4/3$ as an example, even a maximally-spin-polarized state has a quarter of the total number of holes in the other valley, \ie, it is not a fully-spin-polarized state.

In \appref{app:1BPV_ED_results:spin1spinful}, we compare the spinful gaps to the spin-1 gaps for the system sizes of $3\times 4$, $3\times 5$ and $3\times 6$ for both sets of parameters, and find that the spin-1 gap is typically equal to or similar to the spinful gap, where the spin-1 gap is the gap between the lowest-energy state in $S_z=S_{\rm max}-1$ sector and that in the $S_z=S_{\rm max}$ sector (former minus latter) with the same particle numbers.
This trend is also shown in \refcite{reddy2023fractional} for $\nu>-1$ for system sizes of 12 and 15 unit cells, while we further show it for sizes reaching $3\times 6$.
In the rest of this section, we will thus use the spin-1 gap as a proxy for the spinful gap, unless specified otherwise.

Our 1BPV ED calculations were done for four system sizes: $3\times 4$, $3\times 5$, $3\times 6$ and $4\times 6$.
We show the main results on $3\times 6$ and $4\times 6$ in Figs.\,\ref{fig:ED_1BPV_Main_Reddy_etal}-\ref{fig:ED_1BPV_Main_Wang_etal}, while leaving $3\times 4$ and $3\times 5$ results to \appref{app:1BPV_ED_results}.
Throughout the work, we determine the FCI region by the following criterion:
\begin{proposition}
\label{prop:FCI}
The system is in the FCI region if (i) the three lowest states are maximally-spin-polarized, (ii) the momenta of the three lowest states match the momenta of FCI\cite{1961AnPhy..16..407L,2014PhRvB..89o5113W}, and (iii) the spread of the three lowest states is smaller than the gap between the 3rd lowest state and 4th lowest state in maximally-spin-polarized sector.
\end{proposition}
We note that \propref{prop:FCI} is a necessary condition for FCI states.
In addition to this criterion, we will also probe the quality of FCI based on the standard deviation of the particle density in the momentum space---perfect FCI states have uniform particle density in the momentum space (zero standard deviation).

As shown in Figs.\,\ref{fig:ED_1BPV_Main_Reddy_etal} and \ref{fig:ED_1BPV_Main_Wang_etal}, the finite-size effects on the FCI region at $\nu=-2/3$ are non-negligible. Hence we choose the largest system size $4\times 6$ to determine the $\nu=-2/3$ FCI region, which is roughly $\epsilon\in[5,6.25]$ for the parameters in \refcite{reddy2023fractional} and nearly the whole phase diagram for the parameters of \refcite{wang2023fractional}.
Based on the particle density in the momentum, we find that the quality of the FCI states in the $\nu=-2/3$-FCI region are much worse for the parameters in \refcite{reddy2023fractional} (standard deviation of particle density larger than 0.136) than that for \refcite{wang2023fractional} (standard deviation of particle density can be as small as 0.002).
In the ``not FCI" regions of Figs.\,\ref{fig:ED_1BPV_Main_Reddy_etal} and \ref{fig:ED_1BPV_Main_Wang_etal}, we did not see clear signatures for FCI or maximally-spin-polarized CDW, and the signature of metal is also not clear enough (perhaps due to finite-size effect).
(See \appref{app:1BPV_ED_results} for details.)

In the parameter region where $\nu=-2/3$ hosts FCIs, the 1BPV ED calculation shows that the ground states at $\nu=-1/3$, $-2/3$ and $-4/3$ all exhibit a positive spin-1 gap, indicating maximally-spin-polarized ground states.
We note that although the finite-size effect in the FCI region is considerable, the finite size effect of the spin-1 gap is small as the spin-1 gap does not change significantly for all four system sizes we consider (See \appref{app:1BPV_ED_results:GS} for full details). We now compare the magnetic stability of the ground states at different fillings.

We compare $\nu=-4/3$ to $\nu=-2/3$ first.
In the FCI region for $\nu=-2/3$,  the filling fraction $\nu=-4/3$ is in a maximally-spin-polarized charge density wave phase with wave-vector $K_M$  (noted as  $K_M$-CDW) for the parameter values in \refcite{reddy2023fractional}  (see \figref{fig:Fullspingap_oneband_theta_3.7} of \appref{app:1BPV_ED_results}), while $\nu=-4/3$ is mostly in the FCI phase for the parameter values in \refcite{wang2023fractional} (see \figref{fig:ED_1BPV_Main_Wang_etal_app_StateOverlap} of \appref{app:1BPV_ED_results}).
As shown in \figref{fig:ED_1BPV_Main_Reddy_etal}, the spin-1 gap at $\nu=-4/3$ is larger than that at $\nu=-2/3$ for the parameter values in \refcite{reddy2023fractional}, indicating more robust magnetism in the former than the latter, which is contrary to the experiment. Specifically, in the $\nu=-2/3$ FCI region, the ratio of the spin-1 gap at $\nu=-4/3$ and $\nu=-2/3$ takes a value in the range [1.59,1.72] for $3\times 4$, [1.49,1.59] for $3\times 5$, [1.62,1.77] for $3\times 6$, and [1.37,1.64] for $4\times 6$.
In comparison, the same ratios for \refcite{wang2023fractional} are in [0.60,0.92] for $3\times 4$, [0.61,0.93] for $3\times 5$, [0.74,1.00] for $3\times 6$, and [0.71,0.93] for $4\times 6$.
In the top left corner of the phase diagram ($\theta\in[3.5^\circ,3.7^\circ]$ and $10/\epsilon \in [0.8,1.0]$), the similar spin-1 gaps (shown in \figref{fig:ED_1BPV_Main_Wang_etal}) for the parameter values in \refcite{wang2023fractional} can be understood from the approximate PH symmetry between $\nu=-4/3$ and $\nu=-2/3$ in this 1BPV case, as discussed in \secref{sec:PH_sym} and \appref{app:1BPV_ED_results:4third2third}.
Thus, within the 1BPV approximation for existing models\cite{reddy2023fractional,wang2023fractional}, these results show that not only are the many-body states found at $\nu=-4/3$ magnetic, but that their magnetism is far from being significantly less robust than that of the FCI at $\nu=-2/3$. This should be contrasted with the non-magnetic state at $\nu=-4/3$ and the magnetic FCI at $\nu=-2/3$ observed in the experiments~\cite{cai2023signatures,zeng2023integer,park2023observation,Xu2023FCItMoTe2}.

On the other hand, comparing $\nu=-1/3$ and $\nu=-2/3$, we find that the spin-1 gap at $\nu=-1/3$ is indeed considerably smaller than that at $\nu=-2/3$ for both sets of parameter values.
\refcite{reddy2023fractional} finds an order-of-magnitude difference between spinful gaps at $\nu=-1/3$ and $\nu=-2/3$ for the system size of 12 unit cells at $\theta=3.5^\circ$ and $\epsilon=5$, which is consistent with our $3\times 4$ results on both spinful and spin-1 gaps (see \appref{app:1BPV_ED_results:1third2third}). \refcite{wang2023fractional} finds the spinful gap at $\nu=-2/3$ is about 5 times that at $\nu=-1/3$ for $\theta=3.5^\circ$, $\epsilon=15$ and the system size of $3\times 4$, which is also consistent with our $3\times 4$ results on both spinful and spin-1 gaps (see \appref{app:1BPV_ED_results:1third2third}).
(\refcite{reddy2023fractional} also studied the spinful gap at $15$ unit cells, but for interaction strengths that do not give FCIs within the experimentally relevant angle region; nevertheless, their results are consistent with our $3\times 5$ results.)
We find that increasing the system size reduces the difference in the magnetic behavior between $\nu=-1/3$ and $\nu=-2/3$, but this difference still remains considerable (see \appref{app:1BPV_ED_results:1third2third} for more details).

In summary, the spin-1 gap 1BPV ED results indicate that the magnetic stability at $\nu=-4/3$ is far from being significantly weaker than that at $\nu=-2/3$, which is inconsistent with the experiments, whereas the difference between $\nu=-1/3$ and $\nu=-2/3$ has the same trend as the robust absence of ferromagnetic order at $\nu=-1/3$ in experiments (though the ground states at $\nu=-1/3$ are still maximally-spin-polarized).
In the following, we will show that our 2BPV calculations, which include one additional band per valley, the spinful and/or spin-1 gaps are reduced at $\nu=-1/3$ and $-2/3$ (even having sign changes) and the spin of the ground state at $\nu=-4/3$ is greatly changed.
In this 2BPV case, we can capture the difference between $-2/3$ and $-4/3$ only for the parameter values in \refcite{wang2023fractional}. In addition, the difference in the stability of magnetism between (the weakly magnetic) $\nu=-1/3$ state and (the robustly magnetic) $\nu=-2/3$ state increases for both sets of parameter values.

\section{Two-band-per-valley Exact Diagonalization Calculations}
\label{sec:2BPV_ED}

We perform 2BPV ED calculations to study $ \nu = -1/3 $, $ \nu = -2/3 $ and $ \nu = -4/3 $ for $ \xi = 20 nm $ and for the hole interaction in \eqnref{eq:H_h}.
We will first discuss $\nu=-1/3$ and $\nu=-2/3$, and then compare $\nu=-4/3$ to $\nu=-2/3$.
We note that although the remote bands were included in the fully spin-polarized sector in \refcite{Xu2023MLWOFCItTMD}, the effect of remote bands in spinful calculations have not yet been considered.
Our 2BPV calculations will involve the systems sizes of $3\times 3$, $3 \times 4$ and $3\times 5$, since the partially-spin-polarized and spin-unpolarized sectors at larger sizes have very large Hilbert dimensions (taking $3\times 6$ as an example, the $S_z=S^{max}_{2BPV} -1 $ sector at $\nu=-2/3$ has Hilbert space dimension about $1.2 \times 10^9$ per momentum).

\begin{figure*}[t]
\centering
\includegraphics[width=1.8\columnwidth]{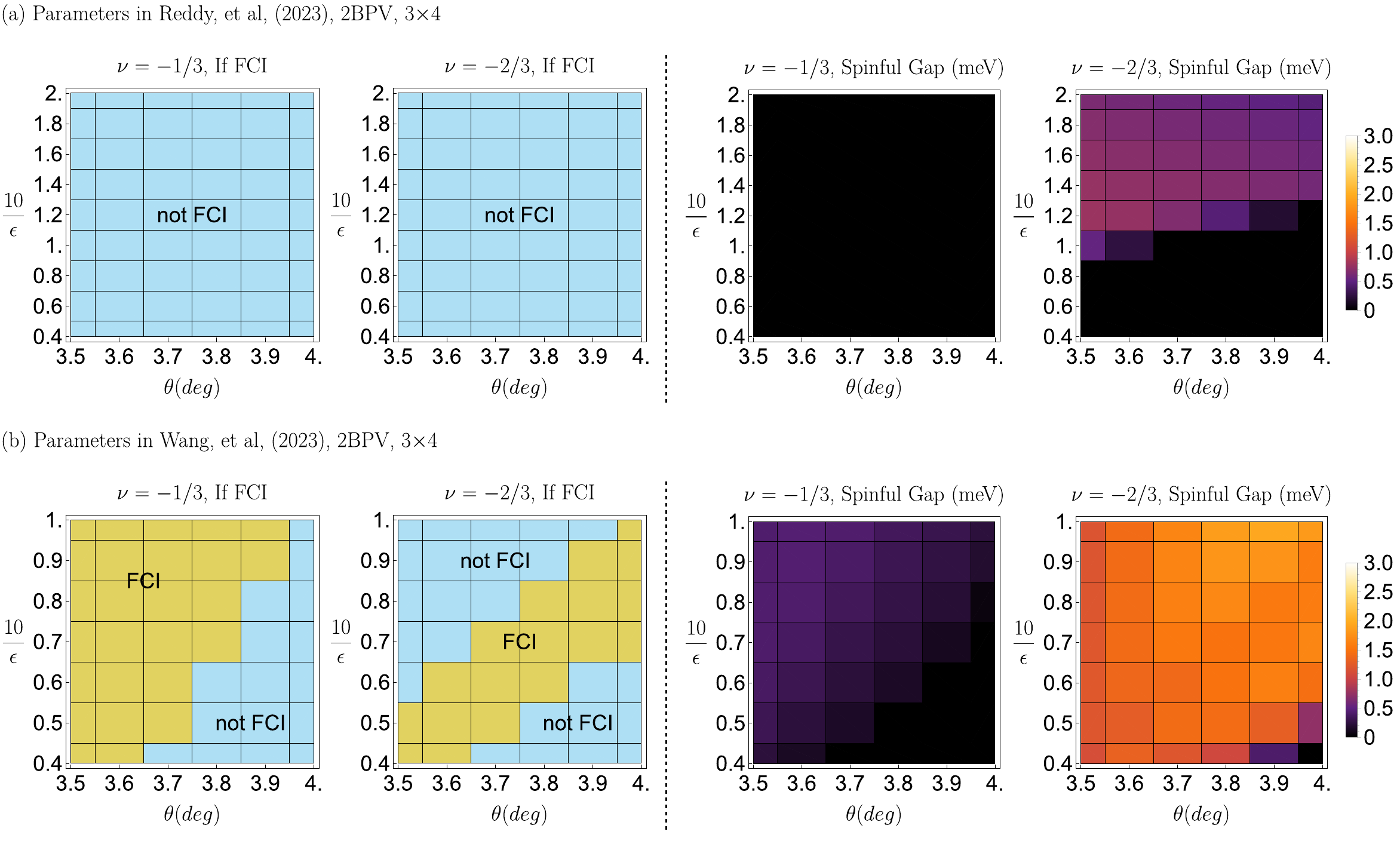}
\caption{$3\times 4$ 2BPV ED calculation at $\nu=-1/3,-2/3$ for the parameter values in Refs.\,\cite{reddy2023fractional,wang2023fractional} using $H_h$ in \eqnref{eq:H_h}.
In the left most two figures of each row, green (``FCI") labels the region that satisfies the criterion in \propref{prop:FCI}, and blue (``not FCI") means that we do not see clear signatures of FCI or maximally-spin-polarized CDW.
In rightmost two figures of (a) or (b), the spinful gap is shown with the same color functions for all plots. Conventions are the same as in \Figs{fig:ED_1BPV_Main_Reddy_etal}{fig:ED_1BPV_Main_Wang_etal}.
}
\label{fig:ED_2BPV_Main_x3y4_1third2third}
\end{figure*}

\subsection{$\nu=-1/3$ and $\nu=-2/3$}

In our $3\times 4$ 2BPV calculations, the spin-1 gaps are considerably larger than the spinful gaps for $\nu=-1/3$, though the spin-1 gap is a good approximation of the spinful gap for $\nu=-2/3$ (See \appref{app:2BPV_ED_results}).
The spinful gap being smaller than the spin-1 gap at $\nu=-1/3$ implies that the spin-zero states actually have lower energy than the partially-spin-polarized $(S_z=1)$ states, since we only have three possible values $|S_z|=0,1,2$ at $\nu=-1/3$  at the system size of $3\times 4$.
Therefore, we will compare the spinful gaps at $\nu=-1/3$ and $\nu=-2/3$ as shown in \figref{fig:ED_2BPV_Main_x3y4_1third2third}.
By comparing to the spinful gaps at the system size $3\times 4$ in the 1BPV case (see \appref{app:1BPV_ED_results}), we can see that including one extra band per valley generally decreases the spinful gaps (and spin-1 gaps) at both fillings.
For example, the spinful gap at $\nu=-2/3$ does not exceed $2$meV in the 2BPV case, while it can be as large as $\sim 8$meV with 1BPV.

For the parameter values in \refcite{reddy2023fractional}, we see the absence of FCIs at $\nu=-2/3$ in \figref{fig:ED_2BPV_Main_x3y4_1third2third}(a) for the $3\times 4$ system size. However, we cannot exclude the possibility that this is due to finite-size effects,  because for the same parameters, the FCI also does not appear in the 1BPV calculation for the $3\times 4$ system size (see \appref{app:1BPV_ED_results}).
If we change the system size to $3\times 5$ at $(\theta, \epsilon) = (3.7^\circ, 5)$, the ground state at $\nu=-2/3$ is still not an FCI.

The magnetic properties appear to not be afflicted by finite size effects. At $(\theta, \epsilon) = (3.7^\circ, 5)$, the ground state at $\nu=-1/3$ is spin-unpolarized on $3\times 4$ systems (\figref{fig:ED_2BPV_Main_x3y4_1third2third}(a)) and is minimally polarized for $3\times 5$ (the total spin cannot be zero for 5 holes), and the spin-1 gaps at $\nu=-2/3$ are similar for the system sizes of $3 \times 4$ and $3\times 5$ ($0.56$meV and $0.75$meV respectively, see \appref{app:2BPV_ED_results}). Therefore, we can see the spin properties here are much more robust against finite-size effects than the FCI phase boundary at $\nu=-2/3$.
In particular, for interaction strengths $\epsilon\in [5,6.25]$ (\ie, $10/\epsilon\in [1.6,2]$ which can give FCIs at $\nu=-2/3$ in 1BPV $4\times 6$ calculations), the zero-spin ground states at $\nu=-1/3$ and the maximally-spin-polarized ground states at $\nu=-2/3$ on the $3\times 4 $ systems (\figref{fig:ED_2BPV_Main_x3y4_1third2third}(a)) for the parameters of \refcite{reddy2023fractional} are consistent with the nonmagnetic $\nu=-1/3$ and magnetic $\nu=-2/3$ in experiments~\cite{cai2023signatures,zeng2023integer,park2023observation,Xu2023FCItMoTe2} after adding the remote bands.

For the parameter values in \refcite{wang2023fractional}, FCI states at $\nu=-2/3$ are clearly present in a considerable ``diagonal-shaped" parameter region in the phase diagram for the $3\times 4$ system size, as shown in \figref{fig:ED_2BPV_Main_x3y4_1third2third}(b).
The quality of the FCI states deep in the $\nu=-2/3$ FCI region is very good (standard deviation of particle density can be as small as 0.006 compared to the filling $|\nu|=2/3\approx 0.67$). (See \appref{app:2BPV_ED_results:xiao} for details.)
Compared to the $\nu=-2/3$ FCI region for the 1BPV case in \figref{fig:1BPV_Wang_etal_app}(a) for the $3\times 4$ system size (\appref{app:1BPV_ED_results}), the remote bands suppress the FCI at larger interaction when fixing the twist angle, which is consistent with the remote bands suppressing CIs at $\nu=-1$ at larger interaction shown in \figref{fig:HF_x18y18_main}.
The diagonal shape of FCI region at $\nu=-2/3$ is consistent with the experimental report\cite{park2023observation} of $\nu=-2/3$ FCIs at $\theta=3.7^\circ$ but not at $\theta=3.5^\circ$ and $\theta=3.9^\circ$.
The fully-spin-polarized 2BPV calculation in \refcite{Xu2023MLWOFCItTMD} finds that the largest-gap FCI at $\nu=-2/3$ occurs at a larger angle for a stronger interaction outside the experimental angle region $\theta\in[3.5^\circ, 4.0^\circ]$, which shows the same trend as our results.

Among the 18 points in the phase diagram \figref{fig:ED_2BPV_Main_x3y4_1third2third}(b) that give FCIs at $\nu=-2/3$ on $3\times 4$ systems, there is one point that favors spin-unpolarized (\ie, spin zero) ground states at $\nu=-1/3$.
For the other 17 points, we find fully-spin-polarized ground states at $\nu=-1/3$ (FCI states for 11 of them), but ratio between the spinful gaps at $\nu=-1/3$ and $\nu=-2/3$ (former divided by latter) takes roughly uniform values in $[0.03, 0.27]$
(see \appref{app:2BPV_ED_results} for details). Therefore, when $\nu=-2/3$ features a fully-spin-polarized FCI, the state at $\nu=-1/3$ is either spin-unpolarized (at one point) or more often is fully-spin-polarized with magnetic stability much weaker than that at $\nu=-2/3$. This is loosely consistent with the experiments, though the ground states at $\nu=-1/3$ are still mostly fully-spin-polarized.
We emphasize that this near consistency relies on $\epsilon >10$ within $\epsilon\in[5,25]$. In fact, a weaker magnetic stability (\ie, smaller spinful gap) at $\nu=-1/3$ tends to happen at larger $\epsilon$ values when fixing the twist angle as shown in \figref{fig:ED_2BPV_Main_x3y4_1third2third}(b) and in \appref{app:2BPV_ED_results}.

\subsection{$\nu=-4/3$ versus $\nu=-2/3$}

\begin{figure}[t]
\centering
\includegraphics[width=\columnwidth]{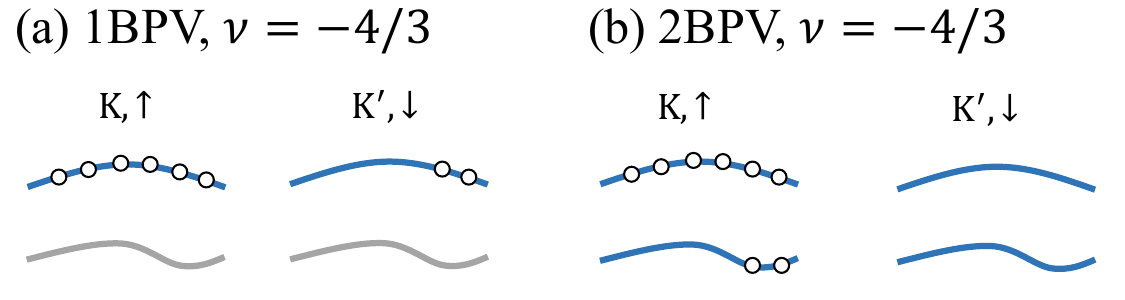}
\caption{
Schematic plots of maximally-spin-polarized states in the 1BPV (a) and 2BPV (b) cases, respectively, for 6 unit cells.
The remote bands (gray) are frozen in (a), and the configuration in (b) is fully-spin-polarized.
We note that the case of (a) is allowed in the 2BPV case as a partially-spin-polarized state.
}
\label{fig:ED_2BPV_Main_m4over3_MSP}
\end{figure}

\begin{figure}[t]
\centering
\includegraphics[width=\columnwidth]{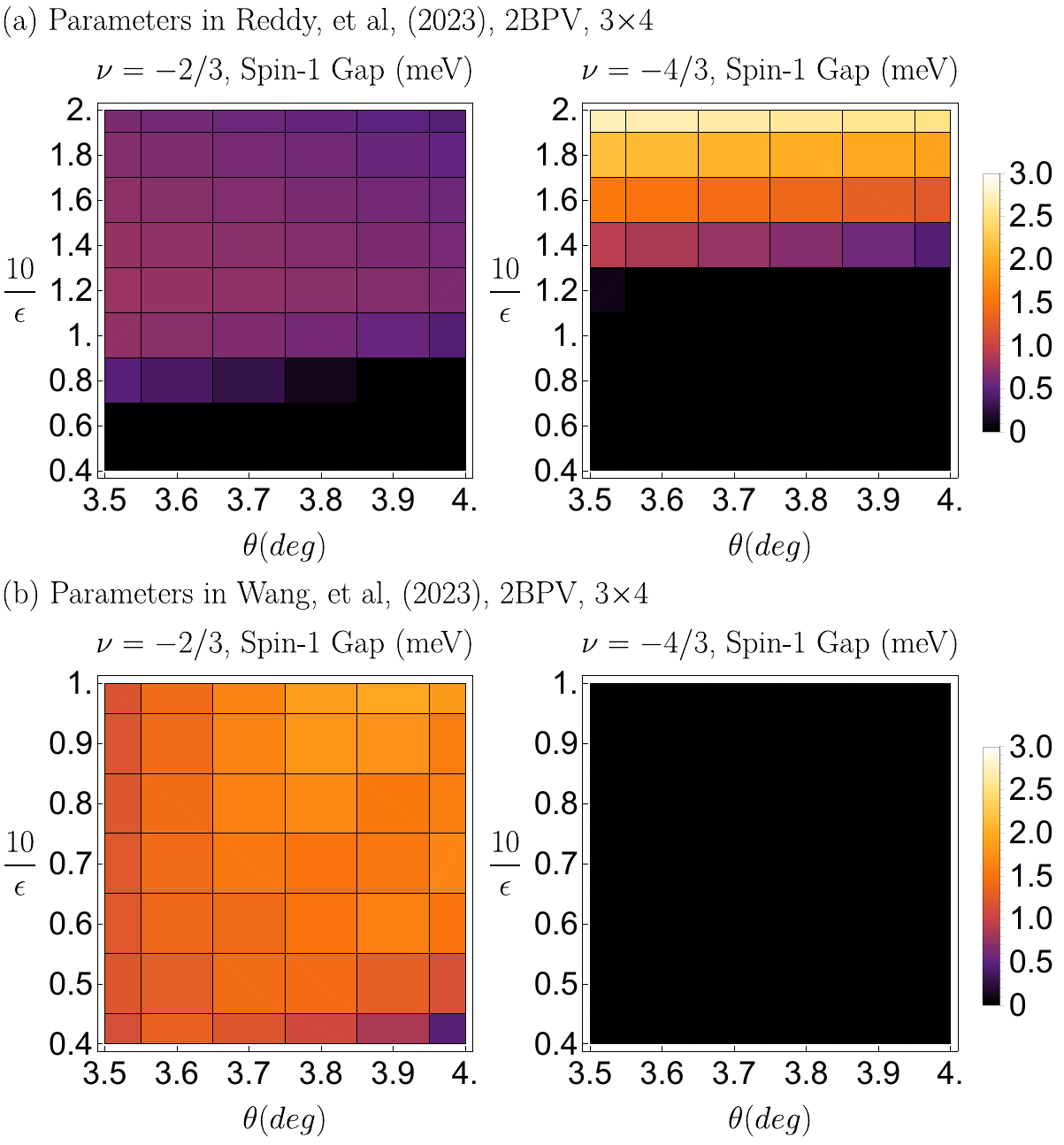}
\caption{
Spin-1 gaps for $3\times 4$ 2BPV at $\nu=-2/3,-4/3$ for parameter values in \refcite{reddy2023fractional} (a) and \refcite{wang2023fractional}(b) using $H_h$.
The spin-1 gap is shown with the same color functions for all plots.
If the spin-1 gap is negative, we set it to zero in the plots.
}
\label{fig:ED_2BPV_Main_m2over3_m4over3_spin1gap}
\end{figure}

At $\nu=-4/3$, the spinful gap is difficult to calculate for a $3\times 4$ system due to the large Hilbert space dimension, which is $\sim 4.5\times 10^{10}$ per momentum in the spin-zero sector. (The largest Hilbert space dimension per spin per momentum in our 2BPV calculations is about $1.09\times 10^8$.) Therefore, we can only compare the spin-1 gap at $\nu=-4/3$ to the spin-1 gap at $\nu=-2/3$ on the $3\times 4$ system.

There is a subtlety regarding the spin-1 gap in the 2BPV case.
Unlike for $\nu=-1/3$ and $\nu=-2/3$, the maximally-spin-polarized state for $\nu=-4/3$ is different in the 1BPV and 2BPV approximations: the maximally-spin-polarized state ($S_z=S^{\rm 1BPV}_{\rm max}$) at $\nu = -4/3$ in the 1BPV case corresponds to a partially-spin-polarized state in the 2BPV case as schematically shown in \figref{fig:ED_2BPV_Main_m4over3_MSP}(a). This partially-spin-polarized sector at $\nu=-4/3$ has the same total spin as the maximally-spin-polarized states at $\nu = -2/3$: it might potentially host FCIs if the remote bands are negligible as shown in the 1BPV calculations in \figref{fig:ED_1BPV_Main_Wang_etal}. But the 2BPV case allows for fully-spin-polarized states ($S_z=S^{\rm 2BPV}_{\rm max}$) $\nu=-4/3$ as depicted in \figref{fig:ED_2BPV_Main_m4over3_MSP}(b). Since the remote bands also have nonzero Chern numbers, this situation might also potentially lead to an FCI phase, namely the product of a CI and a Laughlin-like FCI (potentially with an opposite chirality for two sets of parameters in Refs.\,\cite{reddy2023fractional,wang2023fractional} since they have opposite Chern numbers in the remote bands).

The spin-1 gap at $\nu=-4/3$ is the energy difference between the lowest-energy states in the $S_z=S^{\rm 2BPV}_{\rm max}$ and $S_z=S^{\rm 2BPV}_{\rm max}-1$ sectors, which we first focus on for system sizes of $3\times 4$ and $3\times 5$. Note that even if the spin-1 gap is negative, it is conceivable that partially-spin-polarized states like \figref{fig:ED_2BPV_Main_m4over3_MSP}(a) might still be favored at $\nu=-4/3$ in the 2BVP calculation. However, the Hilbert space dimension of the partially spin-polarized states in \figref{fig:ED_2BPV_Main_m4over3_MSP}(a) is beyond our computational capabilities for $3\times4$ (dimension $\sim 2.4\times 10^9$ per momentum) and $3\times5$ (dimension $\sim 1.5\times 10^{12}$  per momentum) systems and thus will be addressed later in this part by considering smaller $3\times 3$ systems.

As shown in \figref{fig:ED_2BPV_Main_m2over3_m4over3_spin1gap}(a) for the parameters in \refcite{reddy2023fractional}, our 2BPV ED results point toward a spin-1 gap at $\nu=-4/3$ that is larger than the one at $\nu=-2/3$, when the interaction is strong enough ($ \epsilon\in [5,6.25]$) to give an FCI at $\nu=-2/3$ in 1BPV $4\times 6$ calculations. We also see this larger spin-1 gap at $\nu=-4/3$ on $3\times 5$ systems (see \appref{app:2BPV_ED_results:fu}) which seems to suggest that the large-spin states are favored at $\nu=-4/3$.
For the parameters of \refcite{wang2023fractional}, \figref{fig:ED_2BPV_Main_m2over3_m4over3_spin1gap}(b) shows that the ground states are not  fully-spin-polarized at $\nu=-4/3$; this trend persists to the system size of $3\times 5$ as discussed in \appref{app:2BPV_ED_results:xiao}.

\begin{figure}[t]
\centering
\includegraphics[width=\columnwidth]{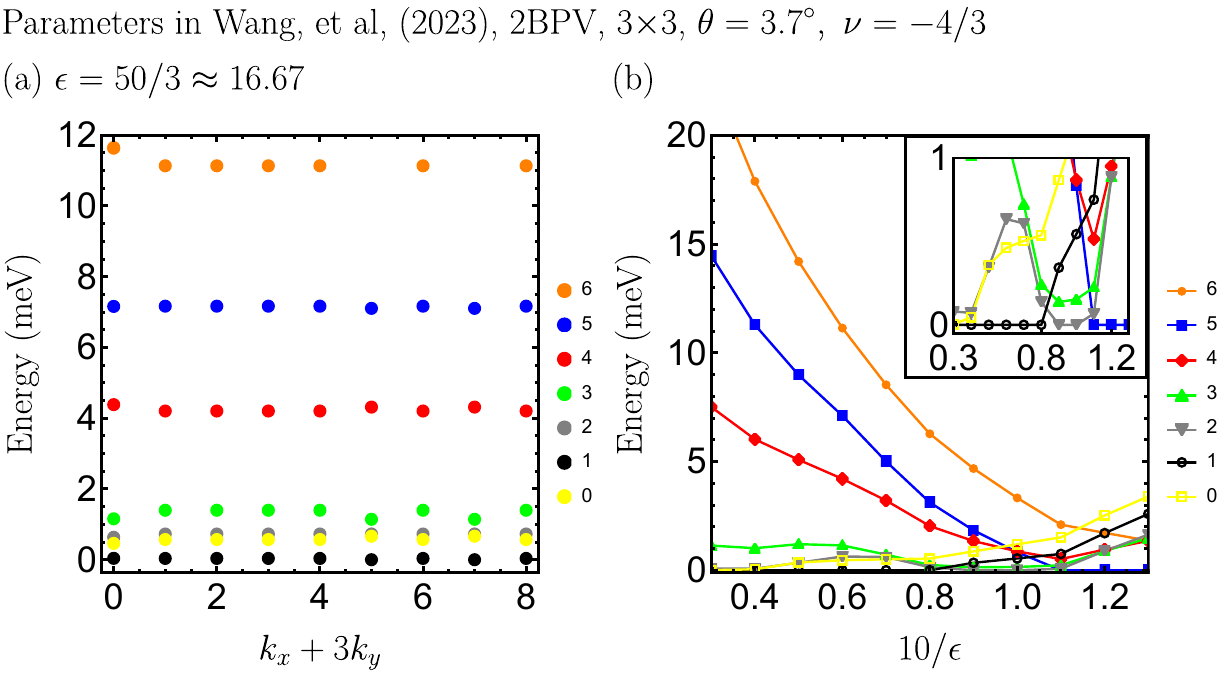}
\caption{
(a) The 2BPV many-body energy spectrum for the parameters in \refcite{wang2023fractional} at $
\theta=3.7^\circ$ and $\epsilon =50/3\approx 16.67$ and $\nu=-4/3$.
Here we only include the lowest-energy state in per momentum and spin sector.
The energy of the ground state is set to zero.
(b) At each value of $\epsilon$, we show the lowest energy of each spin sector for the parameters specified in the plot.
As a comparison, the total spin of the fully-spin-polarized state at $\nu=-2/3$ is $|S_z| = 3$ for the size of $3\times 3$.
The energy of the ground state is set to zero at each value of $\epsilon$.
$k_x$ and $k_y$ are defined as $\bsl{k}=(k_x/L_1) \bsl{b}_1 + (k_y/L_2) \bsl{b}_2 $, where $L_1\times L_2$ is the system size, and $\bsl{b}_1$ and $\bsl{b}_2$ are defined in \eqnref{eq:b_1_b_2}.
}
\label{fig:ED_2BPV_x3y3_Wangetal_theta3p7_m4over3}
\end{figure}

To address the subtlety of partially-spin-polarized states of \figref{fig:ED_2BPV_Main_m4over3_MSP}(b), we resort to the $3\times 3$ system.
For the parameters in \refcite{reddy2023fractional} and $3\times 3$ systems, the large-spin states ($S_z=S^{\rm 2BPV}_{\rm max}, S^{\rm 2BPV}_{\rm max}-1$ with $S^{\rm 2BPV}_{\rm max}=6$) are indeed strongly favored (see \appref{app:2BPV_ED_results}) at $\nu=-4/3$, though the spin-1 gap at $3\times 3$ becomes negative. (For comparison, the fully-spin-polarized states at $\nu=-2/3$ have $S_z=3$.) The preference for large-spin states at $\nu=-4/3$ is inconsistent with experiment~\cite{cai2023signatures,zeng2023integer,park2023observation,Xu2023FCItMoTe2}.
For the parameters in \refcite{wang2023fractional}, the $3\times 3$ system has a similar FCI region at $\nu=-2/3$ and similar spinful gap at $\nu=-2/3$ and $\nu=-4/3$ as those of the system size $3\times 4$, as shown in \appref{app:2BPV_ED_results}.
The $3\times 3$ results in \figref{fig:ED_2BPV_x3y3_Wangetal_theta3p7_m4over3} show that for $\theta=3.7^\circ$, the partially-spin-polarized $S_z =3$ states (which have the same spin as the 1BPV maximally-spin-polarized states at $\nu=-4/3$ and as the fully-spin-polarized state at $\nu=-2/3$) are not favored at $\nu=-4/3$ for the interaction strengths that give FCIs at $\nu=-2/3$ (roughly $10/\epsilon=0.5\sim0.7$ at $\theta=3.7^\circ$). Instead it is the small-spin ($S_z=0,1$) states that dominate at $\nu=-4/3$, in agreement with experiments~\cite{cai2023signatures,zeng2023integer,park2023observation,Xu2023FCItMoTe2}.
As the ground states at $\nu=-2/3$ are always fully-spin-polarized ($S_z=3$) up for $\epsilon\in[10,25]$ at $\theta=3.7^\circ$, the fact that the ground states never have $S_z=3$ in \figref{fig:ED_2BPV_x3y3_Wangetal_theta3p7_m4over3} indicates the clear breaking of the approximate PH symmetry owing to the remote bands for the parameters in \refcite{wang2023fractional}.

\subsection{Summary of 2BPV Results}

After including remote bands, the parameters in \refcite{wang2023fractional} exhibit mostly a fully-spin-polarized state at $\nu=-1/3$ with significantly weakened magnetic stability than that at $\nu=-2/3$ while maintaining spin-polarized FCI states at $\nu=-2/3$, which is nearly consistent with the experiments. These parameters can also capture the observed difference between $\nu=-4/3$ and $\nu=-2/3$, though this fit to the experimental phase diagram relies on a weaker interaction, namely $\epsilon \geq 10$.
For the same system sizes, the parameter values in \refcite{reddy2023fractional} are able to capture the difference in magnetism between $\nu=-1/3$ and $\nu=-2/3$, but we have not seen clear indications that they capture the difference between $\nu=-4/3$ and $\nu=-2/3$. We find that the desired FCI phase at $\nu=-2/3$ is missing, although this is possibly due to finite size effects.

\section{Conclusion}
\label{sec:conclusion}

We have studied the continuum moir\'e model in \refcite{Wu2019TIintTMD}, including remote bands in our HF and ED calculations. We performed a comparative study of the parameter values in both \refcite{reddy2023fractional} and \refcite{wang2023fractional}.
For the experimental twist angles $\theta\in[3.5^\circ,4.0^\circ]$, the 2BPV HF and ED calculations using the parameters of \refcite{wang2023fractional} show FCIs at $\nu=-2/3$, CIs with the quantum anomalous Hall effect at $\nu=-1$,  fully-spin-polarized states at $\nu=-1/3$ with much weaker stability than that at $\nu=-2/3$, and weakly-magnetic (very small total spin) states at $\nu=-4/3$. This phenomenology is in near agreement with experiments, provided that a dielectric constant of $\epsilon > 10$ (larger than $\epsilon\sim 6$ estimated from h-BN~\cite{laturia_dielectric_2018}) can be accounted for.
On the other hand, we have not succeeded in capturing the significant difference in magnetism between $\nu=-2/3$ and $\nu=-4/3$ with the parameter values in \refcite{reddy2023fractional} for the experimental angles $\theta\in[3.5^\circ,4.0^\circ]$ and $\epsilon\in[5,25]$.

Our results suggest that the key new magnetic features (beyond the usual FCI~\cite{neupert, sheng, regnault} phases) of the experimental phase diagram depend on remote bands. Our work demonstrates that the minimal model first proposed in \refcite{Wu2019TIintTMD} with the parameters of \refcite{wang2023fractional} is a reasonable starting point for describing the realistic phase diagram of $t$MoTe$_2$, though it still mostly gives spin-fully-polarized ground states at $\nu=-1/3$ and it needs a larger value of the dielectric constant $\eps$ than estimated values in bulk h-BN.
One potential explanation for a weaker interaction could lie in modifications of the minimal model of \refcite{Wu2019TIintTMD}. In materials like graphene, lattice relaxation \cite{PhysRevLett.99.256802, 2019arXiv190800058F,PhysRevB.105.125127,PhysRevB.107.075408,PhysRevB.107.075123} and strain \cite{PhysRevLett.127.027601,PhysRevX.11.041063,PhysRevLett.128.156401,PhysRevB.105.245408,2022arXiv221102693W,2022arXiv220908204W,2023arXiv230300024N} are understood to play a key role in extending the original Bistrizter-MacDonald model \cite{2011PNAS..10812233B}. Such an effect is a typical feature of moir\'e engineering \cite{2021NatRM...6..201A,2020NatMa..19.1265A,2022NatNa..17..686M,PhysRevB.100.035448,2021NatCo..12.6730D,2021EleSt...3a4004H}. They lead to higher order terms in the single-particle model that could increase the bandwidth, effectively shrinking the interaction strength.

The now-accessible cornucopia of moir\'e platforms and tuning parameters promises to enrich the already large family of FCI states \cite{Liu12,Sterdyniak13,Udagawa14,Wu15,Jaworowski19, 2022arXiv220808449L}, further supported by the experimental characterization of lattice effects in fractional quantum Hall systems \cite{2022Sci...375..321L,2023arXiv230805789H,2023arXiv230316993F,2023arXiv230802821Z}. The successful prediction of these more exotic FCI phases in addition to other material platforms rests on a faithful model of the underlying Hamiltonian. To that end, we have shown that multi-band physics plays a crucial role in $t$MoTe$_2$, and that its inclusion can accurately hew out the FCI region of the experimental phase diagram.

\section{Acknowledgments}
The authors thank Allan H. MacDonald, Kin-Fai Mak, Jie Shan, and Xiaodong Xu for helpful discussions.
N.R. is grateful to Valentin Cr\'epel for fruitful discussions.
J.Y. thanks Yves H Kwan, Ramanjit Sohal, Prashant Kumar, and Yang Zhang for helpful discussions.
J.H.-A. is appreciative of stimulating discussions with Jiaqi Cai and Trithep Devakul.
M.W. and O.V. thank Xiaoyu Wang for helpful discussions.
This work is partly supported by a project that has received funding from the European Research Council (ERC) under the European Union’s Horizon 2020 Research and Innovation Programme (Grant Agreement No. 101020833).
B. A. B.’s work was primarily supported by the DOE Grant No. DE-SC0016239, the Simons Investigator Grant No. 404513. N.R. also acknowledges support from the QuantERA II Programme that has received funding from the European Union’s Horizon 2020 research and innovation programme under Grant Agreement No 101017733. O.V. was funded by the Gordon and Betty Moore Foundation’s EPiQS Initiative Grant GBMF11070, National High Magnetic Field Laboratory
through NSF Grant No. DMR-1157490 and the State of Florida. J. H.-A. is supported by a Hertz Fellowship, with additional support from DOE Grant No. DE-SC0016239 by the Gordon and Betty Moore Foundation through Grant No. GBMF8685 towards the Princeton theory program, the Gordon and Betty Moore Foundation’s EPiQS Initiative (Grant No. GBMF11070), Office of Naval Research (ONR Grant No. N00014-20-1-2303), BSF Israel US foundation No. 2018226 and NSF-MERSEC DMR. J. Y. is supported by the Gordon and Betty Moore Foundation through Grant No. GBMF8685 towards the Princeton theory program.

\bibliographystyle{apsrev4-1}
\bibliography{fcitmds}

\onecolumngrid
\newpage

\tableofcontents

\appendix

\section{Single-Particle Models}
\label{app:SP}

In this Appendix, we discuss the single-particle continuum models of the twisted transition metal dichalcogenides (TMD) homo-bilayer Hamiltonians with an emphasis on their symmetries and topology.

\begin{figure}
\centering
(a) \includegraphics[height=0.26\columnwidth]{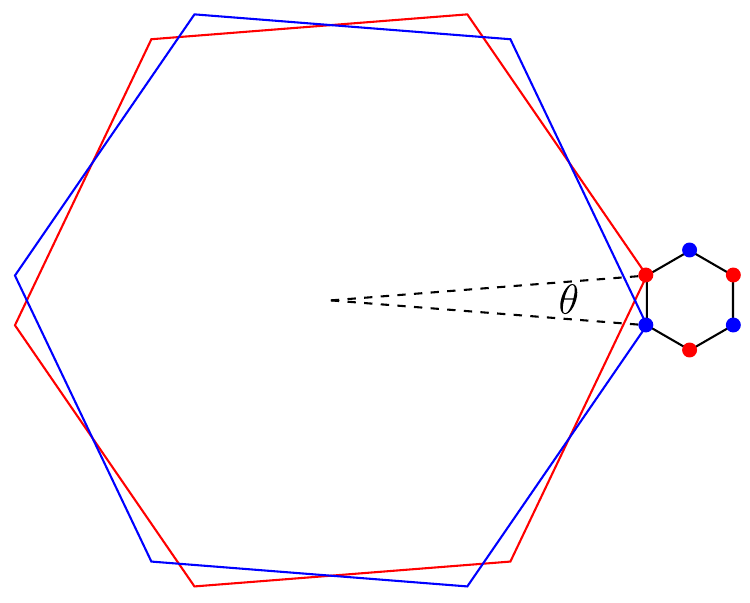}
(b) \includegraphics[height=0.26\columnwidth]{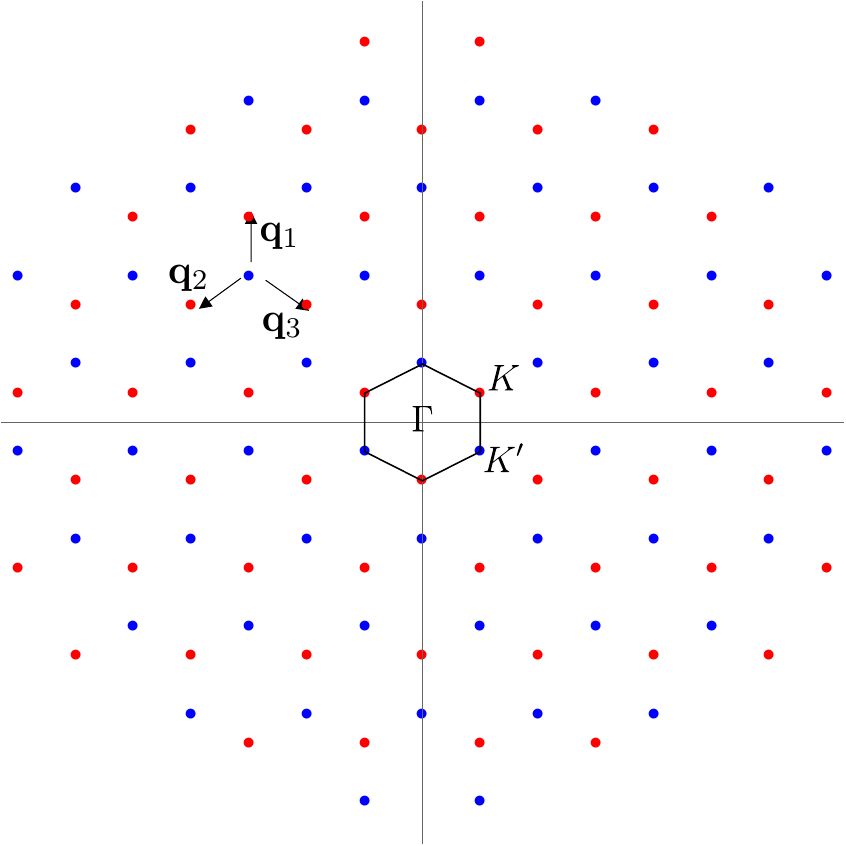}
(c) \includegraphics[height=0.26\columnwidth]{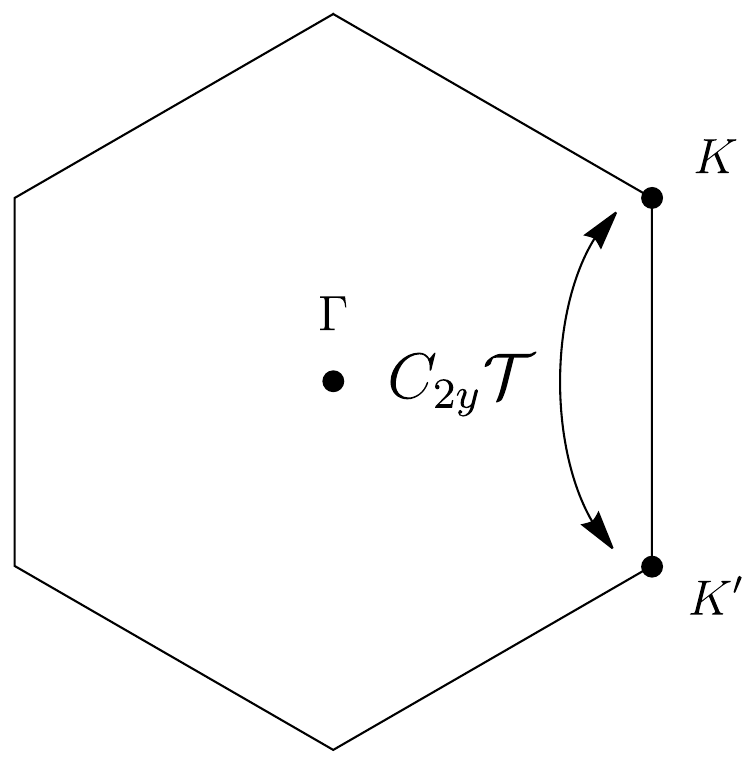}
\caption{(a)  Brillouin zones of the top (red) and bottom (blue) layers with a relative twist angle $\th$, and the emergent moir\'e BZ in the TMD $K$ valley.  (b) Momentum space lattice with red sites (moir\'e $K$ point) corresponding states in the upper layer and blue sites (moir\'e $K'$ point) in the lower layer. There is a single degree of freedom per site. (c) Moir\'e Brillouin zone with $C_{3z}$-symmetric points, with $K$ and $K'$ points exchanged by $C_{2y} \mathcal{T}$.
}
\label{fig:model}
\end{figure}

\subsection{Momentum Space Model}

In TMD materials, strong spin-orbit coupling breaks spin $SU(2)$ and leads to spin polarization: the low energy states in the $\K$ valley are spin-polarized, and those in the $\K'$ valley are oppositely spin-polarized by spinful time-reversal symmetry. Thus the full TMD Hamiltonian only has $U(1)$ valley symmetry (locked to spin), and no $SU(2)$ spin symmetry.

\begin{figure}
\centering
\includegraphics[width=1.\columnwidth]{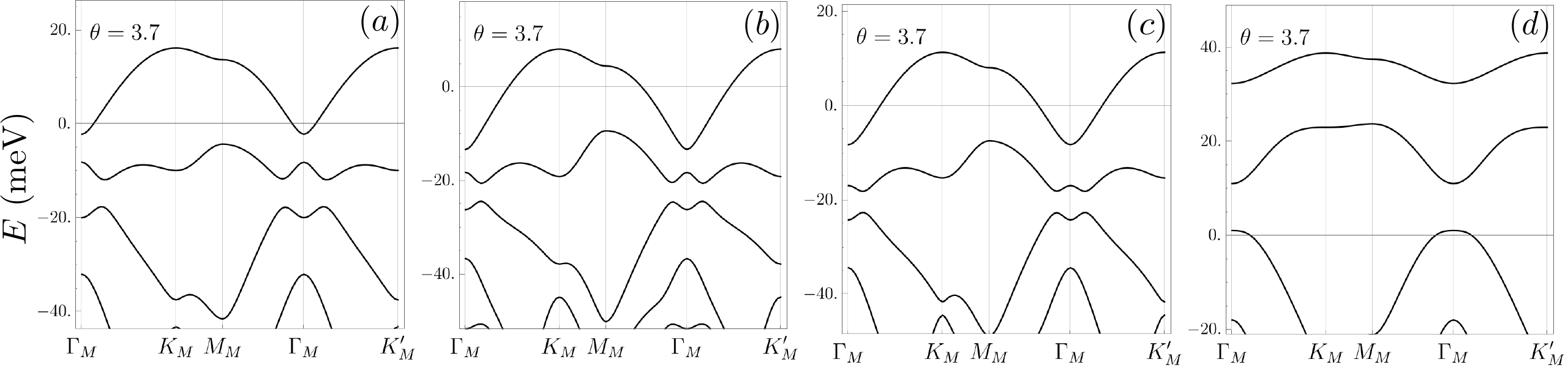}
\caption{Comparison of band structures for the parameter values in \Tab{tab:paramtable}, where $(a)$-$(d)$ correspond to rows 1-4. We see that $(a),(b),(c)$, which are in the $\Ch= (-1,-1)$ phase, are quite similar (see Main Text), whereas $d$ in the $\Ch= (-1,+1)$ phase (see Main Text) has a significantly different dispersion. For this reason, we focus our comparative study on the parameters in $a$ and $d$.}
\label{fig:4plots}
\end{figure}

This paper focuses specifically on twisted MoTe$_2$. The bulk material typically grows in the 2H configuration \cite{PhysRevLett.108.196802} although other configurations are possible \cite{2014arXiv1406.2749Q,2016NatCo...711038Q}. When exfoliated  down to a monolayer (1H), this configuration results in a gapped 2D band structure with peaked and opposite Berry curvature in each valley due to explicit inversion symmetry breaking from the Te atom positions. (Note that the 1H monolayer band gap is opened due to a trivial inversion-breaking mass, so the model is not a topological insulator \cite{PhysRevLett.108.196802}).  The band near the $K,K'$ points are well described by a gapped Dirac Hamiltonian written in the orbital basis of the Mo $d$ electrons \cite{Wu2019TIintTMD}. Explicitly, the $k \cdot p$ theory near the $K$ point is $v_F \mbf{k} \cdot \pmb{\sigma} + \frac{1}{2}\Delta \sigma_3$ where $\Delta \sim 1$eV is the gap, and the basis is $d_{z^2}, d_{x^2-y^2} + i d_{xy}$ orbitals (for a single spin species due to spin-orbital coupling). The $d_{z^2}$ orbitals are at high energy, and Ref. \cite{Wu2019TIintTMD} obtained an effective quadratic Hamiltonian on only the $d_{x^2-y^2} + i d_{xy}$ basis element (in both layers) which reads
\bea
\label{eq:Hkmacdonaldapp}
H_K(\bsl{r}) &= \bpm \frac{\hbar^2\nabla^2}{2m_*} + V_{+}(\bsl{r}) & t(\bsl{r}) \\ t^*(\bsl{r}) &  \frac{\hbar^2\nabla^2}{2m_*} + V_{-}(\bsl{r}) \epm \ .
\eea
The matrix acts on the $d_{x^2-y^2} + i d_{xy}$ orbital in each layer. The Hamiltonian acts on wavepacket states in the upper $(+)$ and lower $(-)$ layers with momenta centered around, respectively,
\bea
\label{eq:momentum_convention}
\pmb{\kappa}_{\pm} &= \frac{\bsl{b}_1}{2} \pm \frac{\bsl{q}_1}{2}
\eea
where
\eq{
\qquad \bsl{b}_1 = k_\th (1,0),  \quad \bsl{q}_1 = k_\th (0, \frac{1}{\sqrt{3}}), \quad \bsl{b}_{i+1} = C_{3z} \bsl{b}_{i}, \quad \bsl{q}_{i +1} = C_{3z} \bsl{q}_1, \quad k_\th =  \frac{4\pi}{\sqrt{3}} \frac{2 \sin \frac{\th}{2}}{a_0}
}
and $a_0 = 3.52 \AA$ is the MoTe$_2$ lattice constant such that the moir\'e reciprocal lattice is spanned by $\bsl{b}_1,\bsl{b}_2$ corresponding to a moir\'e real space lattice $\bsl{a}_i$ obeying $\bsl{a}_i \cdot \bsl{b}_j = 2\pi \delta_{ij}$. Hence, similar to twisted bilayer graphene \cite{2011PNAS..10812233B}, the TMD Hamiltonians are defined by a honeycomb momentum space lattice, but with only a scalar degree of freedom at each site (the single $d_{x^2-y^2} + i d_{xy}$ orbital) as opposed to a sublattice spinor. We obtain the momentum-space Hamiltonian by Fourier transforming into the variables ($N$ is the number of moir\'e unit cells)
\bea
\psi_{\bsl{k}}(\bsl{r}) &= \frac{1}{\sqrt{N}} \sum_{\bsl{k},\bsl{b},l} e^{i (\bsl{k}+\bsl{b} + \pmb{\kappa}_l) \cdot \bsl{r}} U_{\bsl{b},l}(\bsl{k})
\eea
which naturally lie on a honeycomb lattice given by $\bsl{Q} = \bsl{b}  + \pmb{\kappa}_l$, where $\bsl{b}$ ranges over the whole reciprocal lattice. This leads to the momentum space Hamiltonian
\bea
\label{eq:hkK}
H^K_{\bsl{Q},\bsl{Q}'}(\bsl{k}) &= -\frac{\hbar^2}{2m_*}(\bsl{k}+\bsl{Q})^2 \delta_{\bsl{Q} \bsl{Q}'} + V \sum_{i} (e^{i \psi \zeta_{\bsl{Q}}} \delta_{\bsl{Q} - \bsl{Q}',\bsl{b}_i}+e^{-i \psi \zeta_{\bsl{Q}}} \delta_{\bsl{Q} - \bsl{Q}',-\bsl{b}_i}) + w \sum_{i} (\delta_{\bsl{Q} - \bsl{Q}',\bsl{q}_i} + \delta_{\bsl{Q} - \bsl{Q}',-\bsl{q}_i})
\eea
where $\zeta_{\bsl{Q}} = \pm 1$ for $\bsl{Q} = \bsl{b} + \pmb{\kappa}_l$ in the $l = \pm$ layer. The eigenvectors of $H^K_{\bsl{Q},\bsl{Q}'}(\bsl{k})$ in the $n$th band at momentum $\bsl{k}$ are denoted $U_{\bsl{Q},n}(\bsl{k})$, leading to the eigenstates
\bea
\psi_{\bsl{k},n}(\bsl{r}) &= \frac{1}{\sqrt{N}} \sum_{\bsl{Q}} e^{i (\bsl{k}+\bsl{Q}) \cdot \bsl{r}} U_{\bsl{Q},n}(\bsl{k}) \ .
\eea
Note that the spectrum of $H^K(\bsl{k})$ is periodic over the moir\'e BZ due to
\bea
H^K_{\bsl{Q},\bsl{Q}'}(\bsl{k}+\bsl{b}) &= H^K_{\bsl{Q}+\bsl{b},\bsl{Q}'+\bsl{b}}(\bsl{k})  = [V_{\bsl{b}} H^K(\bsl{k}) V_{\bsl{b}}^\dag]_{\bsl{Q},\bsl{Q}'}, \qquad [V_{\bsl{b}}]_{\bsl{Q},\bsl{Q}'} = \delta_{\bsl{Q}+\bsl{b},\bsl{Q}'}
\eea
where $V_{\bsl{b}}$ is the unitary embedding matrix. (Imposing a finite momentum space cutoff breaks the momentum space periodicity and spoils the unitarity of $V_{\bsl{b}}$ due to the cutoff boundary. However, since the low energy eigenstates $U(\bsl{k})$ decay exponentially for large $\bsl{b}$, this effect is negligible.) Thus we employ a periodic convention where
\bea
U(\bsl{k}+\bsl{b}) &= V_{\bsl{b}} U(\bsl{k}), \quad U_{\bsl{Q},n}(\bsl{k}+\bsl{b}) =  U_{\bsl{Q}+\bsl{b},n}(\bsl{k}) \\
\psi_{\bsl{k}+\bsl{b},n}(\bsl{r}) &= \frac{1}{\sqrt{N}} \sum_{\bsl{Q}} e^{i (\bsl{k}+\bsl{b}+\bsl{Q}) \cdot \bsl{r}} U_{\bsl{Q}+\bsl{b},n}(\bsl{k})  =  \psi_{\bsl{k},n}(\bsl{r}) \ .
\eea
In practice, this means one need only compute $U(\bsl{k})$ in the first BZ. Band structures at the experimentally relevant angle $\th = 3.7^\circ$ are show in \Fig{fig:4plots} for all sets of parameters in \Tab{tab:paramtable}.

\subsection{Space Group Symmetries and Topology}

We now discuss the intra-valley space group $G$ of the model. The symmetry representations are obtained from the transformation properties of the $d_{x^2-y^2} + i d_{xy}$ orbitals in the top and bottom layers for valley $K$. A $C_3$ operation takes
\bea
C_3 \bpm d_{x^2-y^2} \\ d_{xy} \epm = \left(
\begin{array}{cc}
-\frac{1}{2} & \frac{\sqrt{3}}{2} \\
-\frac{\sqrt{3}}{2} & -\frac{1}{2} \\
\end{array}
\right) \bpm d_{x^2-y^2} \\ d_{xy} \epm
\eea
so that the $d_{x^2-y^2} + i d_{xy}$ orbital has $C_3$ eigenvalue $e^{i 2\pi/3}$. Similarly $C_{2y} \mathcal{T}$ acts trivially on the $d_{x^2-y^2} + i d_{xy}$ orbital, but interchanges the layer. Thus we have the representations
\bea
D[C_{3z}] = e^{\frac{2\pi i}{3}} \sigma_0, \quad D[C_{2y}\mathcal{T}] = \sigma_1 \mathcal{K}
\eea
where $\sigma_i$ are the Pauli matrices, $\mathcal{K}$ is complex conjugation, and the representations obey $D[g] H_K(\bsl{r}) D^\dag[g] = H_K(g\bsl{r})$ (see \Eq{eq:Hkmacdonaldapp}). We take $C_{3z} \mbf{r} = R(2\pi/3) \mbf{r}$ and $C_{2y}\mathcal{T}(x,y) =(-x,y)$. Remarkably, the Hamiltonian inherits another symmetry due to the lowest order harmonic approximation in \Eq{eq:Hkmacdonaldapp}: a pseudo-inversion symmetry $D[\mathcal{I}] = \sigma_1$ which obeys
\bea
D[\mathcal{I}] H_K(\bsl{r}) D^\dag[\mathcal{I}] = H_K(-\bsl{r})  \ .
\eea
We emphasize that there is no microscopic inversion symmetry in the moir\'e system. Indeed, keeping higher order terms will break $D[\mathcal{I}]$. One example is $t(\bsl{r}) \to t(\bsl{r})  + t'(\bsl{r})$ where (with $w'$ a real number)
\bea
t'(\bsl{r}) = i w' \sum_{n=1}^3 e^{i (-2 \mbf{q}_n) \cdot \mbf{r} }
\eea
which preserves $C_{3z}$ and $C_{2y}\mathcal{T}$. The \emph{ab-initio} spectrum in \refcite{wang2023fractional} does indeed show small $\mathcal{I}$-breaking in the band structure, but these effects appear to be less than $\sim1$meV and can be neglected at leading order.

Using the symmetries of the model, the \emph{generic} appearance of Chern bands can be predicted from topological quantum chemistry\cite{2017Natur.547..298B,2021NatCo..12.5965E}. Consider the large $V$ limit where inter-layer hopping $w$ and kinetic energy are sub-leading terms. Then we expect the valence band eigenstates to be formed from ``atomically" localized wavefunctions at $\bsl{r}_\pm$, the minima of $-V_\pm(\bsl{r})$ in each layer. We find that
\bea
\bsl{r}_\pm =
\begin{cases}
\pm \frac{1}{3}(\mbf{a}_1+\mbf{a}_2), & -\pi < \psi < -\pi/3 \\
\mbf{0}, & |\psi| < \pi/3 \\
\mp \frac{1}{3}(\mbf{a}_1+\mbf{a}_2), & \pi/3 < \psi < \pi \\
\end{cases}
\eea
meaning that, if $|\psi| > \pi/3$, the atomic states are centered at opposite corners of the moir\'e unit cell related by $C_{2y}\mathcal{T}$. This is the 2b Wyckoff position. Consulting the Bilbao crystallographic server\cite{Aroyo:firstpaper,Aroyo:xo5013}, we find that any atomic states induced from this position realize a decomposable elementary band representation formed by two disconnected bands with Chern number $\pm 1 \mod 3$ \cite{PhysRevE.96.023310,2018PhRvB..97c5139C}. This is in agreement with the numerically calculated Chern numbers for the parameters of \refcite{wang2023fractional} where $V/w > 1$, as shown in the Main Text. This shows that topology is inherent in this model, no matter what terms are added (as long as they are small), in a similar way as in twisted bilayer graphene.

We now turn to the representations in momentum space. A space group symmetry $g \in G$ has a matrix representation $D[g]$ on the momentum-space Hamiltonian \Eq{eq:hkK} obeying
\bea
D[g] H^K(\bsl{k}) D[g]^{-1} &= H^K(g \bsl{k})  \ .
\eea
Similarly, the (unitary and anti-unitary) momentum space representations of the true space group symmetries are
\bea
D_{\bsl{Q},\bsl{Q}'}[C_{3z}] = \delta_{\bsl{Q}, C_{3z} \bsl{Q}'}, \quad D_{\bsl{Q},\bsl{Q}'}[C_{2y} \mathcal{T}] = \delta_{\sigma_3 \bsl{Q},\bsl{Q}'} \cc, \quad D_{\bsl{Q},\bsl{Q}'}[\mathcal{I}] = \delta_{\bsl{Q}, -\bsl{Q}'}
\eea
where we abuse notation by taking $C_{2y} \mathcal{T} (k_x,k_y) = (k_x,-k_y) = \sigma_3 \bsl{k} $ as the momentum vector representation (the position space representation is $C_{2y} \mathcal{T} (x,y) = (-x,y)$ since $\mathcal{T}$ is local), and $\cc$ is complex conjugation. Note that $C_{2y} \mathcal{T}$ and pseudo-inversion $\mathcal{I}$ flip the layers, while $C_{3z}$ does not. These symmetries form the 3D magnetic space group 164.89 ($P\bar{3}m'1$), which is isomorphic in 2D to the magnetic wallpaper group $6m'm'$ where the six-fold pseudo-rotation is $C_{3z}^{-1} \mathcal{I}$. We refer to $\mathcal{I}$ as a pseudo-inversion because it is intra-valley, while a true inversion symmetry would exchange the $K$ and $K' = -K$ valleys. Finally, the Hamiltonian in the $K'$ valley can be obtained from the spinful time-reversal symmetry of the underlying TMD model, which acts as $i \tau_2 \mathcal{K}$ and maps e.g. the $d_{x^2-y^2} + i d_{xy}$ spin $\uparrow$ orbitals in valley $K$ to $d_{x^2-y^2} - i d_{xy}$ spin $\downarrow$ orbitals at valley $K'$. (Here $\tau$ is the valley matrix.) Thus the Hamiltonian at $K'$ is
\bea
H^{K'}_{\bsl{Q},\bsl{Q}'}(\bsl{k}) = H^K_{-\bsl{Q},-\bsl{Q}'}(-\bsl{k})^* \ .
\eea
Since the two valleys are related by time-reversal and the model preserves $U(1)$ valley symmetry, we can choose a gauge $U^\eta_{\bsl{Q},n}(\bsl{k}) = \overline{U}^\eta_{\eta \bsl{Q},n}(\eta\bsl{k})$ where $\eta = \pm$ denotes the $K/K'$ a valley, and $\overline{x}^+ = x, \overline{x}^- = x^*$.

\begin{figure}
\centering
\includegraphics[width=1.\columnwidth]{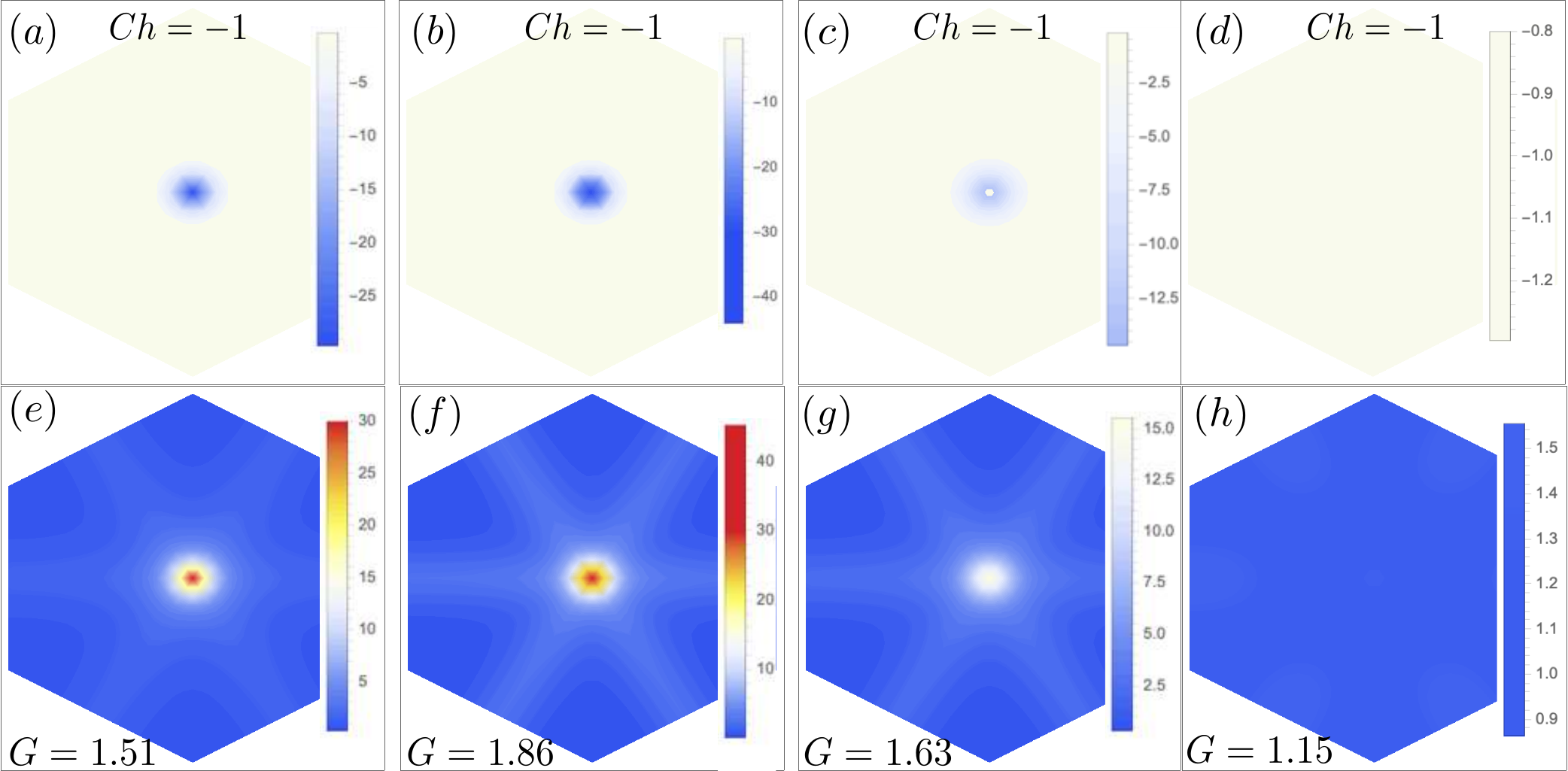}
\caption{Comparison of the highest valence band Berry curvature ($a-d$) and Fubini-Study metric ($e-h$) for the parameter values in \Tab{tab:paramtable},  corresponding to rows 1-4. Again we see strong similarity between the first three rows where the $\Gamma_M$ point shows a strong peak in the quantum geometry, compared to the nearly uniform features in $d,h$. Color bars are consistent across all plots.}
\label{fig:4plotsGQ1}
\end{figure}

To compute the Berry curvature and quantum geometry of the bands, we use the integrals \cite{PhysRevLett.98.046402}
\bea
\label{eq:Cberry}
\frac{\Ch}{2\pi} &= \int \frac{d^2k}{(2\pi)^2} \frac{i}{2} \eps_{ij} \Tr  P(\mbf{k})[\del_i P(\mbf{k}), \del_j P(\mbf{k})] \\
\frac{G}{2\pi} &= \int \frac{d^2k}{(2\pi)^2} \frac{1}{2} \Tr  \del_i P(\mbf{k})\del_i P(\mbf{k}) \\
\eea
where $P(\mbf{k}) = U(\mbf{k})U^\dag(\mbf{k})$ is the gauge-invariant projector on the single-particle eigenvector $U(\mbf{k})$. The advantage of the projector formalism is that no smooth gauge is necessary, and discretization can be done efficiently \cite{2022PhRvL.128h7002H}. The integrands of \Eq{eq:Cberry} are the Berry curvature and Fubini-Study metric respectively. We report the Chern numbers $\Ch$ and integrated Fubini-Study metric $G$, along with their structure over the moir\'e Brillouin zone in \Figs{fig:4plotsGQ1}{fig:4plotsGQ2} for the two bands included in our interacting calculations.

\begin{figure}
\centering
\includegraphics[width=1.\columnwidth]{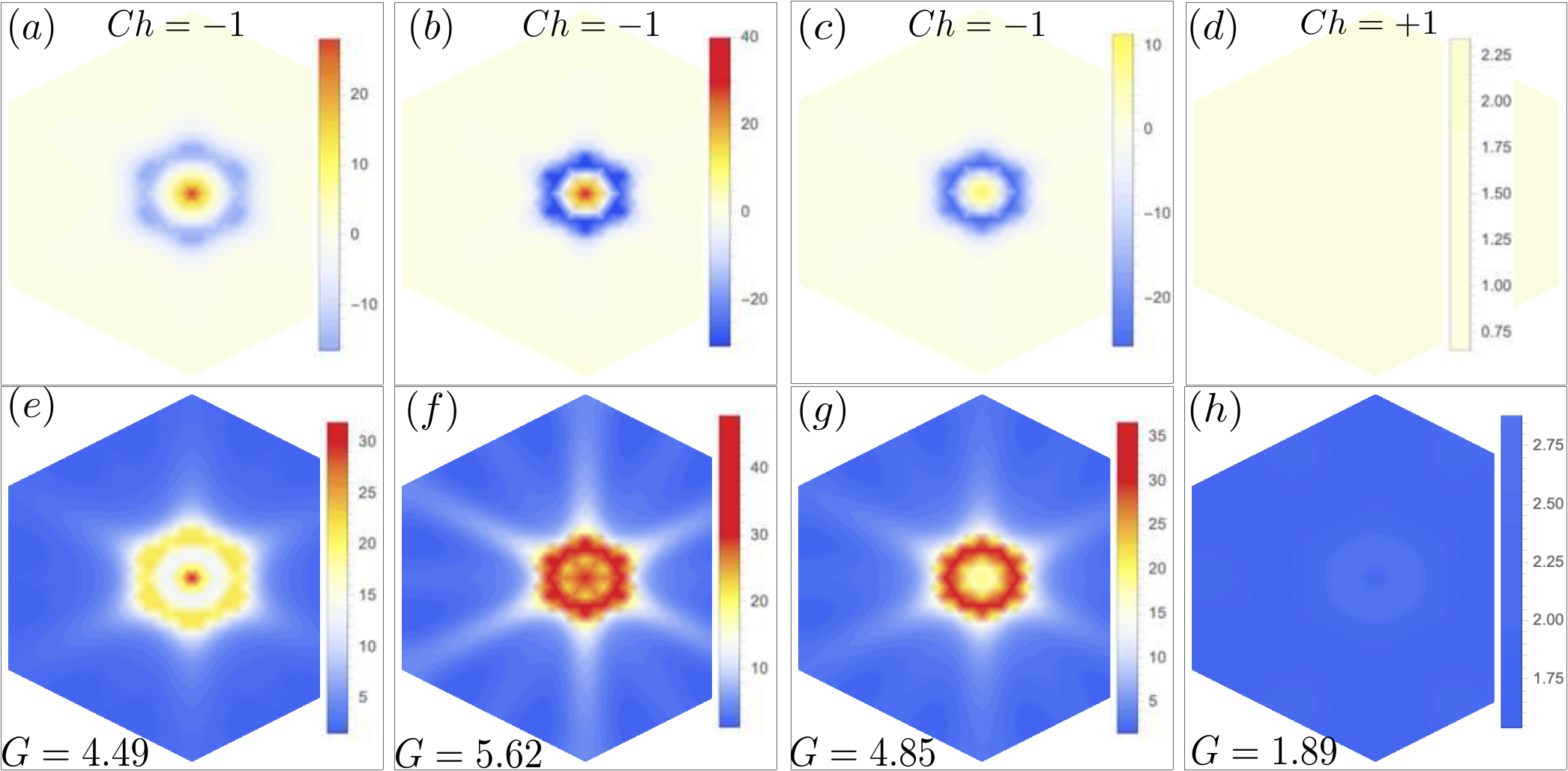}
\caption{Comparison of the second highest valence band Berry curvature ($a-d$) and Fubini-Study metric ($e-h$) for the parameter values in \Tab{tab:paramtable},  corresponding to rows 1-4. Color bars are consistent across all plots.}
\label{fig:4plotsGQ2}
\end{figure}

\section{More details on Interaction and particle-hole symmetry}
\label{app:int}

In this appendix, we discuss two types of the interaction and the PH transformations of the projected Hamiltonians.

\subsection{Interaction in the electron and hole language}

In the previous theoretical works
~\cite{Li2021FCItTMD,reddy2023fractional,wang2023fractional,Dong2023CFLtMoTe2,Goldman2023CFLtMoTe2,Reddy2023GlobalPDFCI,Xu2023MLWOFCItTMD,Zaletel2023tMoTe2FCI} on FCIs in {\tmt}, two types of interactions are mainly used. One is the Coulomb interaction among holes, used in \refcite{reddy2023fractional}, for which the explicit Hamiltonian is
\eqa{
H_{h} & = \int d^2 r \sum_{\eta,ll'} \widetilde{c}^\dagger_{\eta,l,\bsl{r}} \left[ - H_{\eta} (\bsl{r}) \right]_{ll'} \widetilde{c}_{\eta, l',\bsl{r}}  + \frac{1}{2}\sum_{l \eta,l' \eta'} \int d^2 r d^2 r' V(\bsl{r}-\bsl{r}')  \widetilde{c}^\dagger_{\eta, l, \bsl{r}} \widetilde{c}^\dagger_{\eta', l', \bsl{r}'} \widetilde{c}_{\eta', l', \bsl{r}'} \widetilde{c}_{\eta, l, \bsl{r}}  \ ,
}
where $\widetilde{c}_{\eta, l,\bsl{r}}^\dagger$ creates a hole at position $\bsl{r}$ in the $l$th layer in the $\eta$ valley and $H_{\eta} (\bsl{r})$ is the single-particle electron Hamiltonian, and $V(\bsl{r})$ is the double-gated screened Coulomb potential with gate distance  $\xi$:
\bea
V(\bsl{r}) =\int_{\dsR^2} \frac{d^2 p }{ (2\pi)^2} V(\bsl{p}) e^{\ii \bsl{p}\cdot\bsl{r}}, \qquad V(\bsl{p}) = \pi \xi^2 V_{\xi} \frac{\tanh(\xi |\bsl{p}|/2)}{\xi |\bsl{p}|/2}, \quad V_{\xi} = \frac{e^2 }{4\pi \epsilon \epsilon_0 \xi} \ .
\eea
Here $\epsilon_0$ is the vacuum permitivity, and $\epsilon$ is the dielectric constant.
The vacuum state of $\widetilde{c}_{\eta, l, \bsl{r}}$ is the charge neutrality point$\ket{\nu = 0}$ with $\widetilde{c}_{\eta, l, \bsl{r}} \ket{\nu = 0} = 0$, which corresponding to fully filling all the electron states of the single-particle Hamiltonian, and the Hilbert space is just given by acting various powers of $\widetilde{c}_{\eta, l, \bsl{r}}^\dagger$ on $\ket{\nu = 0}$.

The other common Hamiltonian is the Coulomb interaction among electrons, which is used in Refs.\,\cite{Li2021FCItTMD,wang2023fractional}:
\eqa{
H_{e} & = \int d^2 r \sum_{\eta,ll'} c^\dagger_{\eta,l, \bsl{r}}   \left[ H_{\eta} (\bsl{r}) \right]_{ll'}  c_{\eta, l',\bsl{r}} + \frac{1}{2}\sum_{l\eta,l' \eta'} \int d^2 r d^2 r' V(\bsl{r}-\bsl{r}')   c^\dagger_{\eta, l, \bsl{r}} c^\dagger_{\eta', l', \bsl{r}'} c_{\eta', l', \bsl{r}'} c_{\eta, l, \bsl{r}}  \ ,
}
where $c_{\eta,l,\bsl{r}}^\dagger$ creates an electron at position $\bsl{r}$ in the $l$th layer in the $\eta$ valley.
Importantly, charge neutrality $\ket{\nu = 0}$ is not the vacuum of $c_{\eta, l, \bsl{r}}$, \ie, $c_{\eta, l, \bsl{r}}\ket{\nu = 0} \neq 0$. Instead, the Fock vacuum obeys $c_{\eta, l, \bsl{r}} \ket{0} = 0$.

The Hilbert space of $H_{e}$ is the same as $H_{h}$, as $c_{\eta,l,\bsl{r}}^\dagger$ is related to  $\tilde{c}_{\eta,l,\bsl{r}}$ by
\eq{
\label{eq:c_ctilde_complexconjugate}
\cc c_{\eta,l,\bsl{r}}^\dagger \cc^{-1} =  \widetilde{c}_{\eta,l,\bsl{r}} \ ,
}
where $\cc$ is a complex conjugate.
As a result, $H_e$ differs from $H_h$ by the complex conjugate and a single-body term:
\eqa{
\cc    H_{e}  \cc^{-1} & = H_{h} + \Delta H + const. \ ,
}
where
\eqa{
\Delta H & =  \left( V(\bsl{0}) -4 \int d^2 r' V(\bsl{r}') \delta(0)   \right) \sum_{l\eta} \int d^2 r   \widetilde{c}^\dagger_{\eta, l, \bsl{r}}   \widetilde{c}_{\eta, l, \bsl{r}}  \ ,
}
and we have used that $V(\bsl{r})$ is real.
Naively, $\Delta H$ is proportional to the particle-number operator, and thus one may think in a fixed-particle-number sector, $\Delta H $ becomes a constant, and the two Hamiltonian are related by a complex conjugate up to a constant.
However, in reality, we can never use the continuous Hamiltonian to do HF or ED calculations; instead, we will project the Hamiltonian into a finite set of bands.
We will see in the following that after the projection, the two Hamiltonians are not necessarily related due to one-body terms that appear from the different normal orderings.

As just mentioned, we project the Hamiltonian into a finite set of band spanned by the operators
\eq{
\widetilde{\gamma}^\dagger_{n, \eta,\bsl{k}} = \sum_{\bsl{Q}} \widetilde{c}^\dagger_{\eta,\bsl{k},\bsl{Q}} \left[ U_{n, \eta,\bsl{k}} \right]_{\bsl{Q}}\ ,
}
where the band index $n$ only takes a finite set of values.
The projected Hamiltonian in this basis is
\eqa{
H_{h} & = - \sum_{\bsl{k},\eta,n}\widetilde{\gamma}^\dagger_{\eta,n,\bsl{k}} \widetilde{\gamma}_{\eta,n,\bsl{k}}  E_{\eta,n}(\bsl{k}) + \frac{1}{2} \sum_{\bsl{k},\bsl{k}',\bsl{q}} \sum_{\eta,\eta'} \sum_{m, m', n', n}V_{\eta \eta', m m' n' n}(\bsl{k},\bsl{k}',\bsl{q}) \widetilde{\gamma}^\dagger_{\eta,m,\bsl{k}+\bsl{q}} \widetilde{\gamma}^\dagger_{\eta',m',\bsl{k}'-\bsl{q}} \widetilde{\gamma}_{\eta', n',\bsl{k}'}
\widetilde{\gamma}_{\eta,n,\bsl{k}}\ ,
}
where
\bea
V_{\eta \eta', m m' n' n}(\bsl{k},\bsl{k}',\bsl{q}) &= \frac{1}{N}\sum_{\bsl{b}} \frac{V(\bsl{q}+\bsl{b})}{\Omega}  M_{\eta, m n }(\bsl{k},\bsl{q}+\bsl{b})  M_{\eta', m' n'}^*(\bsl{k}'-\bsl{q},\bsl{q}+\bsl{b}) \\
M_{\eta, m n}(\bsl{k},\bsl{q}) &= U_{\eta,m, \bsl{k}+\bsl{q}}^\dagger U_{\eta, n, \bsl{k}} \\
\Omega^{-1} &= \bsl{b}_{1} \times \bsl{b}_{2} / (2\pi)^2
\eea
are the matrix elements, form-factors, and (inverse) moir\'e unit cell area respectively.

Hermiticity requires
\eq{
V^*_{\eta\eta',nn'm'm}(\bsl{k}+\bsl{q},\bsl{k}'-\bsl{q},-\bsl{q}) = V_{\eta\eta', m m' n' n}(\bsl{k},\bsl{k}',\bsl{q})\ ,
}
and the fermion statistics gives
\eq{
V_{\eta'\eta,m'mnn'}(\bsl{k}',\bsl{k},-\bsl{q}) = V_{\eta\eta', m m' n' n}(\bsl{k},\bsl{k}',\bsl{q})\ .
}
On the other hand, after projection into the band basis $\gamma^\dagger_{n, \eta,\bsl{k}} = \sum_{\bsl{Q}} c_{\eta,\bsl{k},\bsl{Q}}^\dagger \left[ U_{n, \eta,\bsl{k}} \right]_{\bsl{Q}}$, $H_{e}$ becomes
\eqa{
\label{eq:H_int_edagedagee}
H_{e} = \sum_{\bsl{k},\eta,n} \gamma^\dagger_{\eta,n,\bsl{k}} \gamma_{\eta,n,\bsl{k}} E_{\eta,n}(\bsl{k}) + \frac{1}{2} \sum_{\bsl{k},\bsl{k}',\bsl{q}} \sum_{\eta,\eta'} V_{\eta \eta' , m m' n' n}(\bsl{k},\bsl{k}',\bsl{q}) \gamma^\dagger_{m, \eta,\bsl{k}+\bsl{q}} \gamma^\dagger_{m', \eta',\bsl{k}'-\bsl{q}} \gamma_{n',\eta',\bsl{k}'}
\gamma_{n, \eta,\bsl{k}}\ .
}

Now we show the relation between the two Hamiltonian after the projection.
According to \eqnref{eq:c_ctilde_complexconjugate}, we have $\widetilde{c}_{\eta,\bsl{k},\bsl{Q}}^\dagger = \mathcal{K} c_{\eta,\bsl{k},\bsl{Q}} \mathcal{K}$, and then $\gamma_{n,\eta,\bsl{k}}$ and $\widetilde{\gamma}_{n,\eta,\bsl{k}}^\dagger$ are related by
\eq{
\label{eq:gamma_gammatilde_complexconjugate}
\cc \gamma_{n,\eta,\bsl{k}} \cc^{-1} =  \widetilde{\gamma}_{n,\eta,\bsl{k}}^\dagger \ .
}
Under the $\cc$, now the projected Hamiltonians are related as
\eqa{
\label{eq:He_Hh_relation_projected}
\cc H_{e} \cc^{-1} & = \sum_{\eta,\bsl{k},n m}\widetilde{\gamma}_{\eta,n,\bsl{k}}^\dagger \widetilde{\gamma}_{\eta,m,\bsl{k}} \epsilon_{\eta,n m}(\bsl{k})  + H_{h}\ ,
}
where
\eqa{
\label{eq:epsilon_PH}
\epsilon_{\eta,n m}(\bsl{k}) &= -\frac{1}{2}\sum_{\bsl{k}',\eta',n'}\left[ V^*_{\eta\eta',m n' n' n}(\bsl{k},\bsl{k}',0)+V_{\eta\eta', n n' n' m}(\bsl{k},\bsl{k}',0) \right] \\
& \quad + \frac{1}{2}\sum_{\bsl{q} n'} \left[ V_{\eta\eta, n' n n' m}(\bsl{k},\bsl{k}+\bsl{q},\bsl{q}) + V_{\eta\eta, n' m n' n}^*(\bsl{k},\bsl{k}+\bsl{q},\bsl{q})\right]\ ,
}
which in general is not zero as we only consider a finite sets of $n$ and $m$.
The non-vanishing $\epsilon_{\eta,n m}(\bsl{k})$ is consistent with the different HF phase diagrams given by $H_e$ and $H_h$ in \figref{fig:HF_x18y18_main}. Physically, $\epsilon_{\eta,n m}(\bsl{k})$ represents the different background potential felt by the electron interaction

\subsection{Particle-Hole Symmetry}\label{app:int:phsymmetry}

In this part, we only talk about $H_h$ and $H_e$ after projection to a finite set of bands.
Neither of the Hamiltonians $H_h$ and $H_e$ preserves the PH symmetry exactly in general after the projection.
Here the PH transformation for $H_e$ is defined as
\eq{
\label{eq:PHtransformation_app}
\C \gamma^\dagger_{\eta,n,\bsl{k}} \C^{-1} = \gamma_{\eta,n,\bsl{k}}\ ,
}
where $\C$ is an anti-unitary operator.
Under this PH transformation, $H_{e}$ would gain two extra terms:
\eqa{
\label{eq:H_e_PH_app}
\C H_{e} \C^{-1} &  = H_e  + \sum_{\eta,\bsl{k},n}\gamma_{\eta,n,\bsl{k}}^\dagger \gamma_{\eta,n,\bsl{k}} (-2 E_{\eta,n}(\bsl{k})) + \sum_{\eta,\bsl{k},n m}\gamma_{\eta,n,\bsl{k}}^\dagger \gamma_{\eta,m,\bsl{k}} \epsilon_{\eta,n m}(\bsl{k}) + const.\ ,
}
where $(-2 E_{\eta,n}(\bsl{k}))$ accounts for the sign flipping of the single-particle dispersion, and $\epsilon_{\eta,n m}(\bsl{k}) $ is given in \eqnref{eq:epsilon_PH}.
The reason that $\epsilon_{\eta,n m}(\bsl{k}) $ in \eqnref{eq:H_e_PH_app} is the same as that in \eqnref{eq:He_Hh_relation_projected} is that the transformation in \eqnref{eq:gamma_gammatilde_complexconjugate} will become the PH transformation in \eqnref{eq:PHtransformation_app} if we identify $\widetilde{\gamma}_{\eta,n,\bsl{k}}$ in \eqnref{eq:gamma_gammatilde_complexconjugate} with $\gamma^\dag_{\eta,n,\bsl{k}}$.

For $H_h$, the PH transformation is given by replacing $\gamma_{\eta,n,\bsl{k}}$ in \eqnref{eq:PHtransformation_app} by $\widetilde{\gamma}_{\eta,n,\bsl{k}}$, and the PH breaking term $\C H_{h} \C^{-1} - H_h$ is just given by performing  $E_{\eta,n}(\bsl{k})\rightarrow - E_{\eta,n}(\bsl{k})$ in the PH-breaking term in \eqnref{eq:H_e_PH_app}:
\eqa{
\label{eq:H_h_PH_app}
\C H_{h} \C^{-1} &  = H_h  + \sum_{\eta,\bsl{k},n}\widetilde{\gamma}_{\eta,n,\bsl{k}}^\dagger \widetilde{\gamma}_{\eta,n,\bsl{k}} (2 E_{\eta,n}(\bsl{k})) + \sum_{\eta,\bsl{k},n m}\widetilde{\gamma}_{\eta,n,\bsl{k}}^\dagger \widetilde{\gamma}_{\eta,m,\bsl{k}} \epsilon_{\eta,n m}(\bsl{k}) + const.\ .
}
The extra one-body term given by the PH transformation is in general non-vanishing, meaning that $H_{e}$ and $H_{h}$ are not invariant under this transformation.

When the PH breaking effect is strong, i.e. $\eps_{\eta,nm}(\mbf{k})$ is large, the two fillings related by PH transformation would have considerable difference.
It is also important to note that as we are talking about the PH transformation of the projected Hamiltonian, the PH transformation only acts on the finite projected subspace, and thus the PH-related fillings depend on the how many bands we keep in the projected model.
If we consider the fully-spin-polarized case where we only keep one valley (spin) and one band in that valley, the PH transformation maps $\nu$ to $-1-\nu$ (\eg, $-1/3$ to $-2/3$) electron filling.
If we keep two valleys and one band per valley, the PH transformation maps $\nu$ to $-2-\nu$ (\eg, $-2/3$ to $-4/3$) electron filling.
In general, if we keep two valleys and $N$ band per valley, the PH transformation maps $\nu$ to $-2N-\nu$.

\section{Additional Hartree-Fock Results at $\nu=-1$}
\label{app:HF_results}

In this appendix, we provide more details on the Hartree-Fock (HF) calculations at $\nu=-1$.
We note that system size $L_1\times L_2$ means that the momenta included in the calculation are $(n/L_1) \bsl{b}_1 + (m/L_2) \bsl{b}_2$ with $n=0,...,L_1-1$ and $m=0,...,L_2-1$, where $\bsl{b}_1$ and $\bsl{b}_2$ are defined in \eqnref{eq:momentum_convention}.

\begin{figure}[H]
\centering
\includegraphics[width=0.9\columnwidth]{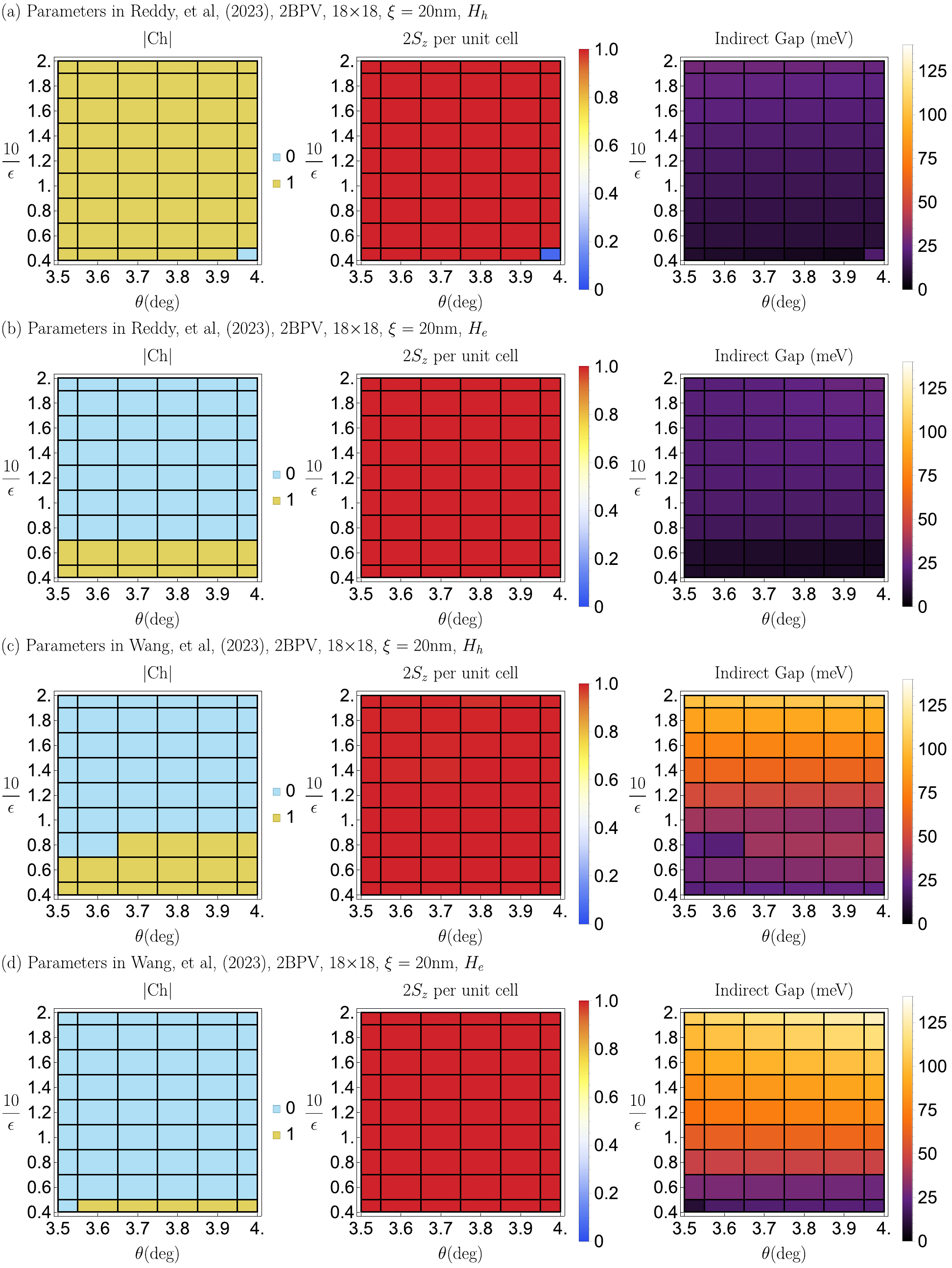}
\caption{
The 2BPV Hartree-Fock results for $\xi=20nm$ at $\nu=-1$ for the parameters in \refcite{reddy2023fractional} (a,b) and the parameters in \refcite{wang2023fractional} (c,d).
$18\times 18$ labels the system size.
$H_h$ (a,c) and $H_e$ (b,d) refer to the Hamiltonian used.
$\Ch$ refers to the Chern number (leftmost panel), the middle panel shows the value of $2S_z$ per unit cell, and the indirect gap (right panel) is the indirect gap of the HF band structure.
The spin-unpolarized state in (a) is inter-valley coherent translationally-breaking state with wavevector $\K_M$.
}
\label{fig:HF_x18y18_xi20_app}
\end{figure}

\begin{figure}[H]
\centering
\includegraphics[width=0.9\columnwidth]{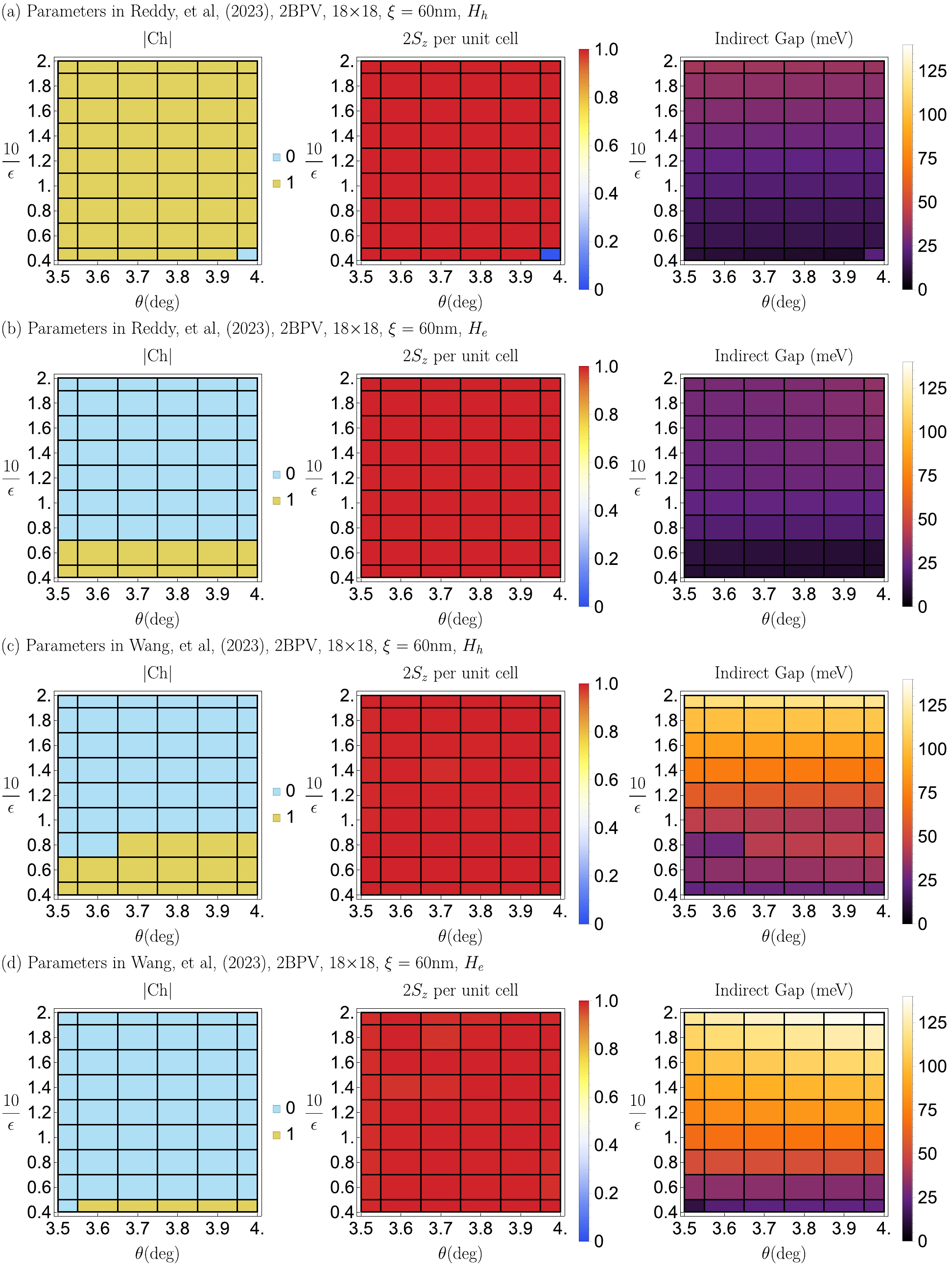}
\caption{
The 2BPV Hartree-Fock results for $\xi=60nm$ at $\nu=-1$ for the parameters in \refcite{reddy2023fractional} (a,b) and the parameters in \refcite{wang2023fractional} (c,d).
The notations are identical to those in \figref{fig:HF_x18y18_xi20_app}.
}
\label{fig:HF_x18y18_xi60_app}
\end{figure}

\begin{figure}[H]
\centering
\includegraphics[width=0.9\columnwidth]{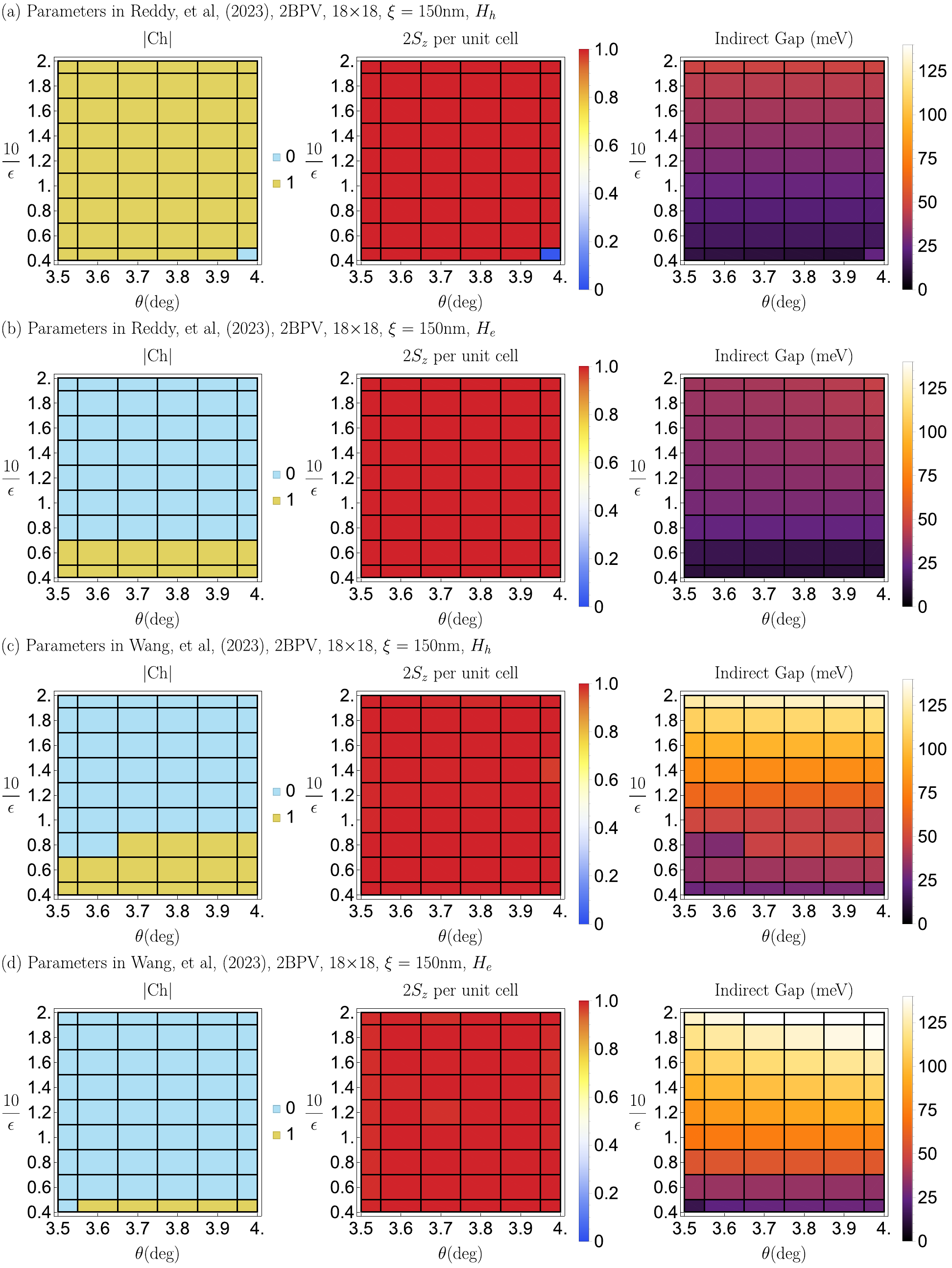}
\caption{
The 2BPV Hartree-Fock results for $\xi=150nm$ at $\nu=-1$ for the parameters in \refcite{reddy2023fractional} (a,b) and the parameters in \refcite{wang2023fractional} (c,d).
The notations are identical to those in \figref{fig:HF_x18y18_xi20_app}.
}
\label{fig:HF_x18y18_xi150_app}
\end{figure}

As mentioned in the Main Text, we have considered the effect of the screening length $\xi$.
We provide the HF phase diagrams for the system size of $18\times 18$ in \figref{fig:HF_x18y18_xi20_app} for $\xi=20$nm, in \figref{fig:HF_x18y18_xi60_app} for $\xi=60$nm and in \figref{fig:HF_x18y18_xi150_app} for $\xi=150$nm (as a reminder, we used $\xi=20$nm in the Main Text).
We choose two types of initial states:(i) translationally invariant initial states (with random band mixing), and (ii) translationally-breaking intervalley-coherent states with wavevector $\K_M$~\cite{Zaletel2023tMoTe2FCI}.
We can see the phase diagrams of the ferromagnetism are the same for all three values of $\xi$: the ground states are always ferromagnetic except the lower right corner of the phase diagram of $H_h$ for the parameters in \refcite{reddy2023fractional}, which is IVC-$\K_M$; IVC-$\K_M$ is also found at small interaction in \refcite{Zaletel2023tMoTe2FCI} for the parameters in \refcite{reddy2023fractional}.
The phase diagrams of Chern number are also the same for all three values of $\xi$ and for the $(\theta,\epsilon)$ mesh that we choose.
The $\Ch=0$ phase that occurs as the interaction increases comes form the band inversion at $K_M$ or $K_M'$ between two bands in one valley.
The only notable effect of the screening length is that as $\xi$ increases, the indirect gap of the HF bands increases, which makes sense as larger screening length means stronger interaction.

\begin{figure}[H]
\centering
\includegraphics[width=0.5\columnwidth]{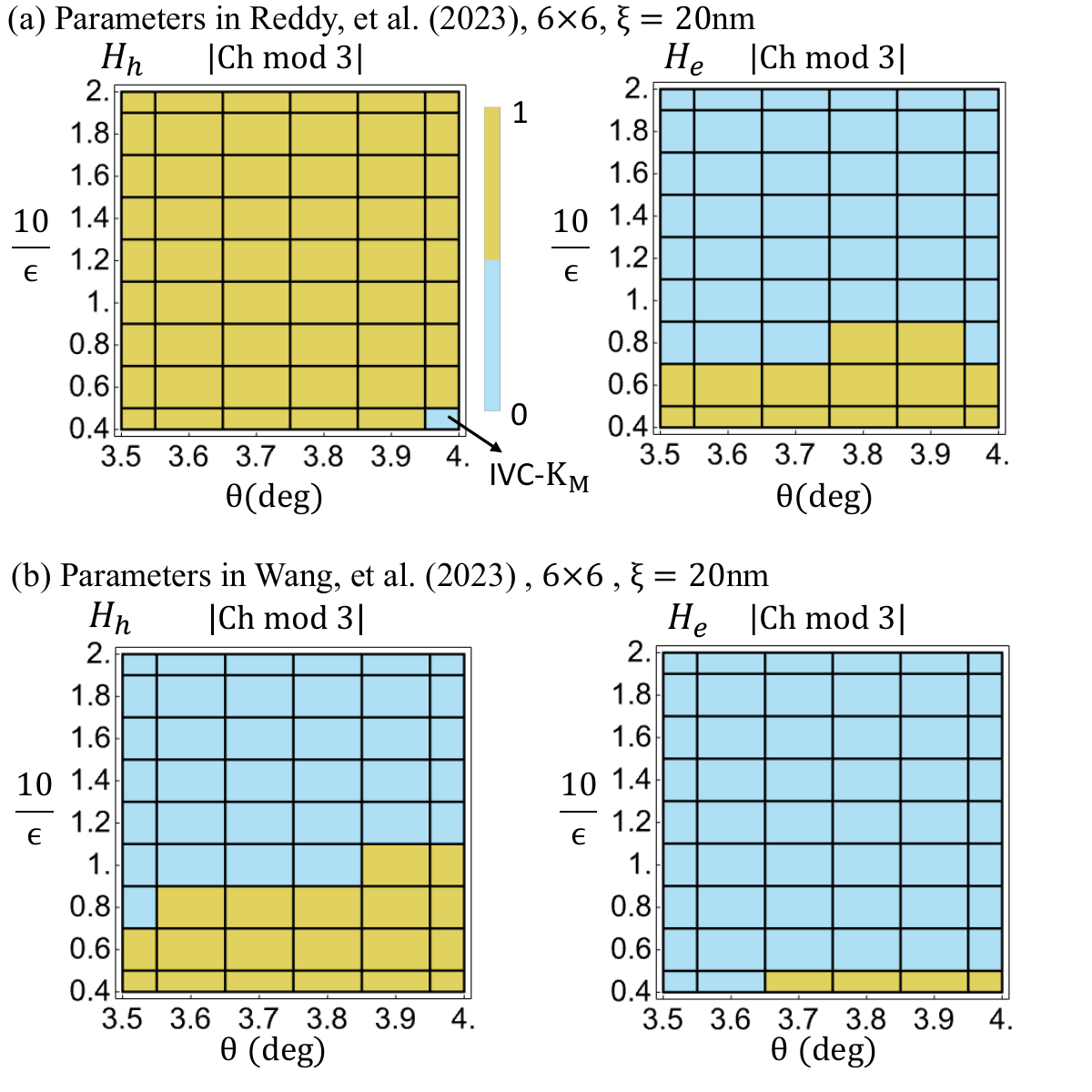}
\caption{
The 2BPV $6\times 6$ results for $\xi=20nm$ at $\nu=-1$.
The notations are analogous to \figref{fig:HF_x18y18_main} in the Main Text.
$H_h$ and $H_e$ refer to the Hamiltonian used, $\Ch$ refers to the Chern number, and we take $\Ch \mod 3 \in \{-1,0,1\}$.
The $|\Ch \mod 3| = 0 $ region for $H_h$ in (a) comes from a \emph{non}-ferromagnetic ground state, which is inter-valley coherent translationally-breaking state with wavevector $\K_M$.
Owing to the small sizes, we only use \Eq{eq:symC3} to compute $\Ch \mod 3 $  as it does not require a Berry curvature integral.
}
\label{fig:HF_x6y6_app}
\end{figure}

\begin{figure}[H]
\centering
\includegraphics[width=0.6\columnwidth]{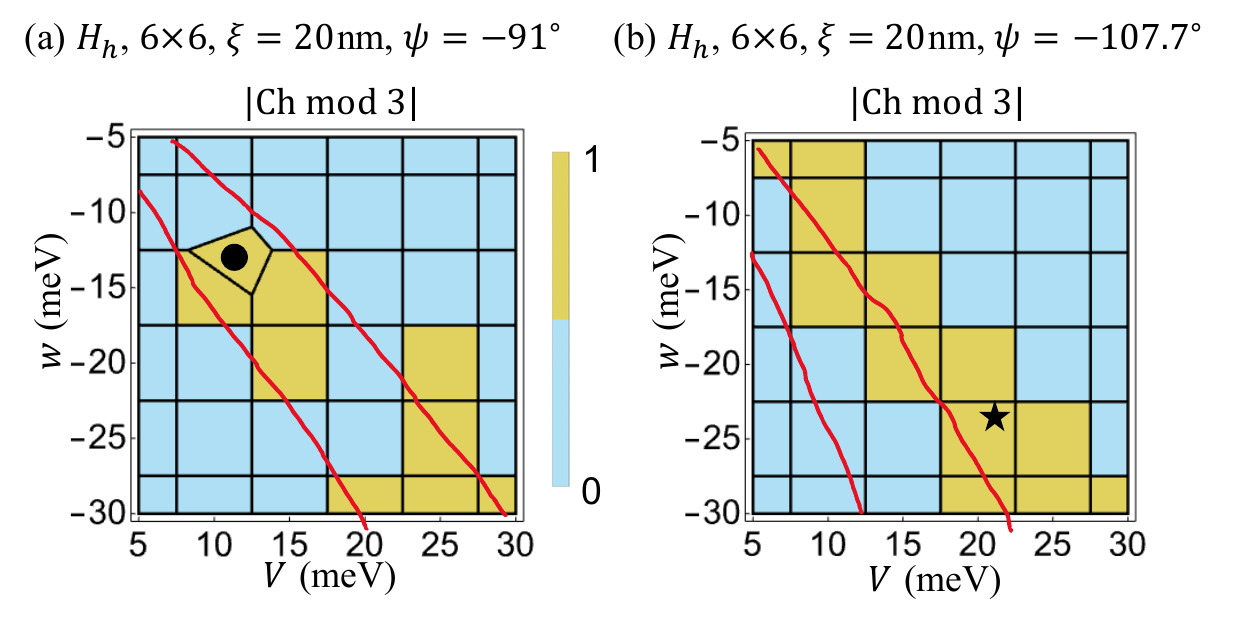}
\caption{
The $6\times 6$ 2BPV Hartree-Fock results for $\xi=20nm$ at $\nu=-1$ for the $m_*, \psi$ parameters in Ref. \cite{reddy2023fractional} (a) and Ref. \cite{wang2023fractional} (b).
We choose  $\epsilon=5$ in (a) and $\epsilon=50/3\approx 16.67$ in (b).
The boundaries of the three single-particle parameter regions are marked in red according to \Fig{fig:phasediagrammain}.
The dot and the star correspond to the parameters in  \refcite{reddy2023fractional} and \refcite{wang2023fractional}, respectively.
}
\label{fig:HF_x6_y6_VW}
\end{figure}

Since our ED calculations are done for much smaller sizes than $18\times 18$, we would like to check whether the phase diagram of the Chern numbers have qualitative changes if we go to much smaller sizes.
For this purpose, we perform $6 \times 6$ HF calculations, and map out the phase diagram for the Chern number in \figref{fig:HF_x6y6_app}.
Compared to the Chern number phase diagrams in \figref{fig:HF_x18y18_xi20_app}, we found that the Chern phase diagrams are very similar to those at $18\times 18$, meaning that the finite size effect of the Chern phase diagrams is very small.
Exploiting this small finite-size effect, we perform HF calculations across various values of  $(V,w)$ for the system size of $6\times 6$, $\theta=3.7^\circ$, and $\xi = 20$nm.
The results are shown in \figref{fig:HF_x6_y6_VW}.
We choose $\epsilon=5$ for $\psi=-91^\circ$~\cite{reddy2023fractional} and $\epsilon\approx 16.67$ for $\psi=-107.7^\circ$~\cite{wang2023fractional}, since $\epsilon=5$ and $\epsilon=50/3\approx 16.67$ are deep in the FCI region at $\nu=-2/3$ in later ED calculations for the parameters in \refcite{reddy2023fractional} and \refcite{wang2023fractional}, respectively.
As shown in \figref{fig:HF_x6_y6_VW}, the CI is robust as long as the parameters do not differ too much from those in \refcite{reddy2023fractional} and \refcite{wang2023fractional}.

\begin{figure}[H]
\centering
\includegraphics[width=0.5\columnwidth]{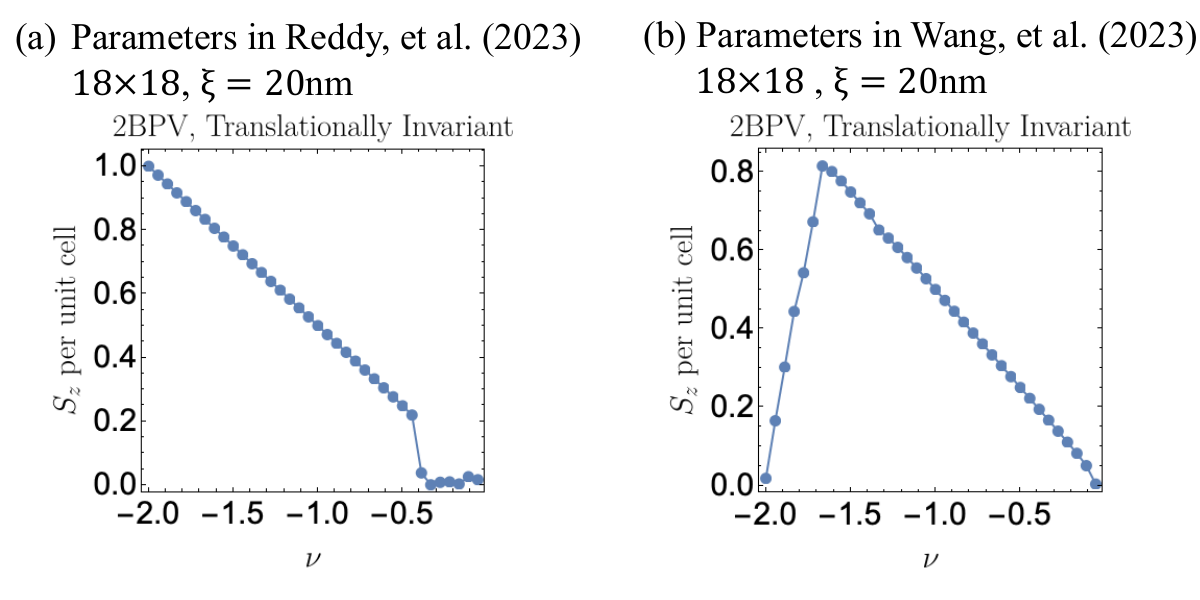}
\caption{
The 2BPV Hartree-Fock results for $H_h$ and $\xi=20nm$ at $\nu=-1$ done by restricting to the translationally invariant subspace.
We choose $\epsilon=5$ in (a) and $\epsilon=50/3\approx 16.67$ in (b).
$18\times 18$ labels the system size.
The spinful gap is smallest energy cost to change the total spin of the ground state according to the HF band structure.
In (a), the nonmagnetic states for $\nu>-0.4$ in (a) are consistent with experiments~\cite{cai2023signatures,zeng2023integer,park2023observation,Xu2023FCItMoTe2}, but the fact that the ferromagnetic ground states persists to $\nu=-2$ differs from the experiments.
In (b), the ferromagnetic ground states starts from nearly-zero $\nu$ and persist to $\nu=-1.6$, which is not consistent with experiments.
Therefore, the translationally-invariant HF cannot fully capture the behaviors of ferromagnetism observed in experiments.
}
\label{fig:HF_x18_y18_xi20_TransInv}
\end{figure}

In \figref{fig:HF_x18_y18_xi20_TransInv}, we perform the translationally-invariant subspace 2BPV Hartree-Fock calculations beyond $\nu=-1$, and reasonably the results are not reliable.
Specifically, for the parameters in \refcite{reddy2023fractional} in \figref{fig:HF_x18_y18_xi20_TransInv}(a), the nonmagnetic states for $\nu>-0.4$ are consistent with experiments~\cite{cai2023signatures,zeng2023integer,park2023observation,Xu2023FCItMoTe2}, but the fact that the ferromagnetism persists to $\nu=-2$ is different from the experiments, meaning that the calculation is not completely reliable.
For the parameters in \refcite{wang2023fractional} in \figref{fig:HF_x18_y18_xi20_TransInv}(b), the ferromagnetic ground states start from nearly-zero $\nu$ and persist to $\nu=-1.6$, which is not consistent with experiments, meaning that the calculation is not reliable.

\begin{figure}[H]
\centering
\includegraphics[width=0.5\columnwidth]{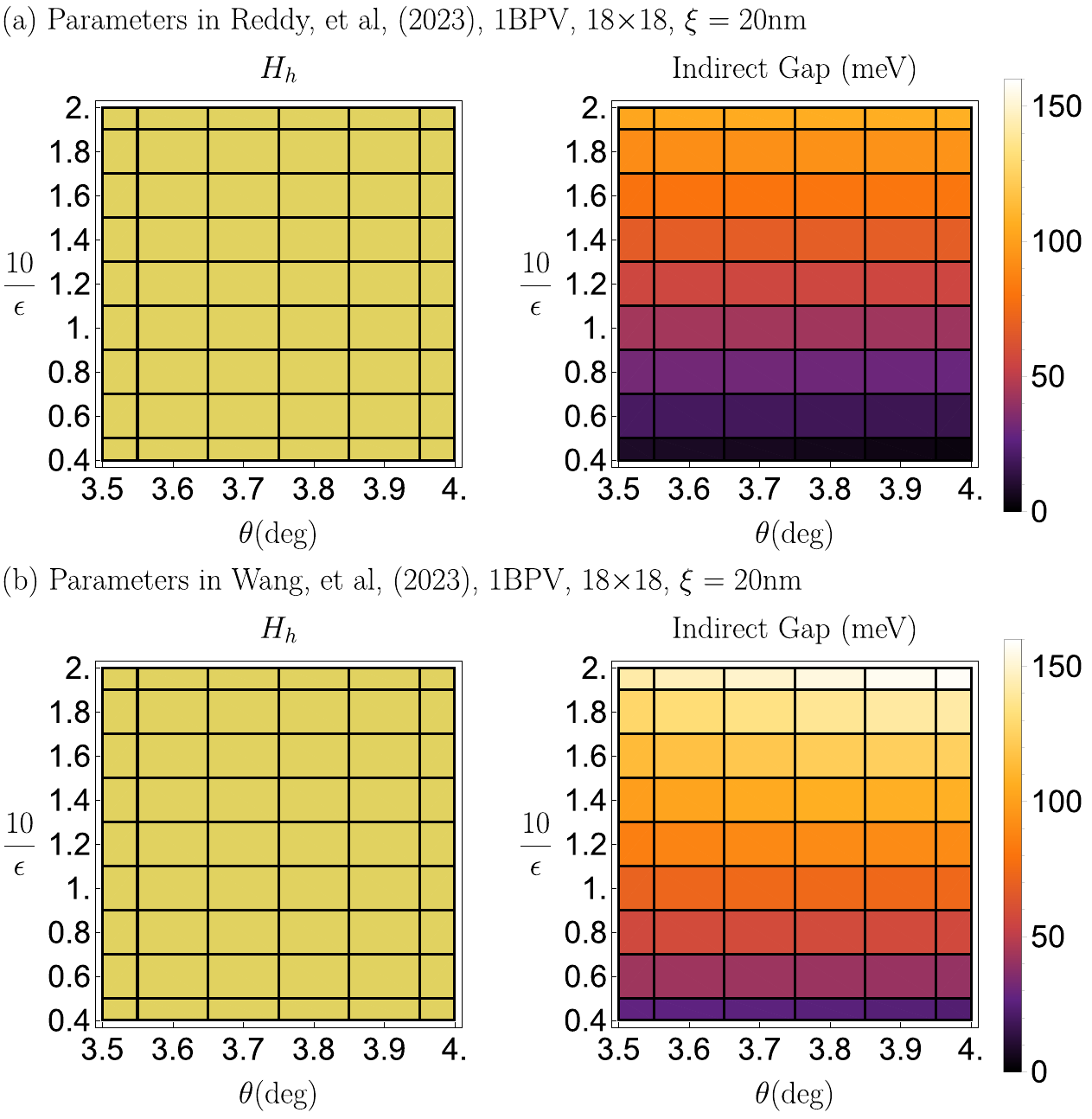}
\caption{
The 1BPV Hartree-Fock results for $H_h$ with $\xi=20nm$ at $\nu=-1$.
$18\times 18$ labels the system size, and and $\Ch$ refers to the Chern number.
The ground states are always spin-polarized and thus has nonzero Chern number; the indirect gap is the indirect gap of the HF band structure.
}
\label{fig:HF_x18y18_1BPV}
\end{figure}

As last, we also perform the 1BPV Hartree-Fock calculations for $H_h$ with $\xi=20nm$ at $\nu=-1$ as shown in \eqnref{fig:HF_x18y18_1BPV}.
The ground states are always spin-polarized and thus has nonzero Chern number for $\theta\in[3.5^\circ,4.0^\circ]$, $\epsilon\in[5,25]$ and both sets of parameters in Refs.\,\cite{reddy2023fractional} and \cite{wang2023fractional}.
CI at $\nu=-1$ was also found in the 1BPV HF calculation in \refcite{Li2021FCItTMD} outsides the experimental angle range.
Therefore, the 1BPV calculations missed the IVC-$\K_M$ states for parameters in \refcite{reddy2023fractional} and the $\Ch=0$ states for the parameters in \refcite{wang2023fractional} in the 2BPV results (\figref{fig:HF_x18y18_xi20_app}), indicating the importance of the remote bands.

\section{1BPV ED Results}
\label{app:1BPV_ED_results}

In this appendix, we provide more in-depth analysis of the 1BPV ED results.
Before going into details, we first present the definitions of the spinful and spin-1 gaps of the model that we study.
\begin{definition}[Spinful Gap]
Given a fixed particle number, the spinful gap of the system with the spin-$\U(1)$ symmetry and TR symmetry is defined as the energy difference between the lowest-energy state in all $ S_z \neq S_{max}$ sectors and the lowest-energy state in the $ S_z = S_{max}$ sector (former minus latter), where we only consider $S_z\geq 0 $ owing to the TR symmetry.
\end{definition}
\begin{definition}[Spin-1 Gap]
Given a fixed particle number, the spin-1 gap of the system with the spin-$\U(1)$ symmetry and TR symmetry is defined as the energy difference between the lowest-energy state in the $ S_z = S_{max}-1$ sector and the lowest-energy state in the $ S_z = S_{max}$ sector (former minus latter), where we only consider $S_z\geq 0 $ owing to the TR symmetry.
\end{definition}
We note that in the 1BPV case, the maximally-spin-polarized sector at $\nu=-4/3$ has the same spin as the maximally-spin-polarized sector at $\nu=-2/3$.

\subsection{Spin-1 gap as an approximation of the spinful gap}
\label{app:1BPV_ED_results:spin1spinful}

We first benchmark that the spin-1 gap is a good approximation of the spinful gap at the fractional fillings of interest in our 1BPV calculations.
\begin{itemize}
\item System size of $3\times 4$ (1BPV): When the spinful gaps are positive, the ratio between the spinful gaps and spin-1 gaps (former divided by latter) in average takes value of 0.36 at $\nu=-1/3$, of 0.93 at $\nu=-2/3$ and of 1 at $\nu=-4/3$ for the parameters of \refcite{reddy2023fractional}, as shown in \figref{fig:1BPV_Reddy_etal_spinful_app}(a) and \figref{fig:1BPV_Reddy_etal_app}(a). For the parameters of \refcite{wang2023fractional}, the same ratio in average takes value of 0.65 at $\nu=-1/3$, of 0.98 at $\nu=-2/3$ and of 0.71 at $\nu=-4/3$, as shown in \figref{fig:1BPV_Wang_etal_spinful_app}(a) and \figref{fig:1BPV_Wang_etal_app}(a).
\item System size of $3\times 5$ (1BPV):  When the spinful gaps are positive, the ratio between the spinful gaps and spin-1 gaps (former divided by latter) takes value of 1 at $\nu=-1/3$,  of 0.95 at $\nu=-2/3$ and of 1 at $\nu=-4/3$ for the parameters of \refcite{reddy2023fractional}, as shown in \figref{fig:1BPV_Reddy_etal_spinful_app}(b)  and \figref{fig:1BPV_Reddy_etal_app}(b). For the parameters of \refcite{wang2023fractional}, the same ratio in average take the value of 0.97 at $\nu=-1/3$, of 1 at $\nu=-2/3$ and of 1 at $\nu=-4/3$, as shown in \figref{fig:1BPV_Wang_etal_spinful_app}(b) and \figref{fig:1BPV_Wang_etal_app}(b).
\item System size of $3\times 6$ (1BPV): We choose $\theta=3.7^\circ$ and $\epsilon=5$ as the representative point for the parameter values in \refcite{reddy2023fractional}, and $\theta=3.7^\circ$ and $\epsilon\approx 16.67$ ($10/\epsilon=0.6$) for the parameters of \refcite{wang2023fractional}, as those points are in the ($\nu=-1$)-CI region shown in \figref{fig:HF_x18y18_main} of the Main Text and in the ($\nu=-2/3$)-FCI region for $4\times 6$ (Figs.\ref{fig:1BPV_Reddy_etal_app} and \ref{fig:1BPV_Wang_etal_app}). As shown by \figref{fig:Fullspingap_oneband_theta_3.7}(a) which is for the parameters in \refcite{reddy2023fractional}, the spinful gap and spin-1 gap are equal for all three fillings---they take values 0.63meV, 7.86meV and 12.84meV for $\nu=-1/3$, $\nu=-2/3$ and $\nu=-4/3$, respectively. As shown by \figref{fig:Fullspingap_oneband_theta_3.7}(b) which is for the parameters in \refcite{wang2023fractional}, the spinful gap and spin-1 gap are equal for $\nu=-1/3$ and $\nu=-2/3$, and they take values 1.31meV and 5.12meV for $\nu=-1/3$ and $\nu=-2/3$, respectively; the spinful gap and spin-1 gap are not identical but similar for $\nu=-4/3$---they take values 3.15meV and 4.28meV, respectively, and the ratio (former divided by latter) reads 0.74.
\end{itemize}
Our results on $3\times 4$ and $3\times 5$ the spin-1 gap becomes better approximation of the spinful gap as sizes grow; the trend persists to our examples on $3\times 6$, except for $\nu=-4/3$ for parameters in \refcite{wang2023fractional}.
Nevertheless, for $\nu=-4/3$ for parameters in \refcite{wang2023fractional}, the ratio between the spin-1 gap and spinful gap is still close to 1 for the largest size $3\times 6$.
Thus, for the 1BPV calculations, our results on $3\times 4$ and $3\times 5$ and our examples on $3\times 6$ clearly show that the spin-1 gap is a good approximation of the spinful gap.

\subsection{Ground states at $\nu=-1/3,-2/3,-4/3$}
\label{app:1BPV_ED_results:GS}

\begin{figure}[H]
\centering
\includegraphics[width=0.9\columnwidth]{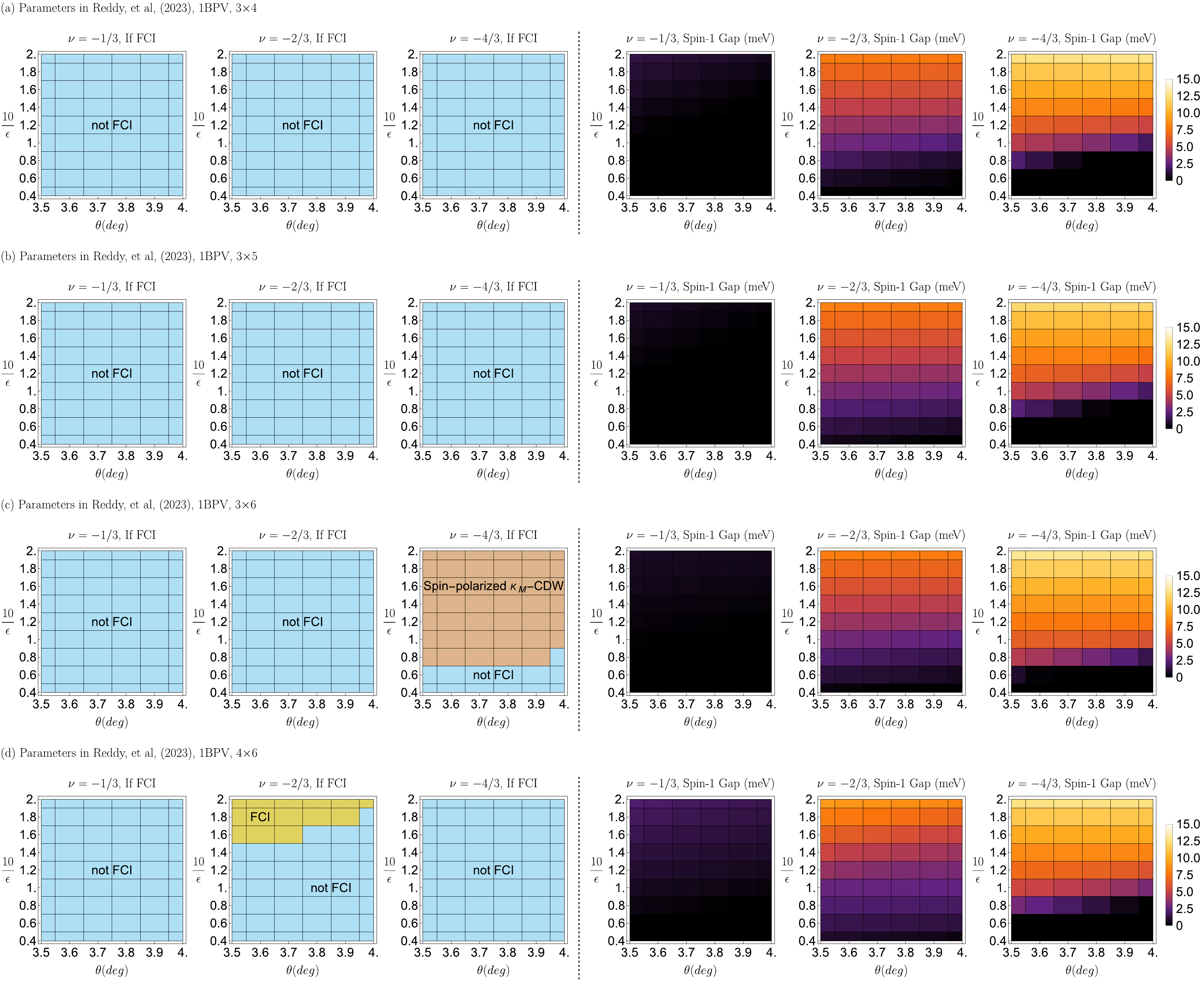}%.jpg}
\caption{
The summary of 1BPV spin-1 gaps for the parameter values of \refcite{reddy2023fractional} for the system sizes of $3\times 4$, $3\times 5$, $3\times 6$ and $4\times 6$.
In the left most three figures of each row, green (``FCI") labels the region that satisfies the criterion in \propref{prop:FCI}, and blue (``not FCI") means that we do not see clear signatures of FCI or maximally-spin-polarized CDW.
For $\nu=-4/3$ in (c) for the system size of $3\times 6$, 1BPV results show that the system is in a maximally-spin-polarized $K_M$-CDW phase for relatively large interaction (brown) and for experimental angles $\theta\in[3.5^\circ,4.0^\circ]$; the $K_M$-CDW phase is suppressed for the system sizes of $3\times 4$, $3\times 5$ and $4\times 6$ since their momentum meshes do not include the $K_M$ or $K_M'$ point.
The rightmost three figures of  each row give the spin-1 gaps, which are shown with the same color functions for all plots.
If the spin-1 gap is negative, it is set to zero in the plot.
}
\label{fig:1BPV_Reddy_etal_app}
\end{figure}

\begin{figure}[H]
\centering
\includegraphics[width=0.9\columnwidth]{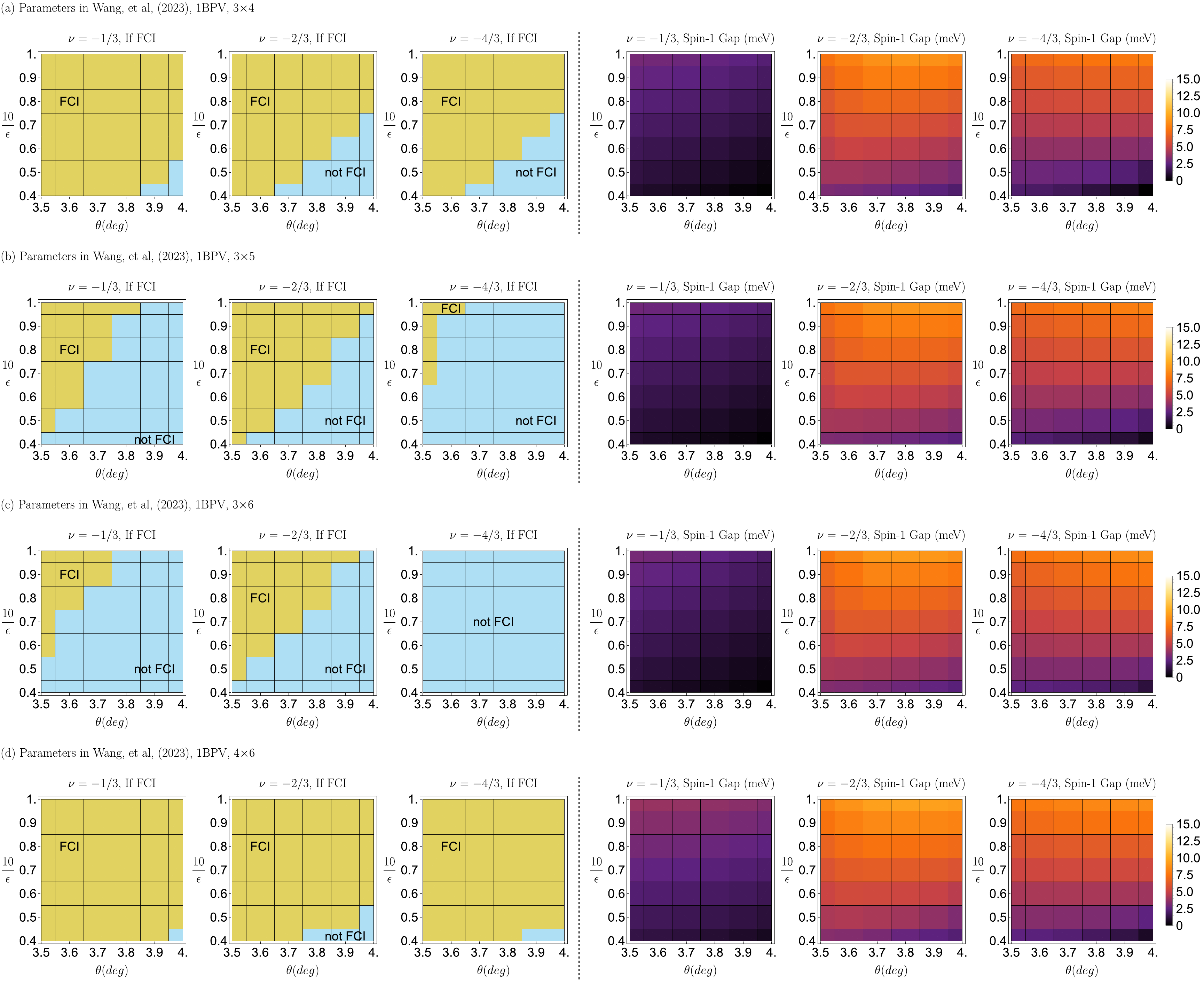}%.jpg}
\caption{
The summary of 1BPV spin-1 gaps for the parameter values of \refcite{wang2023fractional} for the system sizes of $3\times 4$, $3\times 5$, $3\times 6$ and $4\times 6$.
In the left most three figures of each row, green (``FCI") labels the region that satisfies the criterion in \propref{prop:FCI}, and blue (``not FCI") means that we do not see clear signatures of FCI or maximally-spin-polarized CDW.
The rightmost three figures of each row give the spin-1 gaps, which are shown with the same color functions for all plots.
If the spin-1 gap is negative, it is set to zero in the plot.
}
\label{fig:1BPV_Wang_etal_app}
\end{figure}

\begin{figure}[H]
\centering
\includegraphics[width=0.9\columnwidth]{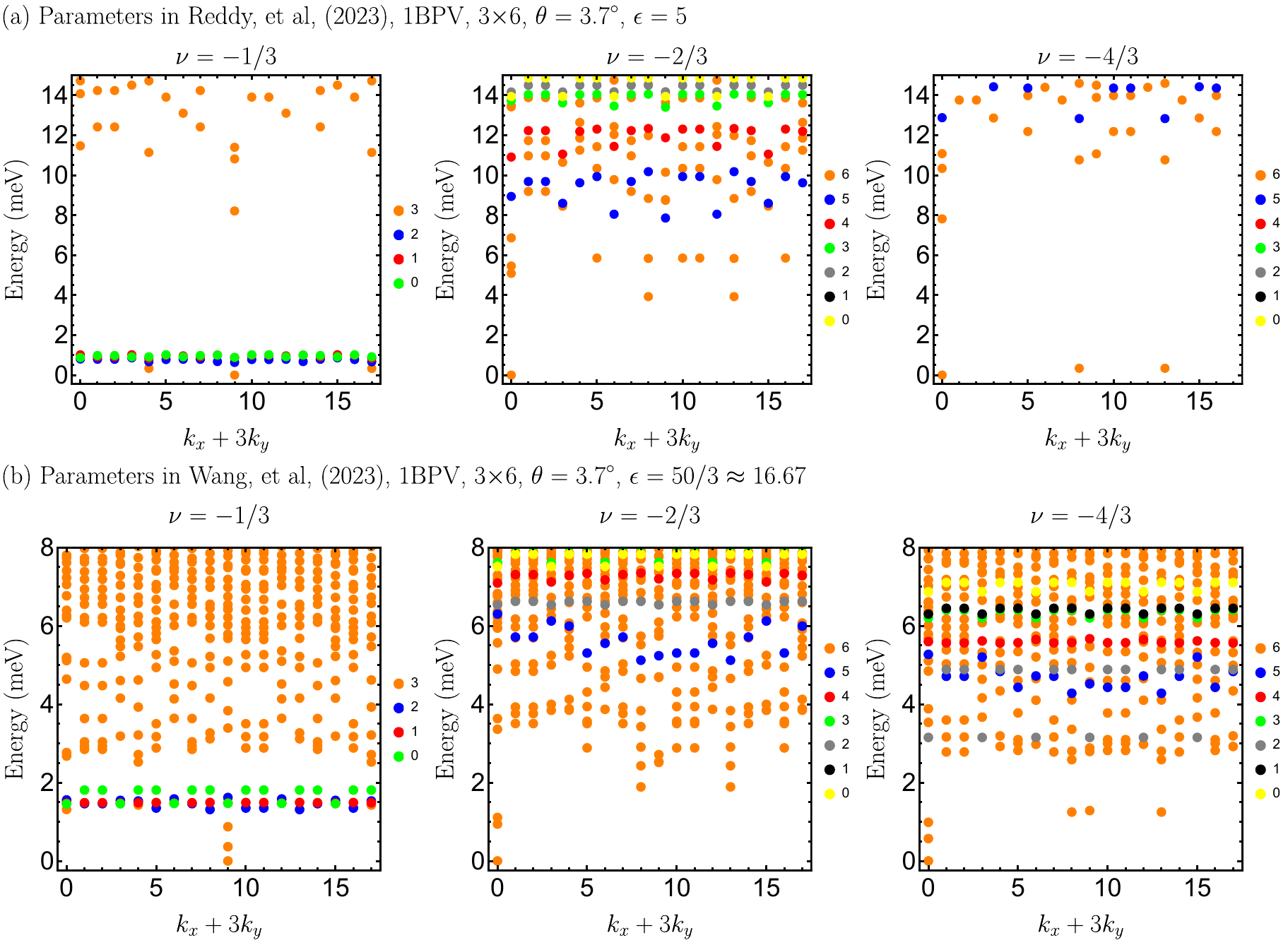}
\caption{
The 1BPV ED many-body spectrum for the system size of $3\times 6$.
The dots of different colors correspond to different spin $S_z$ sectors.
The ground state energy is chosen to be zero.
In the partially-spin-polarized sectors, we only show the lowest-energy state for each momentum and each spin, which are enough to determine the spinful and spin-1 gaps.
Here, the FCI states, if they occur, have momenta at $(k_x,k_y)=(0,3),(0,3),(0,3)$ at $\nu=-1/3$, and have momenta at $(k_x,k_y) = (0,0), (0,0), (0,0)$ for $\nu=-2/3$ and $\nu=-4/3$.
In (a), the maximally-spin-polarized sector at $\nu=-1/3$ shows a CDW with wavevector $\K_M$, since it has three ground states at $(1,1)$, $(0,3)$ and $(2,5)$ which have relative momentum differences $\pm \K_M$.
In (a), the ground states at $\nu=-4/3$ also show a maximally-spin-polarized CDW with wavevector $\K_M$, as it has three ground states at $(0,0)$, $(2,2)$ and $(1,4)$ which have relative momentum differences $\pm \K_M$
In (b), although the lowest-energy three states at $\nu=-1/3,-2/3,-4/3$ have the right FCI momentum, their spread is larger than the gap between the third lowest-energy state and the fourth lowest-energy state in the maximally-spin-polarized sector; thus they are not FCI states according to the criterion in \propref{prop:FCI}.
At $\nu=-4/3$ of (b), if we combine the third lowest-energy state at $(0,0)$ with the lowest energy states at $(2,2)$ and $(1,4)$, this combination would give us a $K_M$-CDW around energy $1$meV; however, such combination is not reliable since we have a state at $(0,3)$ that has close to $1$meV.
}
\label{fig:Fullspingap_oneband_theta_3.7}
\end{figure}

\begin{figure}[H]
\centering
\includegraphics[width=0.8\columnwidth]{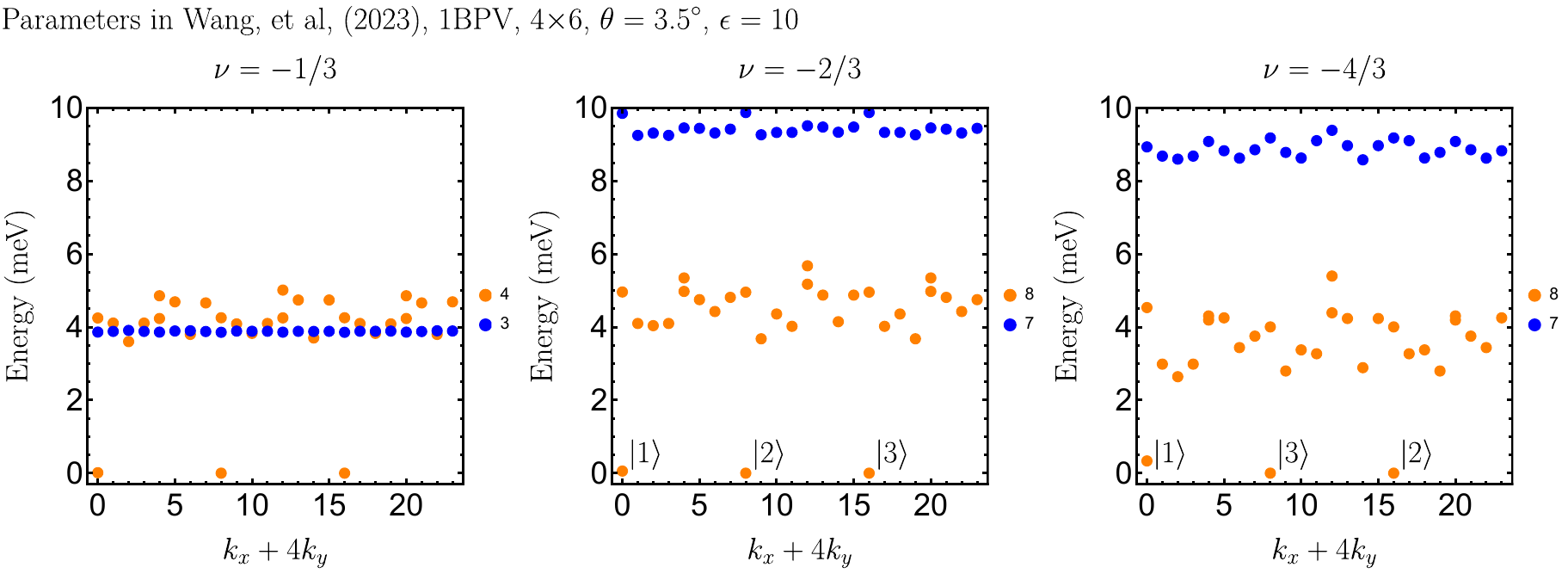}
\caption{
The 1BPV ED many-body spectrum for the system size of $4\times 6$.
The dots of different colors correspond to different spin $S_z$ sectors.
The ground state energy is chosen to be zero.
Here we only show the lowest-energy state per momentum for the sector $S_z=S_{\rm max}-1$ (using blue dots).
For the sector $S_z=S_{\rm max}$ (using orange dots), we show the lowest-energy state per momentum, except for $k_x =0$ where we show the two lowest-energy states per momentum in order to make sure there are two states per momentum for the FCI momenta $(k_x,k_y)=(0,0), (0,2), (0,4)$.
According to the criterion in \propref{prop:FCI}, FCI states exist for all three fillings for the parameters of interest.
$\ket{1}$, $\ket{2}$ and $\ket{3}$ label the three-fold degenerate ground states at $\nu=-2/3$ and $\nu=-4/3$ for the FCI, which are used in verifying the approximate PH symmetry between $\nu=-2/3$ and $\nu=-4/3$ in \eqnref{eq:1BPV_x4y6_FCI_state_overlap}.
}
\label{fig:ED_1BPV_Main_Wang_etal_app_StateOverlap}
\end{figure}

The 1BPV ED calculations for the system sizes of $3\times 4$, $3\times 5$, $3\times 6$ and $4\times 6$ are summarized in Figs.\ref{fig:1BPV_Reddy_etal_app} and \ref{fig:1BPV_Wang_etal_app} for both sets of parameters in Refs.\cite{reddy2023fractional} and \cite{wang2023fractional}.
We first discuss the ground states at $\nu=-1/3,-2/3,-4/3$.
\begin{itemize}
\item $\nu=-2/3$ (1BPV): As shown in Figs.\ref{fig:1BPV_Reddy_etal_app} and \ref{fig:1BPV_Wang_etal_app}, the finite-size effect on the FCI region at $\nu=-2/3$ is non-negligible, so we choose the largest system size $4\times 6$ to determine the $\nu=-2/3$ FCI region.
The $\nu=-2/3$ FCI region is roughly $\epsilon\in[5,6.25]$ for the parameters of \refcite{reddy2023fractional}; the FCI region at $\nu=-2/3$ basically covers the entire phase diagram in \figref{fig:1BPV_Wang_etal_app} for the parameters of \refcite{wang2023fractional}.

\item $\nu=-4/3$ (1BPV): In the parameter region where $\nu=-2/3$ hosts an FCI, the 1BPV ED calculation shows that the ground states at $\nu=-4/3$ exhibit a positive spin-1 gap for both sets of parameters in Refs.\cite{reddy2023fractional} and \cite{wang2023fractional}, indicating maximally-spin-polarized ground states. In this region, $\nu=-4/3$ is in the spin-polarized $K_M$-CDW phase for the parameter values in \refcite{reddy2023fractional},  which is indicated by the fact that we have three nearly degenerate ground states for $\nu=-4/3$ in \figref{fig:Fullspingap_oneband_theta_3.7} with a considerable gap $\sim 7 meV$ and their momentum difference is $\pm K_M= \pm (\frac{1}{3}\bsl{b}_1 +\frac{2}{3} \bsl{b}_2)$ for the system size of $3\times 6$.
It is not an FCI since the three FCI states at the system size of $3\times 6$ are all in the $(0,0)$ sector for $\nu=-4/3$, and such $K_M$-CDW phase cannot be detected for the system sizes of $3\times 4$, $3\times 5$ and $4\times 6$ as their momentum meshes do not include the $K_M$ point.
For the parameter values in \refcite{wang2023fractional}, $\nu=-4/3$ are mostly in the FCI phase for the parameters that give FCI at $\nu=-2/3$, as shown in \figref{fig:ED_1BPV_Main_Wang_etal_app_StateOverlap} for $\theta=3.5^\circ$ and $\epsilon=10$ deep in the FCI region at $\nu=-2/3$.
\item $\nu=-1/3$ (1BPV): In the parameter region where $\nu=-2/3$ hosts an FCI, the 1BPV ED calculation shows that the ground states at $\nu=-1/3$ exhibit a positive spin-1 gap for both sets of parameters in Refs.\cite{reddy2023fractional} and \cite{wang2023fractional}, indicating maximally-spin-polarized ground states.
In the parameter region where $\nu=-2/3$ hosts an FCI, if we focus on the fully-spin-polarized sector of $\nu=-1/3$, we have a $K_M$-CDW phase for the system size of $3\times 6$ for the parameters of \refcite{reddy2023fractional} (\figref{fig:Fullspingap_oneband_theta_3.7}), which is consistent with the results in \refcite{reddy2023fractional}.
However, the partially-spin-polarized sector has energies too low to solidly claim the presence of the fully-spin-polarized $K_M$-CDW as the ground state (finite size effects or disorder might change this).
For the parameter values in \refcite{wang2023fractional}, $\nu=-1/3$ is mostly in the FCI phase for the parameters that give FCI at $\nu=-2/3$, as shown in \figref{fig:ED_1BPV_Main_Wang_etal_app_StateOverlap} for $\theta=3.5^\circ$ and $\epsilon=10$ deep in the FCI region at $\nu=-2/3$, where the $\nu=-1/3$ FCI phase is consistent with the 1BPV ED calculations in \refcite{wang2023fractional}.
\end{itemize}

\begin{figure}[H]
\centering
\includegraphics[width=\columnwidth]{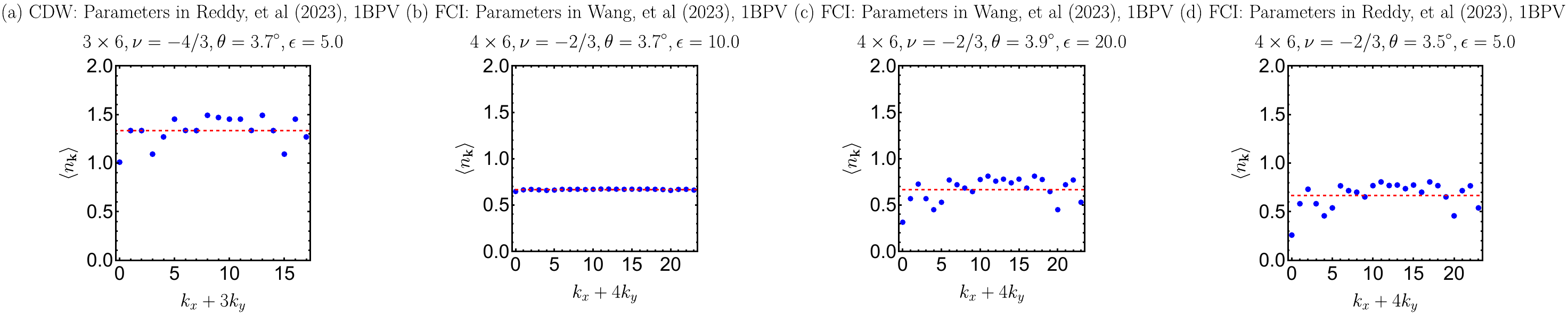}%.jpg}
\caption{
Representative plots of the $\langle n_{\bsl{k}} \rangle$ in \eqnref{eq:n_k}.
In (a), we plot it for the $K_M$-CDW states at $\nu=-4/3$ for $\theta=3.7^\circ$, $\epsilon=5$, the $3\times 6$ system size, and the parameters in \refcite{reddy2023fractional}.
In (a), we consider for $\langle n_{\bsl{k}} \rangle$ the three $K_M$-CDW states---the maximally-spin-polarized lowest-energy states at the CDW momenta $(0,0)$, $(2,2)$ and $(1,4)$  as shown in \figref{fig:Fullspingap_oneband_theta_3.7}(a).
In (b-d), we plot $\langle n_{\bsl{k}} \rangle$ for the FCI states at $\nu=-2/3$, where $\rho$ is the projection operator for the subspace spanned by three-FCI states---the maximally-spin-polarized lowest-energy states at the FCI momenta $(k_x,k_y)=(0,0),(0,2),(0,4)$.
The red dashed line mark the the mean value of $\langle n_{\bsl{k}} \rangle$, which equals to $-\nu$.
}
\label{fig:PESplotApp}
\end{figure}

We now discuss the qualities of the FCI states based on $\langle n_{\bsl{k}} \rangle$, where
\eq{
\label{eq:n_k}
\langle n_{\bsl{k}} \rangle = \frac{\Tr[\rho n_{\bsl{k}} ]}{\Tr[\rho]}\ ,
}
where $\rho$ is the projection operator for the states of interest; for three FCI states $\ket{\psi^{FCI}_i}$ with $i=1,2,3$ (such as $\ket{1}$, $\ket{2}$ and $\ket{3}$ at $\nu=-2/3$ in \figref{fig:ED_1BPV_Main_Wang_etal_app_StateOverlap}), $\rho$ reads
\eq{
\rho = \sum_{i=1}^3 \ket{\psi^{FCI}_i} \bra{\psi^{FCI}_i}\ .
}
$n_{\bsl{k}} = \sum_{n,\eta} \widetilde{\gamma}^\dagger_{\eta,n,\bsl{k}} \widetilde{\gamma}_{\eta,n,\bsl{k}}$ is the hole number operator at $\bsl{k}$ ($n$ only takes one value in this section).
As a benchmark, we show in Fig.\,\ref{fig:PESplotApp}(a) the $\langle n_{\bsl{k}} \rangle$ for the $K_M$-CDW states at $\nu=-4/3$ for $\theta=3.7^\circ$, $\epsilon=5$, the $3\times 6$ system size, and the parameters in \refcite{reddy2023fractional}.
As we can see, $\langle n_{\bsl{k}} \rangle$ the $K_M$-CDW states have considerable fluctuations with standard deviation 0.144.
In Fig.\,\ref{fig:PESplotApp}(b), we show $n_{\bsl{k}}$ of the FCI states deep in the FCI region at $\nu=-2/3$ for the $4\times 6$ system size and the parameters in \refcite{wang2023fractional}, which is very uniform with standard deviation 0.006. If we move close to the boundary of the FCI region, the fluctuations become much stronger---standard deviation increases to 0.134 as shown in Fig.\,\ref{fig:PESplotApp}(c), which is close to the fluctuations of the CDW.
As shown by Fig.\,\ref{fig:PESplotApp}(d), the quality of the FCI states at $\nu=-2/3$ for the $4\times 6$ system size and the parameters in \refcite{reddy2023fractional} (standard deviation of $n_{\bsl{k}}$ is 0.137) is as bad as those close to the boundary of the FCI region for the same size and the parameters in \refcite{wang2023fractional}.

\begin{figure}[H]
\centering
\includegraphics[width=0.5\columnwidth]{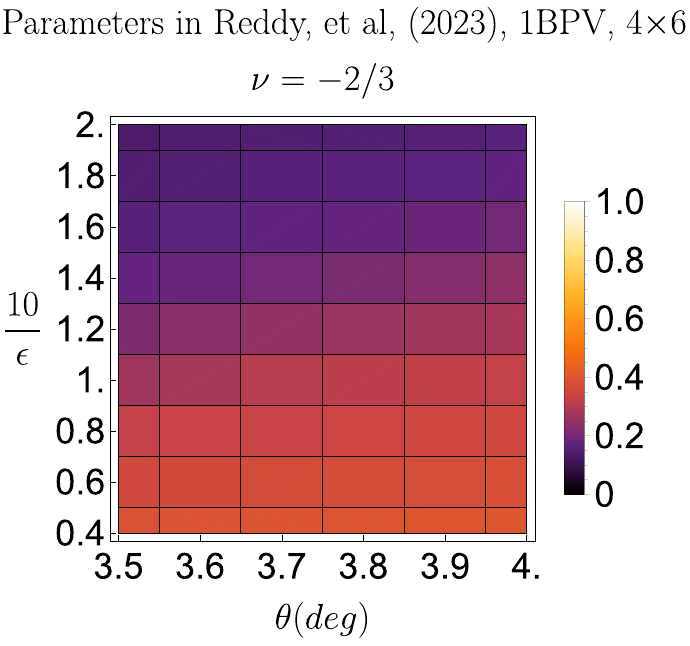}%.jpg}
\caption{
1BPV ED calcualtions of the standard deviation of the $\langle n_{\bsl{k}} \rangle$ in \eqnref{eq:n_k} for the system size of $4\times 6$ by using the parameters in \refcite{reddy2023fractional} at $\nu=-2/3$.
$\langle n_{\bsl{k}} \rangle$ is plotted for the maximally-spin-polarized lowest-energy states at the FCI momenta---$(k_x,k_y)=(0,0),(0,2),(0,4)$.
}
\label{fig:n1PESFluctuation_x4y6_1BPV_Fu}
\end{figure}

\begin{figure}[H]
\centering
\includegraphics[width=0.9\columnwidth]{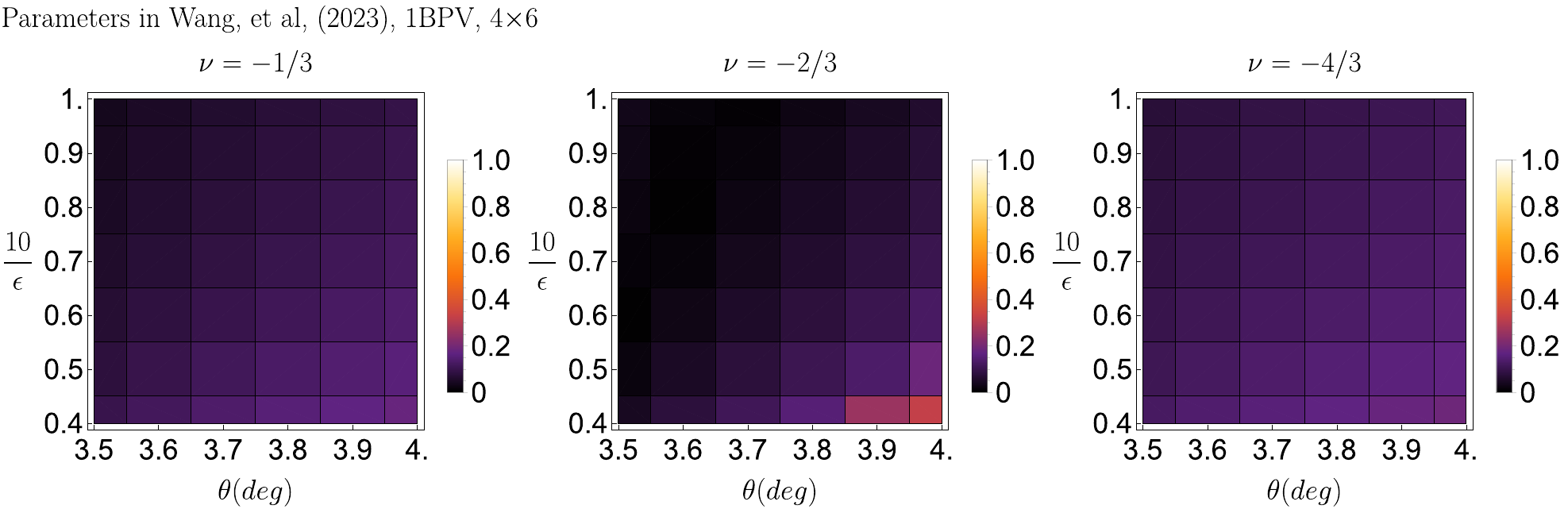}%.jpg}
\caption{
1BPV ED calcualtions of the standard deviation of the $\langle n_{\bsl{k}} \rangle$ in \eqnref{eq:n_k} for the system size of $4\times 6$ by using the parameters in \refcite{wang2023fractional}.
$\langle n_{\bsl{k}} \rangle$ is plotted for the maximally-spin-polarized lowest-energy states at the FCI momenta---$(k_x,k_y)=(0,0),(0,2),(0,4)$ at $\nu=-1/3, -2/3,-4/3$.
}
\label{fig:n1PESFluctuation_x4y6_1BPV_Xiao}
\end{figure}

We further calculate $\langle n_{\bsl{k}} \rangle$ for the maximally-spin-polarized lowest-energy states at the FCI momenta---$(k_x,k_y)=(0,0),(0,2),(0,4)$---in the phase diagrams that involve FCI at the $4\times 6$ system size as shown in Figs.\,\ref{fig:n1PESFluctuation_x4y6_1BPV_Fu} and \ref{fig:n1PESFluctuation_x4y6_1BPV_Xiao}.
As shown in \figref{fig:n1PESFluctuation_x4y6_1BPV_Fu} (compared to Fig.\,\ref{fig:1BPV_Reddy_etal_app}(d)), the standard deviation of $\langle n_{\bsl{k}} \rangle$ at $\nu=-2/3$ does minimize in the region where the three states become FCI according to the criterion in \propref{prop:FCI} for the parameters in \refcite{reddy2023fractional}; the same holds for each for the three fillings $\nu=-1/3, -2/3, -4/3$ for the parameters in \refcite{wang2023fractional} as shown in \figref{fig:n1PESFluctuation_x4y6_1BPV_Xiao}.
The FCI states at $\nu=-1/3, -2/3, -4/3$ for the parameters in \refcite{wang2023fractional} can have good quality---standard deviations of $\langle n_{\bsl{k}} \rangle$ take values in [0.037, 0.166] at $\nu=-1/3$, in [0.002, 0.134] at $\nu=-2/3$ and in [0.077, 0.181] at $\nu=-4/3$.
Yet, the FCI states at $\nu=-2/3$ for the parameters in \refcite{reddy2023fractional} has large fluctuations in $\langle n_{\bsl{k}} \rangle$  with the standard deviation in the range [0.136, 0.169], well beyond the minimum values for the FCI states for the parameters in \refcite{wang2023fractional}.

In the parameter region where $\nu=-2/3$ hosts an FCI, the ground states at $\nu=-1/3$ and $\nu=-4/3$ are maximally-spin-polarized in 1BPV calculations for both sets of parameters in Refs.\cite{reddy2023fractional} and \cite{wang2023fractional}.
Although the finite-size effects on the FCI region are substantial, the finite size effect of the spin-1 gap is small: the spin-1 gap does not change significantly for all four different system sizes.
In the following, we compare the magnetic stability of the ground states at different fillings based on the spin-1 gaps.

\subsection{Magnetic stability at $\nu=-4/3$ and $\nu=-2/3$}
\label{app:1BPV_ED_results:4third2third}

We now compare the magnetic stability of the ground states between $\nu=-4/3$ and $\nu=-2/3$ first in the 1BPV calculations.
\begin{itemize}
\item $\nu=-2/3$ versus $\nu=-4/3$ for the parameters in \refcite{reddy2023fractional} (1BPV): We find that the spin-1 gap at $\nu=-4/3$ is larger than that at $\nu=-2/3$; specifically, in the $\nu=-2/3$ FCI region, the ratio of the spin-1 gap at $\nu=-4/3$ and $\nu=-2/3$ takes values in the range [1.59,1.72] for $3\times 4$, [1.49,1.59] for $3\times 5$, [1.62,1.77] for $3\times 6$, and [1.37,1.64] for $4\times 6$.
\item $\nu=-2/3$ versus $\nu=-4/3$ for the parameters in \refcite{wang2023fractional} (1BPV): We find that the spin-1 gap at $\nu=-4/3$ is similar to that at $\nu=-2/3$; specifically, in the $\nu=-2/3$ FCI region, the ratio of the spin-1 gap at $\nu=-4/3$ and $\nu=-2/3$ takes values in the range [0.60,0.92] for $3\times 4$, [0.61,0.93] for $3\times 5$, [0.74,1.00] for $3\times 6$, and [0.71,0.93] for $4\times 6$.
\end{itemize}
These 1BPV results show that the magnetic stability at $\nu=-4/3$ cannot be significantly weaker than that at $\nu=-2/3$, which is dramatically different from the clearly nonmagnetic state at $\nu=-4/3$ and the magnetic FCI at $\nu=-2/3$ observed in the experiments~\cite{cai2023signatures,zeng2023integer,park2023observation,Xu2023FCItMoTe2}.

\begin{figure}[H]
\centering
\includegraphics[width=0.6\columnwidth]{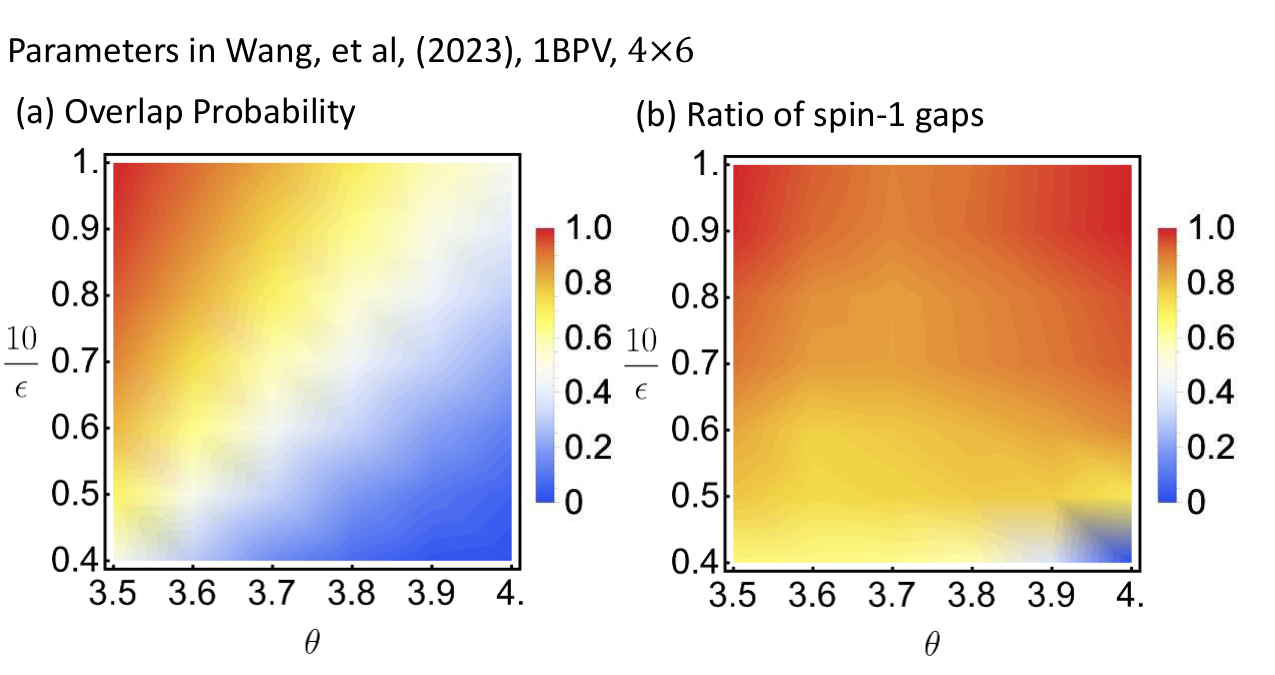}
\caption{
(a) The overlap probability of the rank-3 lowest-energy subspaces with the FCI quantum numbers between $\nu=-2/3$ and $
\nu=-4/3$ (see \eqnref{eq:1BPV_x4y6_FCI_state_overlap}) for the system size of $4\times 6$, the parameters in \refcite{wang2023fractional}, and for the 1BPV calculations.
Explicitly, the rank-3 lowest-energy subspace with the FCI quantum numbers is spanned by the lowest-energy states at the FCI momenta $(k_x,k_y)=(0,0), (0,2), (0,4)$ in the maximally-spin-polarized sector at $\nu=-2/3$ or $\nu=-4/3$, as shown in \figref{fig:ED_1BPV_Main_Wang_etal_app_StateOverlap}.
If the criterion in \propref{prop:FCI} is satisfied, the three states at $\nu=-2/3$ or $\nu=-4/3$ are in the FCI phase, which is true for most parts of the phase diagram, as shown in \figref{fig:1BPV_Wang_etal_app}(d).
(b) The ratio between the spin-1 gaps at $\nu=-4/3$ and $\nu=-2/3$ (former divided by latter) for the system size of $4\times 6$, the parameters in \refcite{wang2023fractional}, and for the 1BPV calculations.
}
\label{fig:ED_1BPV_Wang_etal_app_x4y6_PlotofProbability}
\end{figure}

In the 1BPV results, the similarity between the spin-1 gaps at $\nu=-4/3$ and $\nu=-2/3$ for the parameter values in \refcite{wang2023fractional} can be understood from the approximate PH symmetry roughly for $\theta\in [3.5^\circ,3.7^\circ]$ and for $10/\epsilon \in [0.8, 1]$.
As discussed in \secref{sec:PH_sym} and \appref{app:int:phsymmetry}, $\nu=-4/3$ and $\nu=-2/3$ is allowed to be PH partners in the 1BPV case, as we have two bands in total (PH transforms $\nu$ to $-2-\nu$).
Let us first take $\theta=3.5^\circ$ and $\epsilon=10$ as an example.
As shown in \figref{fig:ED_1BPV_Main_Wang_etal_app_StateOverlap} for $\theta=3.5^\circ$ and $\epsilon=10$, we have the extremely-similar the low-energy spectra (including the $S_z=S_{\rm max}-1$ sector) at $\nu=-2/3$ and $\nu=-4/3$, indicating the approximate (almost exact) PH symmetry.
We can be more quantitative and compute the overlap probability between the ground states at $\nu=-2/3$ and the PH-transformed ground states at $\nu=-4/3$ at the system size $4\times 6$.
We note that the ground states at the system size $4\times 6$ for $\theta=3.5^\circ$ and $\epsilon=10$ are FCI states at both $\nu=-2/3$ and $\nu=-4/3$, and the FCI momenta are $(0,0)$, $(0,2)$ and $(0,4)$ at both $\nu=-2/3$ and $\nu=-4/3$.
We label the three FCI states at $\nu$ by $i=1,2,3$, and making sure that $\ket{\nu=-2/3,i}$ and $\C \ket{\nu=-4/3,i}$ have the same momentum as shown in \figref{fig:ED_1BPV_Main_Wang_etal_app_StateOverlap}, where $\C$ is the PH transformation operator defined in \eqnref{eq:PHtransformation}.
We obtain
\eqa{
\label{eq:1BPV_x4y6_FCI_state_overlap}
& \frac{1}{3} \Tr\left[ \sum_{i=1}^{3} \ket{\nu=-2/3,i}\bra{\nu=-2/3,i} \sum_{j=1}^{3} (\C  \ket{\nu=-4/3,j})(\bra{\nu=-4/3,j} \C^{-1}) \right] \\
& =\frac{1}{3} \sum_{i=1}^{3}|\bra{\nu=-2/3,i}\C \ket{\nu=-4/3,i}|^2 = 0.95\ ,
}
which is close to 1, verifying the approximate PH symmetry.
Here we have used $\bra{\nu=-2/3,i}\C \ket{\nu=-4/3,j}=0$ for $i\neq j$ owing to the different momenta, and we also have used
\eqa{
& |\bra{\nu=-2/3,1}\C \ket{\nu=-4/3,1}|^2 = 0.96 \\
& |\bra{\nu=-2/3,2}\C \ket{\nu=-4/3,2}|^2 = 0.95 \\
& |\bra{\nu=-2/3,3}\C \ket{\nu=-4/3,3}|^2 = 0.95 \ .
}
For completeness, we give in \figref{fig:ED_1BPV_Wang_etal_app_x4y6_PlotofProbability}(a) the phase diagram indicating the overlap quantifying the approximate PH symmetry between $\nu=-2/3$ and $\nu=-4/3$ in the 1BPV calculations.
We note that the three states $\ket{\nu,1}$, $\ket{\nu,2}$ and $\ket{\nu,3}$  with $\nu=2/3$ or $\nu=-4/3$ are chosen to be the lowest-energy states at the FCI momenta $(k_x,k_y)=(0,0),(0,2),(0,4)$ in the maximally-spin-polarized sector. In the phase diagram, $\ket{\nu,1}$, $\ket{\nu,2}$ and $\ket{\nu,3}$ are not always the absolute ground states, but they become the absolute ground states and form FCI for most part of the phase diagram as shown in \figref{fig:1BPV_Wang_etal_app}(d).
In \figref{fig:ED_1BPV_Wang_etal_app_x4y6_PlotofProbability}(b), we can see extreme similarity between the spin-1 gaps at $\nu=-3/4$  and $\nu=-2/3$ in the top left corner, where the approximate PH symmetry is good.
Specifically, for $\theta\in [3.5^\circ,3.7^\circ]$ and for $10/\epsilon \in [0.8, 1]$, the overlap probability in \eqnref{eq:1BPV_x4y6_FCI_state_overlap} is no less than $0.66$, indicating the fairly good approximate PH symmetry (\figref{fig:ED_1BPV_Wang_etal_app_x4y6_PlotofProbability}(a)), and the ratio between the spin-1 gaps at $\nu=-3/4$ and $\nu=-2/3$ (former divided by latter) is no less than $0.84$ (\figref{fig:ED_1BPV_Wang_etal_app_x4y6_PlotofProbability}(b)).
Nevertheless, the extreme similarity between the spin-1 gaps at $\nu=-3/4$ and $\nu=-2/3$ in the top right corner of \figref{fig:ED_1BPV_Wang_etal_app_x4y6_PlotofProbability}(b) (\ie, $\theta$ close to $4.0^\circ$ and $10/\epsilon$ close to 1) happens when the overlap probability in \eqnref{eq:1BPV_x4y6_FCI_state_overlap} is relatively small (smaller than $\sim 0.5$), of which the explanation we leave for future work.

\subsection{Magnetic stability at $\nu=-1/3$ and $\nu=-2/3$}
\label{app:1BPV_ED_results:1third2third}

\begin{figure}[H]
\centering
\includegraphics[width=0.9\columnwidth]{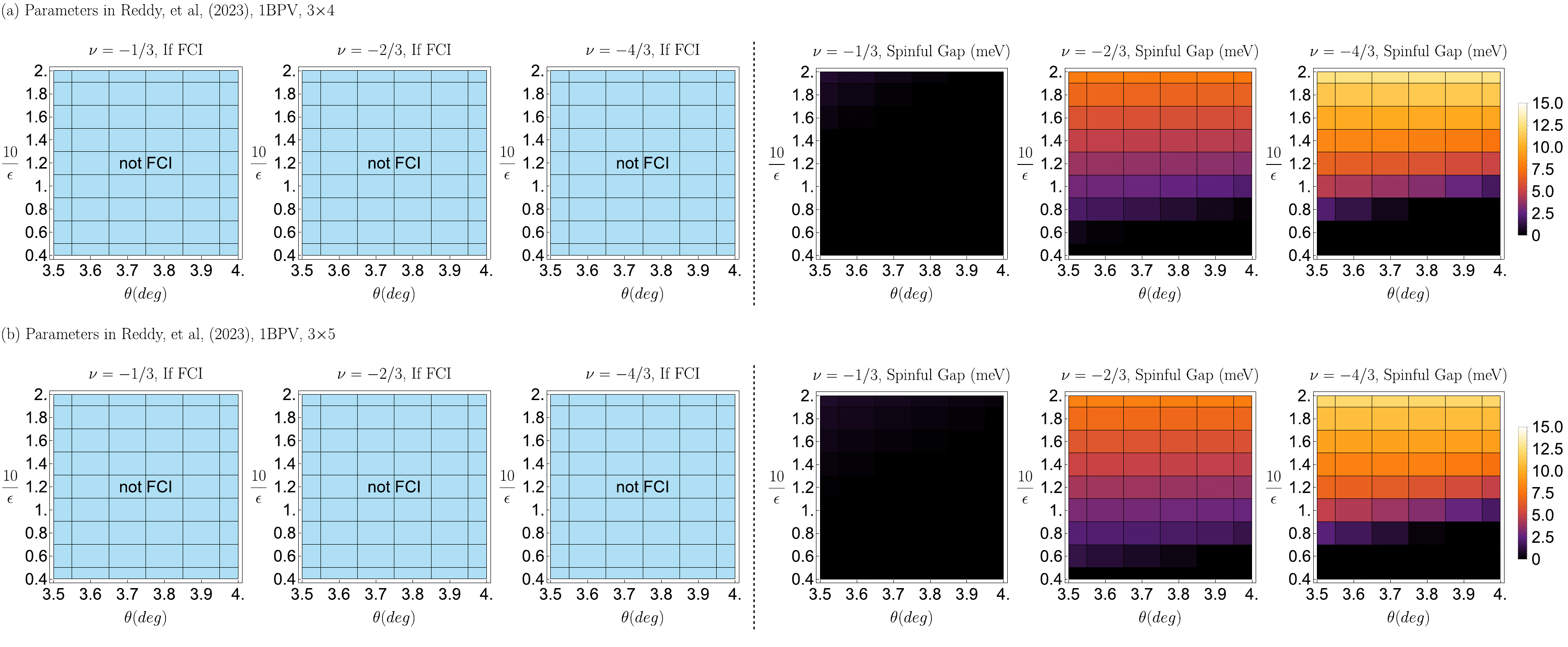}%.jpg}
\caption{
The summary of 1BPV spinful gaps for the parameter values of \refcite{reddy2023fractional} for the system sizes of $3\times 4$ and $3\times 5$.
In the left most three figures, blue (``not FCI") means that we do not see clear signatures of FCI or maximally-spin-polarized CDW.
The rightmost three figures of give the spinful gaps, which are shown with the same color functions for all plots.
If the spinful gap is negative, it is set to zero in the plot.
}
\label{fig:1BPV_Reddy_etal_spinful_app}
\end{figure}

\begin{figure}[H]
\centering
\includegraphics[width=0.9\columnwidth]{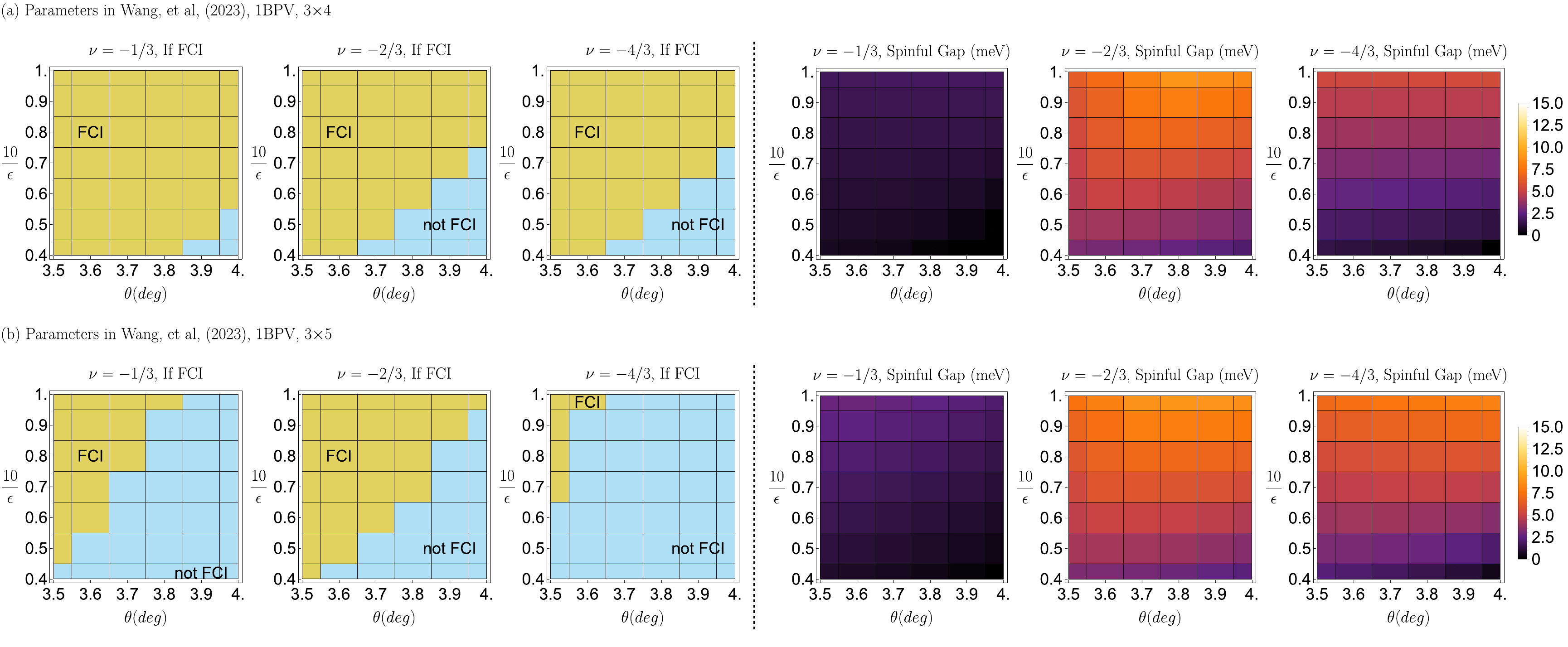}%.jpg}
\caption{
The summary of 1BPV spinful gaps for the parameter values of \refcite{wang2023fractional} for the system sizes of $3\times 4$ and $3\times 5$.
In the left most three figures of each row, green (``FCI") labels the region that satisfies the criterion in \propref{prop:FCI}, and blue (``not FCI") means that we do not see clear signatures of FCI or maximally-spin-polarized CDW.
The rightmost three figures of each row give the spinful gaps, which are shown with the same color functions for all plots.
If the spinful gap is negative, it is set to zero in the plot.
}
\label{fig:1BPV_Wang_etal_spinful_app}
\end{figure}

We now compare the magnetic stability at $\nu=-1/3$ and $\nu=-2/3$ based on the spin-1 gaps in the 1BPV calculations.
As itemized in the following, when $\nu=-2/3$ is an FCI, we find that the spin-1 gap at $\nu=-1/3$ is indeed considerably smaller than that at $\nu=-2/3$ for both sets of parameter values of Refs.\,\cite{reddy2023fractional,wang2023fractional}, though the difference between $\nu=-1/3$ and $\nu=-2/3$ is reduced at larger system sizes, compared to the numbers reported in Refs.\,\cite{reddy2023fractional,wang2023fractional} on smaller systems.
\begin{itemize}
\item $\nu=-2/3$ versus $\nu=-1/3$ for the parameters in \refcite{reddy2023fractional} (1BPV): In the $\nu=-2/3$ FCI region, the ratio between the spin-1 gap at $\nu=-1/3$ and that at $\nu=-2/3$ takes the value in the range [0.07, 0.17] for $3\times 4$, [0.02, 0.10] for $3\times 5$, [0.07, 0.09] for $3\times 6$, and [0.18, 0.23] for $4\times 6$.
\refcite{reddy2023fractional} finds an order-of-magnitude difference between the spinful gaps at $\nu=-1/3$ and $\nu=-2/3$ for the system size of 12 unit cells for $\theta=3.5^\circ$ and $\epsilon=5$, which is consistent with our $3\times 4$ spinful gaps (\figref{fig:1BPV_Reddy_etal_spinful_app}) and spin-1 gaps.

\item $\nu=-2/3$ versus $\nu=-1/3$ for the parameters in \refcite{wang2023fractional} (1BPV): In the $\nu=-2/3$ FCI region, the ratio between the spin-1 gap at $\nu=-1/3$ and that at $\nu=-2/3$ takes the value in the range [0.17,0.41] for $3\times 4$, [0.16,0.38] for $3\times 5$, [0.16,0.39] for $3\times 6$, and [0.35,0.47] for $4\times 6$.
Therefore, increasing the system size does increase the ratio between the spin-1 gaps at $\nu=-1/3$ and $\nu=-2/3$ (former divided by latter), but the effect is not significant enough to rule out the trend that the spin-1 gap at $\nu=-1/3$ is considerably smaller than that at $\nu=-2/3$.
\refcite{wang2023fractional} finds the spinful gap at $\nu=-2/3$ is about 5 times that at $\nu=-1/3$ for $\theta=3.5^\circ$, $\epsilon=15$ and the system size of $3\times 4$, which is consistent with our $3\times 4$ spinful gaps (\figref{fig:1BPV_Wang_etal_spinful_app}) and spin-1 gaps.
\end{itemize}

To summarize, the spin-1 gap 1BPV ED results indicate that the magnetic stability at $\nu=-4/3$ cannot be significantly weaker than that at $\nu=-2/3$, which is inconsistent with the experiments, though the fact that the spin-1 gap at $\nu=-1/3$ is much smaller than that at $\nu=-2/3$ has the same trend as the robust absence of ferromagnetic order at $\nu=-1/3$ in experiments (though the ground states at $\nu=-1/3$ are still maximally-spin-polarized).

\section{2BPV ED Results}
\label{app:2BPV_ED_results}

In this appendix, we provide more details on the 2BPV ED results.

\subsection{Parameters in \refcite{reddy2023fractional}}\label{app:2BPV_ED_results:fu}

First we start with the parameters of \refcite{reddy2023fractional}.
The ground states are fully-polarized at $\nu=-2/3$ in more than 70\% of the 2BPV phase diagram (\figref{fig:ED_2BPV_Main_x3y4_1third2third}(a)) on a $3 \times 4$ system, but the FCI phase is absent at $\nu=-2/3$ in the entire 2BPV phase diagram.
However, as mentioned in the Main Text and from the discussion about the 1BPV, we cannot determine whether this absence of FCI at $\nu=-2/3$ on a $3 \times 4$ system is due to the finite-size effect, since $3 \times 4$ systems do not exhibit FCIs at $\nu=-2/3$ even in the 1BPV calculations for the parameters in \refcite{reddy2023fractional}.
In 1BPV calculations, we found the $\nu=-2/3$ FCI only at $4\times 6$, not for $3\times 6$, $3\times 5$ or $3\times 4$.
2BPV calculations for $4\times 6$ at $\nu=-2/3$ are beyond our numerical accessibility, as the Hilbert space dimension per momentum at $4\times 6$ at $\nu=-2/3$ in the full-spin-polarized sector is larger than 90 billion.
In fact, for all the 1BPV and 2BPV calculations that we did with the parameters in \refcite{reddy2023fractional}, we only find FCI at $\nu=-2/3$ for the system size $4\times 6$ and the interaction strength $\epsilon\in [5,6.25]$ in 1BPV calculations.
Therefore, we will study the 2BPV results for the parameters of \refcite{reddy2023fractional} in the 1BPV $\nu=-2/3$ FCI region within $\epsilon\in [5,6.25]$.

We first discuss $\nu=-1/3$ and $\nu=-2/3$ at the system sizes of $3\times 4$ and $3\times 5$.
\begin{itemize}
\item  The spin-1 gaps are positive but the spinful gaps are negative at $\nu=-1/3$ for the $3\times 4$ system size and for $\epsilon\in [5,6.25]$, as shown in \figref{fig:ED_2BPV_m1over3_m2over3_fullspingap_spin1gap_app}(a).
The negative spinful gaps and the positive spin-1 gaps at $\nu=-1/3$ for $\epsilon\in [5,6.25]$ imply that the ground states are spin-unpolarized for the $3\times 4$ system size, as exemplified in the spectra of \figref{fig:Fullspingap_twoband_theta_3.7}(a).
This is in contrast to the spin-fully-polarized states at $\nu=-2/3$ for $\epsilon\in [5,6.25]$ and the system size $3\times 4$.

\item As a test of the finite-size effects, we calculate the spinful gap at $\nu=-1/3$ and the spinful gap at $\nu=-2/3$ for the system size of $3\times 5$ and $(\theta,10/\epsilon)=(3.7^\circ, 2.0)$ (see \figref{fig:ED_2BPV_m1over3_m2over3_fullspingap_spin1gap_app}(a)), which read $-1.18$meV and $0.75$meV, respectively.
Compared to the corresponding values $-0.39$meV and $0.56$meV for the $3\times 4$ system, we see that the difference between $-1/3$ and $-2/3$ is even stronger at a larger size, indicating different magnetic behaviors for these two filling factors.
\end{itemize}
In contrast to the different the spin-1 and spinful gaps for $\nu=-1/3$ and the system size of $3\times 4$ in first item, the spin-1 gap is equal to the spinful gap at $\nu=-2/3$ for $\epsilon\in [5,6.25]$ for the same system size.
Nevertheless, we should not compare the spin-1 gaps between $\nu=-1/3$ and $\nu=-2/3$; instead we compare the spinful gaps between $\nu=-1/3$ and $\nu=-2/3$ as shown in \figref{fig:ED_2BPV_Main_x3y4_1third2third}(a) of the Main Text.
Our findings clearly show a significant contrast in the spin polarization between $\nu=-1/3$ and $\nu=-2/3$, which is consistent with the experimentally observed difference in magnetism between $\nu=-1/3$ and $\nu=-2/3$.

\begin{figure}[H]
\centering
\includegraphics[width=0.6\columnwidth]{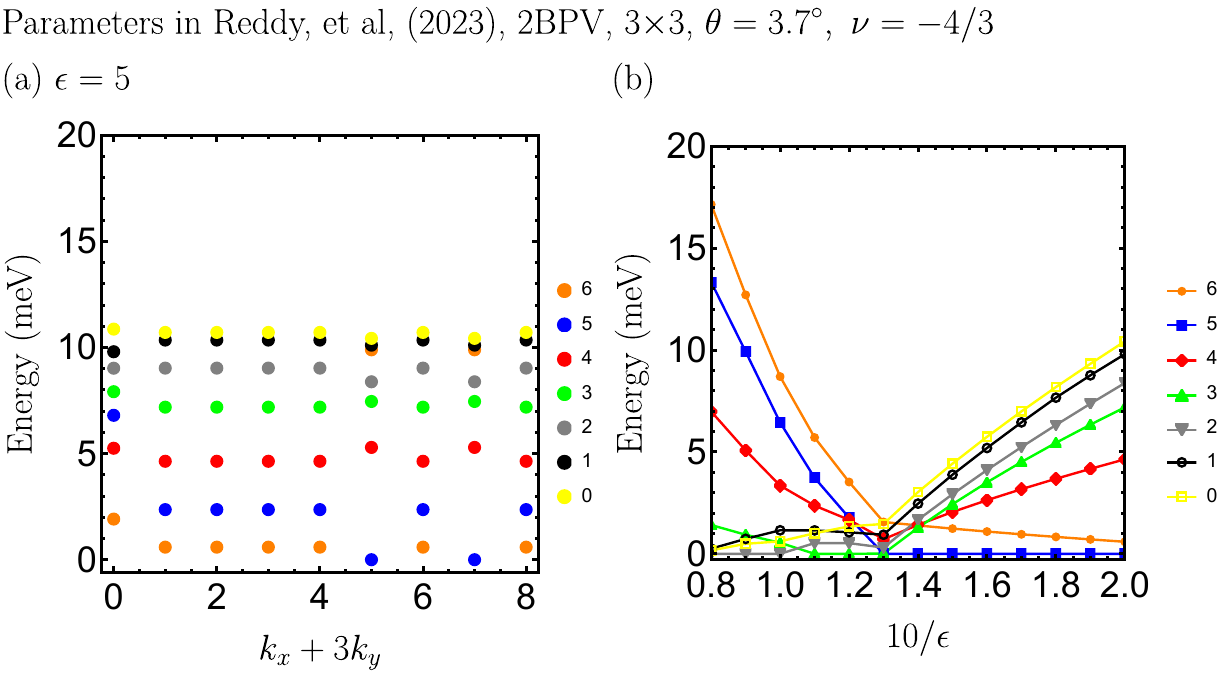}
\caption{Analogue of \figref{fig:ED_2BPV_x3y3_Wangetal_theta3p7_m4over3} in the Main Text but for the parameters in \refcite{reddy2023fractional}.
(a) The many-body 2BPV spectrum at $\nu=-4/3$ for $3\times 3$, the parameters in \refcite{reddy2023fractional}, $\epsilon=5$ and $\theta=3.7^\circ$.
We only include the lowest-energy state per momentum per partially-spin-polarized sector. For this system size, the FCI states (in the $S_{\rm max}^{\rm 2BPV}=6$ and $S_{\rm max}^{\rm 1BPV}=3$ sectors) should appear as an (almost) threefold degenerate ground state manifold at $(k_x,k_y)=(0,0)$.
According to the criterion in \propref{prop:FCI}, we cannot have FCI states here, since the ground states are not even at $(0,0)$ momentum.
The energy of the ground state is set to zero.
(b) At each value of $\epsilon$, we show the lowest energy of each spin sector for the parameters specified in the plot.
As a comparison of the spin, we note that the spin of the fully-spin-polarized state at $\nu=-2/3$ is $S_z = 3$ for the size of $3\times 3$.
The energy of the ground state is set to zero at each value of $\epsilon$.
}
\label{fig:ED_2BPV_App_Fu_m4over3_x3y3}
\end{figure}

Although there is a clear difference between $\nu=-1/3$ and $\nu=-2/3$, there is no obvious indication that the difference between $\nu=-2/3$ and $\nu=-4/3$ can be captured at the system size of $3\times 4$ or $3\times 5$ for the parameters in \refcite{reddy2023fractional}.
\begin{itemize}
\item The spin-1 gap away from the fully-spin-polarized sector at $\nu=-4/3$ is larger than the spin-1 gap at $\nu=-2/3$ for $3\times 4$ systems  (see \figref{fig:ED_2BPV_Main_m2over3_m4over3_spin1gap}(a) in the Main Text).
\item Even on system size $3\times 5$, the spin-1 gap is still much larger at $\nu=-4/3$ than that at $\nu=-2/3$, as shown in \figref{fig:ED_2BPV_app_x3y5}(a).
Specifically, at representative parameters $(\theta,\epsilon)=(3.7^\circ, 5)$, the spin-1 gaps at $\nu=-2/3,-4/3$ respectively read $0.56$meV and $2.65$meV for the system size $3\times 4$, and respectively $0.75$meV and $2.56$meV for the system size $3\times 5$, where we can see that the spin-1 gap is roughly the same for the two sizes.
\item If we consider the $S_z = S_{\rm max}^{\rm 2BPV}-2 = 8 -2 =6 $ sector at $\nu=-4/3$ for $3\times 4$, $\theta=3.7^\circ$ and $\epsilon=5$, the lowest-energy state in  the $S_z = 6 $ sector has energy higher than the lowest-energy state in the $S_z = S_{\rm max}^{\rm 2BPV} = 8$ sector by about $7$meV and higher than the lowest-energy states in the $S_z = S_{\rm max}^{\rm 2BPV} = 7$ sector by about $4.5$meV  as shown in \figref{fig:Fullspingap_twoband_theta_3.7}(a).
\end{itemize}
These findings indicate that the large-spin states might be favored.
Nevertheless, as discussed in \secref{sec:2BPV_ED} of the Main Text, the 1BPV maximally-spin-polarized state at $\nu=-4/3$ corresponds to the 2BPV partially-spin-polarized states at $\nu=-4/3$; we need to check whether such 2BPV partially-spin-polarized states at $\nu=-4/3$ (with $S_z = S_{\rm max}^{\rm 1BPV}$) are also energetically favored.
In all the three items, we have not got close to the  $S_z = S_{\rm max}^{\rm 1BPV}=4$ sector for the system size of $3\times 4$.
To fully address this question, we use $3\times 3$ systems where all sectors are numerically feasible. In this case, the finite-size effect of the spin-1 gap becomes strong (see \figref{fig:ED_2BPV_App_Fu_m4over3_x3y3}(a) for $\theta=3.7^{\circ}$ and $\epsilon=5$), which makes the spin-1 gap at $\nu=-4/3$ negative (about -0.5meV). Nevertheless, the large-spin states are still the ground states for the interaction strength $ \epsilon \in [5,6.25] $ at $\theta=3.7^\circ$ as shown in \figref{fig:ED_2BPV_App_Fu_m4over3_x3y3}(b), which is significantly different from the clear spin-zero ground states at $\nu=-1/3$ and small-spin ground states at $\nu=-4/3$ for the parameters in \refcite{wang2023fractional} discussed in the following.
In fact, the large-spin states at $\nu=-4/3$ are the ground states even for $\epsilon$ going to $\sim 7$ at $\theta=3.7^\circ$, though the smaller spin states start to become more competitive at smaller interaction (see \figref{fig:ED_2BPV_App_Fu_m4over3_x3y3}(b)).
Therefore, we do not see clear numerical evidence that the parameters in \refcite{reddy2023fractional} explain the experimental difference between $\nu=-4/3$ and $\nu=-2/3$ at the available system sizes, even if we include remote bands.

\subsection{Parameters in \refcite{wang2023fractional}}\label{app:2BPV_ED_results:xiao}

\begin{figure}[H]
\centering
\includegraphics[width=0.9\columnwidth]{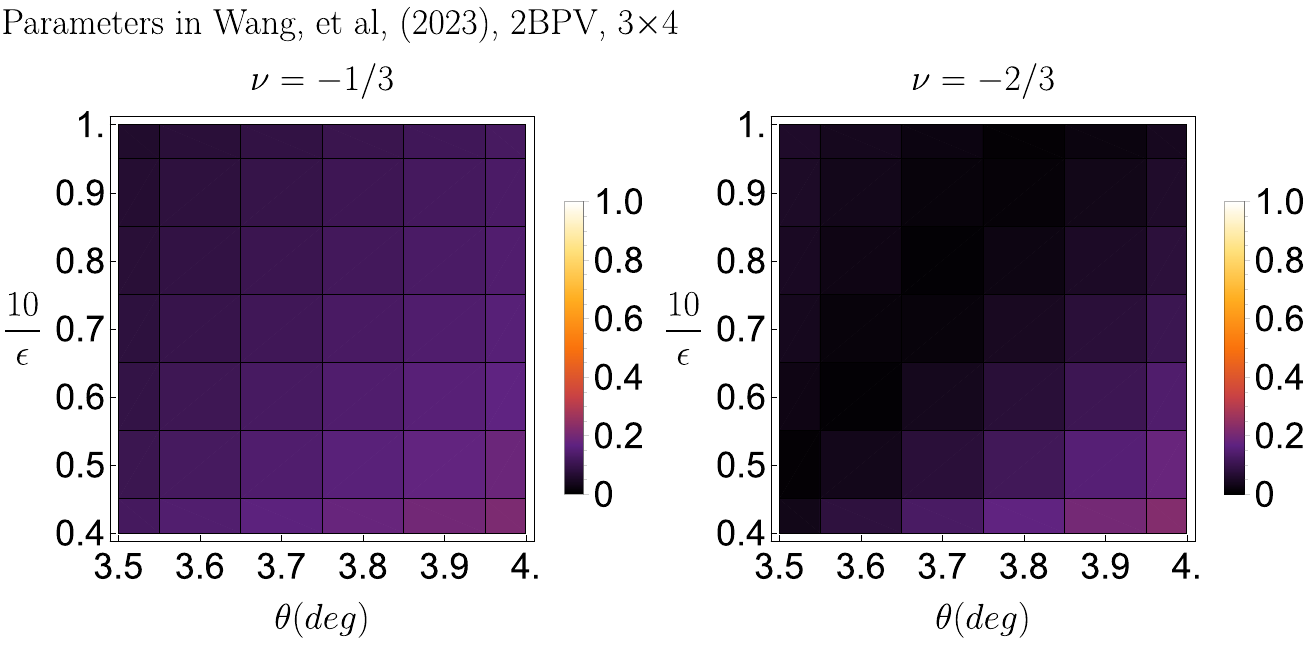}%.jpg}
\caption{
2BPV ED calcualtions of the standard deviation of the $\langle n_{\bsl{k}} \rangle$ (see \eqnref{eq:n_k}) for the parameters in \refcite{wang2023fractional}  the system size of $3\times 4$.
$\langle n_{\bsl{k}} \rangle$ is calculated for the maximally-spin-polarized lowest-energy states at the FCI momenta---$(k_x,k_y)=(0,2),(1,2),(2,2)$ at $\nu=-1/3$ and $(k_x,k_y) = (0,0), (1,0), (2,0)$ for $\nu=-2/3$.
}
\label{fig:n1PESFluctuation_x3y4_2BPV}
\end{figure}

\begin{figure}[H]
\centering
\includegraphics[width=\columnwidth]{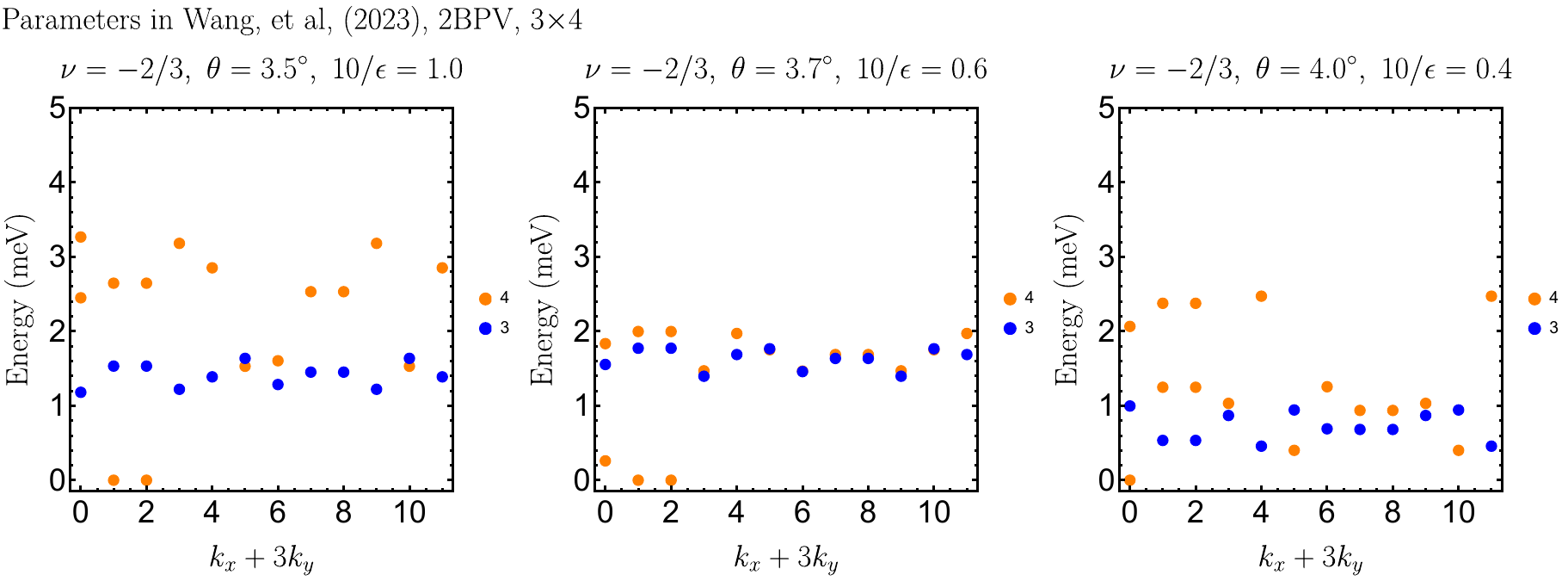}
\caption{The many-body spectrum at $3\times 4$ for 2BPV at $\nu=-2/3$ and for the parameter values of \refcite{wang2023fractional}.     The ground state energy is chosen to be zero.
We only include two spin sectors $S_z=S_{\rm max}^{\rm 2BPV}, S_{\rm max}^{\rm 2BPV}-1$ at $\nu=-2/3$, and the lowest-energy state per momentum in each spin sector except at the momenta where FCI states might occur, namely $(k_x,k_y)=(0,0),(1,0),(2,0)$.
According to the criterion in \propref{prop:FCI}, FCI states only exist in the middle plot.
}
\label{fig:ED_m2over3_Wang_etal_2BPV}
\end{figure}

We now turn to the parameters in  \refcite{wang2023fractional}.
At $\nu=-2/3$ on a $3 \times 4$ system, \figref{fig:ED_2BPV_Main_x3y4_1third2third}(b) in the Main Text shows a diagonal region hosting an FCI phase, with neighboring non-FCI phases (see \figref{fig:ED_m2over3_Wang_etal_2BPV}).
For the system size of $3\times 4$, we again evaluate the standard deviation of $\langle n_{\bsl{k}} \rangle$ (see \eqnref{eq:n_k}) for the maximally-spin-polarized lowest-energy states at the FCI momenta---$(k_x,k_y) = (0,0), (1,0), (2,0)$ for $\nu=-2/3$.
We find that standard deviation at $\nu=-2/3$ can be very small in the FCI region, taking values in $[0.006,0.104]$, which is similar to the 1BPV FCI states at $\nu=-2/3$ for the $4\times 6$ system size and for the parameters in \refcite{wang2023fractional}, as discussed in \eqnref{app:1BPV_ED_results:GS}.
As a side note, the standard deviation $\langle n_{\bsl{k}} \rangle$ at $\nu=-1/3$ minimizes in the FCI region without reaching the extremely small values obtained at $\nu=-1/3$. Indeed, the minimum standard deviation of $\langle n_{\bsl{k}} \rangle$ at $\nu=-1/3$ is 0.058 whereas that at $\nu=-2/3$ is 0.006. (\figref{fig:n1PESFluctuation_x3y4_2BPV})

\begin{figure}[H]
\centering
\includegraphics[width=\columnwidth]{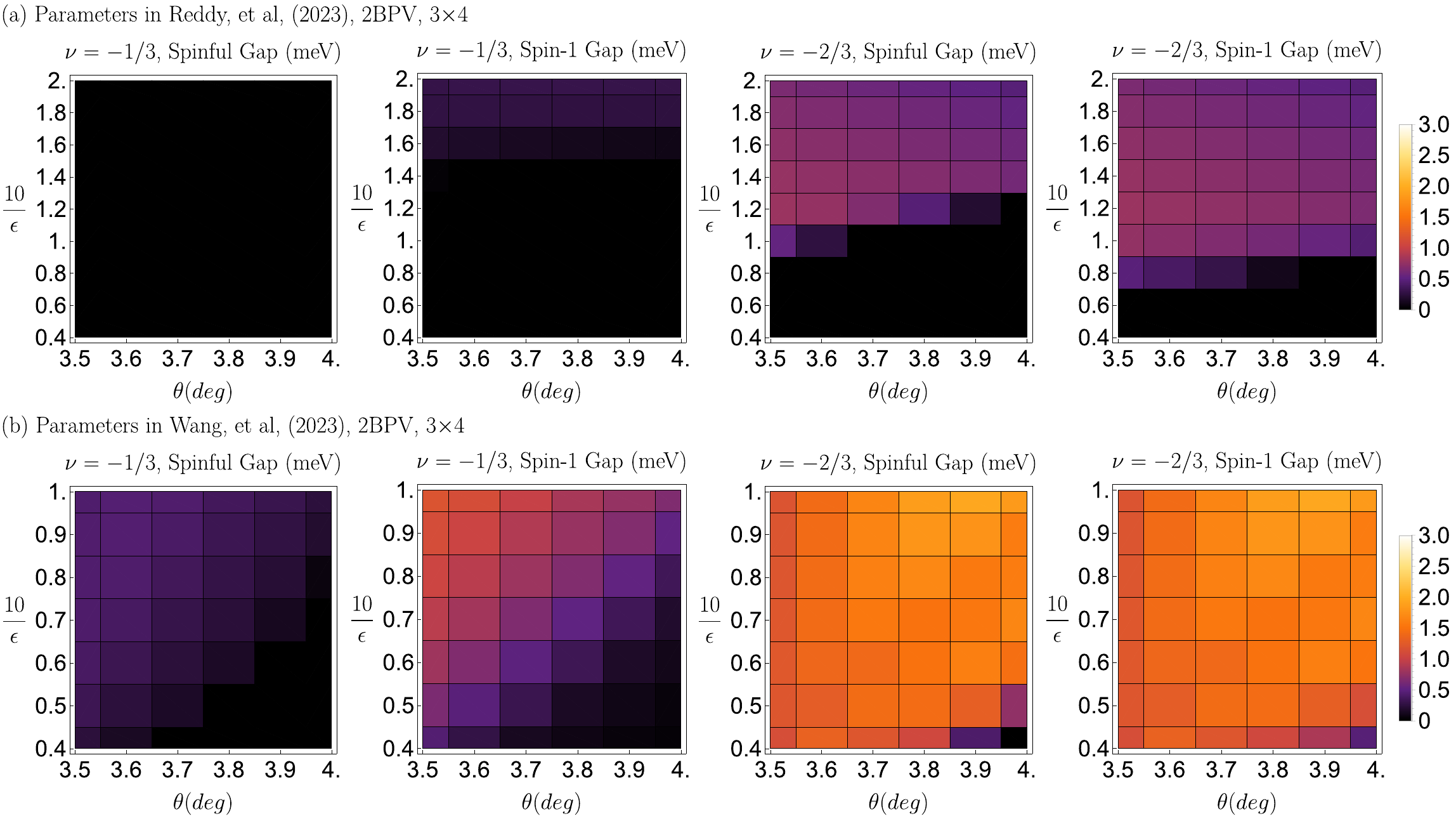}%.jpg}
\caption{
The comparison of the spinful gap and spin-1 gap for $3\times 4$ in 2D 2BPV ED calculations at $\nu=-1/3$ and $\nu=-2/3$ (a) using the parameters of \refcite{reddy2023fractional} and (b) using the parameters of \refcite{wang2023fractional}.
Spinful and spin-1 gaps are set to zero if they are negative.
}
\label{fig:ED_2BPV_m1over3_m2over3_fullspingap_spin1gap_app}
\end{figure}

We focus on the parameter region that gives an FCI phase at $\nu=-2/3$.
\begin{itemize}
\item In the $\nu=-2/3$ FCI region on a $3 \times 4$ system, the spin-1 gaps are mostly about twice the spinful gaps at $\nu=-1/3$, though the spin-1 gap is equal to the spinful gap at $\nu=-2/3$, as shown in \figref{fig:ED_2BPV_m1over3_m2over3_fullspingap_spin1gap_app}(b).
So we should use the spinful gaps (instead of spin-1 gaps) as a measure of the magnetic stability at $\nu=-1/3$ and $\nu=-2/3$ as done in the Main Text.
\item Among the 18 points in the phase diagram \figref{fig:ED_2BPV_Main_x3y4_1third2third}(b) that give FCIs at $\nu=-2/3$ at the $3\times 4$ system size, there is one point that favors spin-unpolarized (\ie, spin zero) ground states at $\nu=-1/3$.
\item For the other 17 points of the 18 points in the last item, we find fully-spin-polarized ground states at $\nu=-1/3$ (FCI states for 11 of them), but ratio between the spinful gaps at $\nu=-1/3$ and $\nu=-2/3$ (former divided by latter) takes values in $[0.03, 0.27]$ with the mean value 0.13, as shown in \tabref{tab:spinful_spin-1gap_2BPV_ED_wang_etal_x3y4}. (For these 17 points, the spin-zero state again has lower energy than the partially-spin-polarized state at $\nu=-1/3$, as exemplified in \figref{fig:ED_2BPV_m1over3_m2over3_fullspingap_spin1gap_app}(b).)
\item As a test of the finite-size effect, the spinful gaps at $\nu=-1/3$ and $\nu=-2/3$ $(\theta,10/\epsilon)=(3.7^\circ, 0.6)$ respectively read $0.22$meV and $1.40$meV for the $3\times 4$ system, where the ratio (former divided by latter) is $0.16$; the spinful gaps at $\nu=-1/3$ and $\nu=-2/3$ $(\theta,10/\epsilon)=(3.7^\circ, 0.6)$ read $0.52$meV and $1.48$meV for the $3\times 5$ system, where the ratio (former divided by latter) is $0.35$.
\end{itemize}
Therefore, when $\nu=-2/3$ features a fully spin polarized FCI, the state at $\nu=-1/3$ is either spin-unpolarized (1/18 of the region) or fully-spin-polarized with magnetic stability much weaker than that at $\nu=-2/3$ (17/18 of the region) at the $3\times 4$ system size.
The difference between the $\nu=-1/3$ and $\nu=-2/3$ spin-1 gap is decreased as the size increases (ratio changes from $0.16$ to  $0.35$), but not significant enough to eliminate the trend that $\nu=-1/3$ has much weaker magnetic stability than $\nu=-2/3$, hinting towards different magnetic behaviors at $\nu=-1/3$ and $\nu=-2/3$.
Such difference between $\nu=-1/3$ and $\nu=-2/3$ is nearly consistent with the experiments, though $\nu=-1/3$ is still mostly fully-spin-polarized.

\begin{figure}[H]
\centering
\includegraphics[width=0.6\columnwidth]{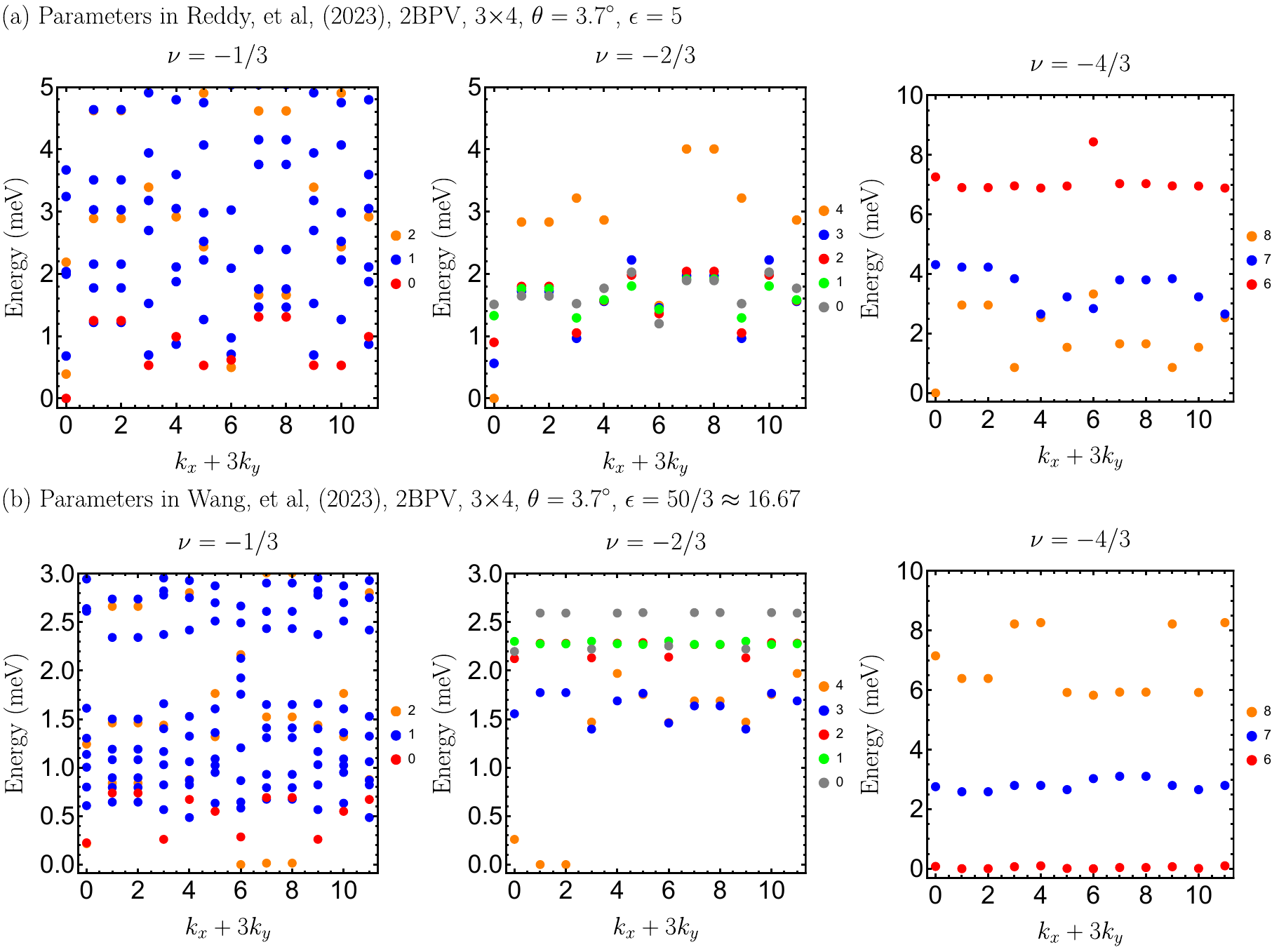}
\caption{
The many-body spectrum at $3\times 4$ for 2BPV at $\nu=-1/3,-2/3,-4/3$ (a) for the parameters of \refcite{reddy2023fractional} and (b) for the parameters of \refcite{wang2023fractional}.
For $\nu=-2/3$ and for the spin-zero sector at $\nu=-1/3$, we only show the lowest-energy state for each momentum.
For $\nu=-4/3$, we only show three spin sectors, and the lowest-energy state for each momentum and for each spin. Note that for this system size $S_{\rm max}^{\rm 2BPV}=8$ and $S_{\rm max}^{\rm 1BPV}=4$; the $S_{\rm max}^{\rm 1BPV}=4$ sector at $\nu=-4/3$ has Hilbert dimension about 2.3 billion per momentum, which is hard to address.
The ground-state energy is chosen to be zero.
Here, the FCI states, if any, should have momenta at $(k_x,k_y)=(0,2),(1,2),(2,2)$ at $\nu=-1/3$, and have momenta at $(k_x,k_y) = (0,0), (1,0), (2,0)$ for $\nu=-2/3$ and $\nu=-4/3$.
According to the criterion in \propref{prop:FCI}, the FCI states exist for both $\nu=-1/3$ and $\nu=-2/3$ in (b). However, one can clearly see the minuscule FCI gap at $\nu=-1/3$ and the large FCI gap at $\nu=-2/3$.
}
\label{fig:Fullspingap_twoband_theta_3.7}
\end{figure}

\begin{figure}[H]
\centering
\includegraphics[width=0.9\columnwidth]{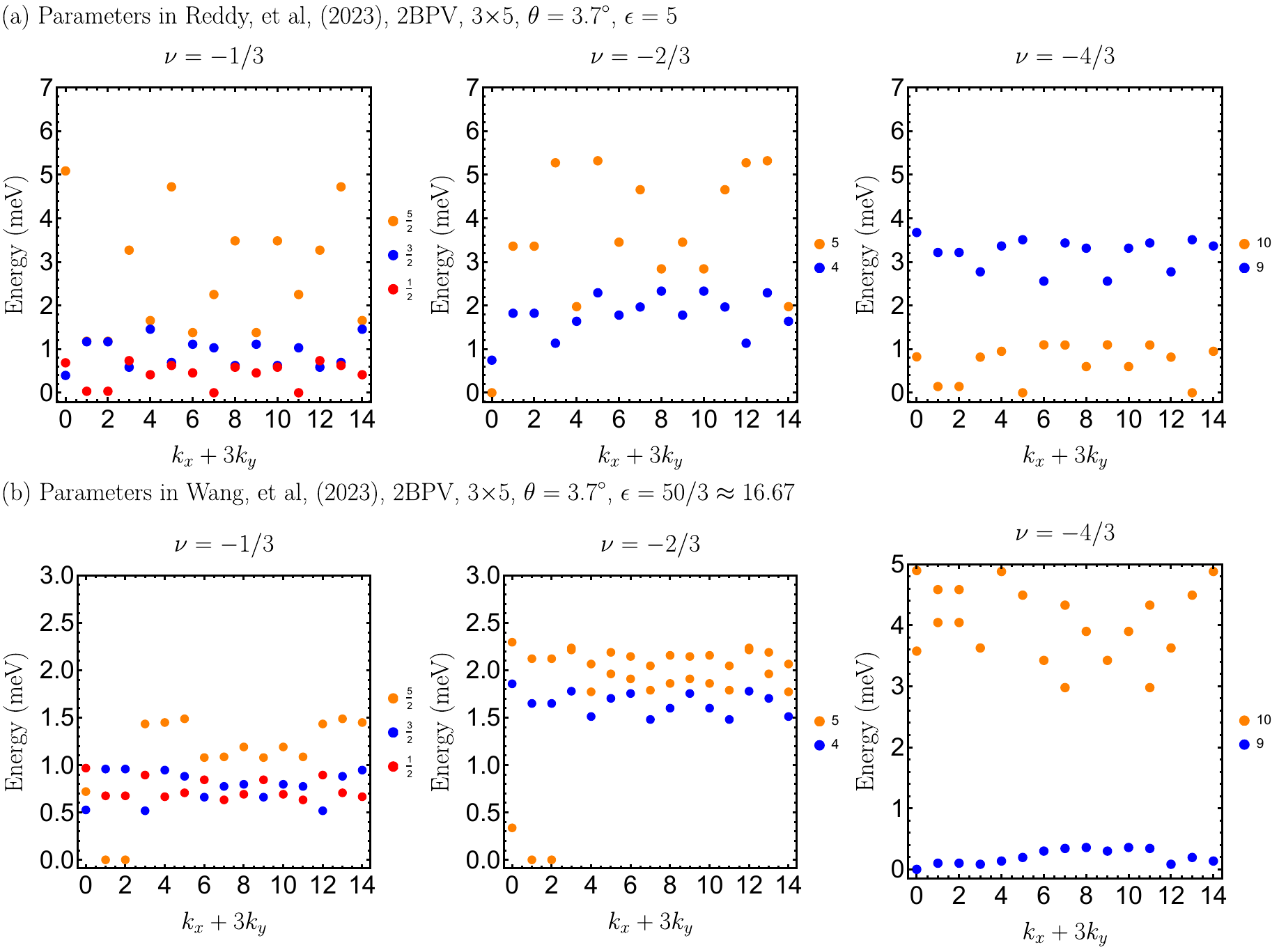}%.jpg}
\caption{The many-body spectrum at $3\times 5$ for 2BPV at $\nu=-1/3,-2/3,-4/3$ and for the parameter values specified.
We include all spin sectors at $\nu=-1/3$ and the lowest-energy state per momentum in each spin sector to obtain the spinful gap.
We only include two spin sectors $S_z=S_{\rm max}^{\rm 2BPV}, S_{\rm max}^{\rm 2BPV}-1$ at $\nu=-2/3,-4/3$ to show the spin-1 gap and the lowest energy state per momentum per spin, except that we keep two lowest-energy states per momentum in $S_z=S_{\rm max}^{\rm 2BPV}$ sector for possible FCIs in (b).
Here, the FCI states, if occur, have momenta at $(k_x,k_y)=(0,0),(1,0),(2,0)$ at $\nu=-1/3$, $\nu=-2/3$ and $\nu=-4/3$.
According to the criterion in \propref{prop:FCI}, FCI states exist for $\nu=-2/3$ in (b).
For $\nu=-1/3$ in (b), it is not an FCI state according to the criterion \propref{prop:FCI} for two reasons.
First, the lowest three states at the FCI momenta $(0,0)$, $(1,0)$ and $(2,0)$ in the fully-spin-polarized sector are not the absolute ground states, since the lowest-energy fully-spin-polarized state at $(0,0)$ has higher energy than the partially-spin-polarized state at the same momentum,  violating the combination of (i) and (ii) in the criterion \propref{prop:FCI}.
Second, the lowest three fully-spin-polarized states at the FCI momenta $(0,0)$, $(1,0)$ and $(2,0)$ for $\nu=-1/3$ have a spread larger than the gap between the 3rd lowest state and the 4th lowest state in the fully-spin-polarized sector, violating (iii) in the criterion \propref{prop:FCI}.
Therefore, even if the partially-spin-polarized states have much higher energies than the three lowest fully-spin-polarized states (as $\nu=-2/3$ in (b)), $\nu=-1/3$ in (b) is still not an FCI due to the second reason.
We note that lowest two states at momenta $(1,0)$ and $(2,0)$ are guaranteed to be exactly degenerate due to the effective inversion symmetry for $\nu=-1/3$ in (b).
}
\label{fig:ED_2BPV_app_x3y5}
\end{figure}

\begin{figure}[H]
\centering
\includegraphics[width=\columnwidth]{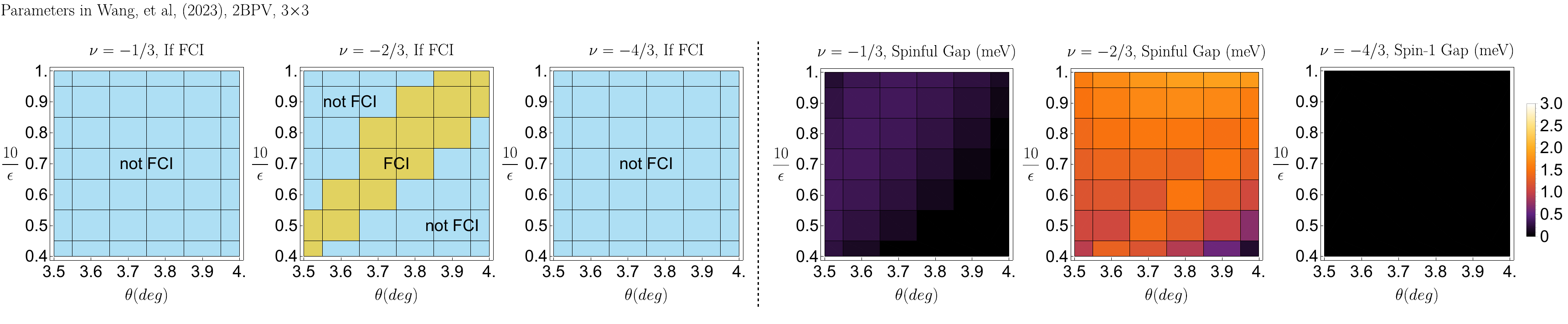}%.jpg}
\caption{
The 2BPV ED calculation done for parameter values in \refcite{wang2023fractional} on a $3 \times 3$ system.
In the left most three figures, green (``FCI") labels the region that satisfies the criterion in \propref{prop:FCI}, and blue (``not FCI") means that we do not see clear signatures of FCI or maximally-spin-polarized CDW.
The rightmost three figures give the spin-1 or spinful gaps, which are shown with the same color scale for all plots.
If the spin-1 or spinful gap is negative, it is set to zero in the plot.
}
\label{fig:ED_2BPV_App_x3y3_Wangetal}
\end{figure}

Now we discuss the difference between $\nu=-4/3$ and $\nu=-2/3$.
\begin{itemize}
\item \figref{fig:ED_2BPV_Main_m2over3_m4over3_spin1gap}(b) in the Main Text for the system size of $3\times 4$ shows that the fully-spin-polarized state is not favored at $\nu=-4/3$ for the parameter values in \refcite{wang2023fractional}, according to the negative spin-1 gaps.
\item The fact that the fully-spin-polarized state is not favored at $\nu=-4/3$ persists to the system size of $3\times 5$ at $(\theta,\epsilon)=(3.7^\circ,50/3)$ (which leads to an FCI at $\nu=-2/3$ for the experimentally relevant angle) as shown in \figref{fig:ED_2BPV_app_x3y5}(b).
\item At $\nu=-4/3$ for $3\times 4$, $\theta=3.7^\circ$ and $\epsilon=50/3\approx 16.67$, the lowest-energy state in  the $S_z = 6 $ sector has energy lower than the lowest-energy state in the $S_z = S_{\rm max}^{\rm 2BPV} = 8$ sector by about $6$meV and lower than the lowest-energy states in the $S_z = S_{\rm max}^{\rm 2BPV} = 7$ sector by about $2.5$meV as shown in \figref{fig:Fullspingap_twoband_theta_3.7}(b).
\end{itemize}
These findings indicate that the small-spin states might be favored. Still, we are away from the $S_z = S_{\rm max}^{\rm 1BPV} =4$ sector for the system size of $3 \times 4$, where a potential FCI phase similar to the 1BPV case could emerge.
Since it is technically difficult to probe lower $S_z$ sectors on a $3 \times 4$ system, we look at the smaller $3\times 3$ case. Despite its moderate size, it has a similar FCI region at $\nu=-2/3$ and similar spinful gap at $\nu=-2/3$ and similar spin-1 gap $\nu=-4/3$ as those of the system size $3\times 4$, as indicated by the comparison of \figref{fig:ED_2BPV_App_x3y3_Wangetal} and Figs.\,\ref{fig:ED_2BPV_Main_x3y4_1third2third}(b) and \ref{fig:ED_2BPV_Main_m2over3_m4over3_spin1gap}(b) in the Main Text.
The results for the system size of $3\times 3$ for $\nu=-4/3$ suggest that the partially-spin-polarized states that correspond to the 1BPV-maximally-spin-polarized states are not favored at $\nu=-4/3$; instead it is the small spin ($S_z=0,1$) states that are favored for the parameters that give FCIs at $\nu=-2/3$ (roughly $10/\epsilon=0.5\sim0.7$ at $\theta=3.7^\circ$).
Therefore, the parameters in \refcite{wang2023fractional} can capture the observed difference between $\nu=-4/3$ and $\nu=-2/3$, though this fit to the experimental phase diagram relies on a weaker interaction, namely $\epsilon > 10$.

\begin{table}[H]
\centering
\begin{tabular}{c|c|c|c|c|c|c}
$\theta$  &   10/ $\epsilon$   &\begin{tabular}{c}$\text{Spinful Gap}_{\nu=-1/3}$\\ (meV)\end{tabular}  & \begin{tabular}{c}$\text{Spin-1 Gap}_{\nu=-1/3}$\\ (meV)\end{tabular} & \begin{tabular}{c}$\text{Spinful Gap}_{\nu=-2/3}$ \\ (meV)\end{tabular}& \begin{tabular}{c}$\text{Spin-1 Gap}_{\nu=-2/3}$\\ (meV)\end{tabular} & $\frac{\text{Spinful Gap}_{\nu=-1/3}}{\text{Spinful Gap}_{\nu=-2/3}}$ \\
\hline
3.5 & 0.40 & 0.23 & 0.44 & 1.12 & 1.12 & 0.21 \\
3.5 & 0.50 & 0.32 & 0.63 & 1.20 & 1.20 & 0.27 \\
3.6 & 0.40 & 0.14 & 0.27 & 1.33 & 1.33 & 0.10 \\
3.6 & 0.50 & 0.23 & 0.47 & 1.27 & 1.27 & 0.18 \\
3.6 & 0.60 & 0.32 & 0.66 & 1.38 & 1.38 & 0.23 \\
3.7 & 0.50 & 0.15 & 0.30 & 1.42 & 1.42 & 0.10 \\
3.7 & 0.60 & 0.22 & 0.48 & 1.40 & 1.40 & 0.16 \\
3.7 & 0.70 & 0.29 & 0.65 & 1.54 & 1.54 & 0.19 \\
3.8 & 0.60 & 0.15 & 0.33 & 1.49 & 1.49 & 0.10 \\
3.8 & 0.70 & 0.22 & 0.50 & 1.49 & 1.49 & 0.15 \\
3.8 & 0.80 & 0.27 & 0.66 & 1.67 & 1.67 & 0.16 \\
3.9 & 0.70 & 0.12 & 0.34 & 1.54 & 1.54 & 0.08 \\
3.9 & 0.80 & 0.21 & 0.51 & 1.55 & 1.55 & 0.13 \\
3.9 & 0.90 & 0.26 & 0.66 & 1.77 & 1.77 & 0.15 \\
4.0 & 0.70 & -0.15 & 0.18 & 1.66 & 1.66 & -0.09 \\
4.0 & 0.80 & 0.06 & 0.34 & 1.57 & 1.57 & 0.04 \\
4.0 & 0.90 & 0.18 & 0.50 & 1.57 & 1.57 & 0.12 \\
4.0 & 1.00 & 0.23 & 0.65 & 1.83 & 1.83 & 0.13 \\
\end{tabular}
\caption{
The summary of the spin-1 gap and spinful gap values for the parameters in \refcite{wang2023fractional} and for system size $3\times 4$ at $\nu=-1/3,-2/3$ in 2BPV calculations.
We only include the angles and dielectric constant values in the phase diagram that leads to an FCI phase at $\nu=-2/3$.
}
\label{tab:spinful_spin-1gap_2BPV_ED_wang_etal_x3y4}
\end{table}

\end{document}